\PassOptionsToPackage{pdfpagelabels=false}{hyperref} 

\documentclass[preprint,authoryear,12pt]{elsarticle}  

\usepackage[T1]{fontenc}
\usepackage{placeins}

\usepackage{graphicx}	
\usepackage{amsmath}	
\usepackage{amssymb}	
\usepackage{hyperref}
\usepackage{natbib}
\usepackage{multirow} 
\usepackage{lineno}

\title{Impact Excitation of a Seismic Pulse and Vibrational Normal Modes 
on Asteroid Bennu and Associated Slumping of Regolith}

\author[a1]{Alice C. Quillen\corref{cor1}\fnref{fn1}}
\ead{alice.quillen@rochester.edu}
\author[a2]{Yuhui Zhao}
\ead{zhaoyuhui@pmo.ac.cn}
\author[a2]{YuanYuan Chen}
\ead{chenyy@pmo.ac.cn} 
\author[a3]{Paul S\'anchez}
\ead{diego.sanchez-lana@colorado.edu}
\author[a4]{Randal C. Nelson}
\ead{nelson@cs.rochester.edu}
\author[a5]{Stephen R. Schwartz$^{\rm e,}$}
\ead{srs@lpl.arizona.edu}

\cortext[cor1]{Corresponding author}
\fntext[fn1]{2017 Simons Fellow in Theoretical Physics}

\address[a1]{Department of Physics and Astronomy, University of Rochester, Rochester, NY 14627, USA}
\address[a2]{Key Laboratory of Planetary Sciences, Purple Mountain Observatory, Chinese Academy of Sciences, Nanjing 210008, China}
\address[a3]{Colorado Center for Astrodynamics Research, The University of Colorado Boulder, UCB 431, Boulder, CO 80309-0431, United States}
\address[a4]{Department of Computer Science, University of Rochester, Rochester, NY 14627, USA}
\address[a5]{Lunar and Planetary Laboratory, University of Arizona, Tucson, AZ, USA}
\address[a5]{Laboratoire Lagrange, Universit\'e C\^ote d'Azur, Observatoire de la C\^ote d'Azur, CNRS, C.S. 34229, 06304 Nice Cedex 4, France}

\begin{document}


\begin{abstract}
We consider an impact on an asteroid that is energetic enough to cause resurfacing by seismic reverberation and just below the catastrophic disruption threshold, assuming that seismic waves are not rapidly attenuated.  In asteroids with diameter less than 1 km we identify a regime where rare energetic impactors can excite seismic waves with frequencies near those of the asteroid's slowest normal modes.  In this regime, the distribution of seismic reverberation is not evenly distributed across the body surface.   With mass-spring model elastic simulations, we model impact excitation of seismic waves with a force pulse exerted on the surface and using three different asteroid shape models.  The simulations exhibit antipodal focusing and normal mode excitation.  If the impulse excited vibrational energy is long lasting, vibrations are highest at impact point, its antipode  and at high surface elevations such as an equatorial ridge.   A near equatorial impact launches a seismic impulse on a non-spherical body that can be focused  on two additional points on an the equatorial ridge.  We explore simple flow models for the morphology of vibration induced surface slumping.   We find that the initial seismic pulse is unlikely to cause large shape changes. Long lasting seismic reverberation  on Bennu caused by a near equatorial impact  could have raised the height of its equatorial ridge by a few meters and raised two peaks on it, one near impact site and the other near its antipode.  
\end{abstract}


\maketitle


\section{Introduction}

Impact induced seismic waves and associated seismic shaking can modify
the surface of an asteroid.   Impact induced seismicity is a surface modification
process that is particularly important on small asteroids due to their 
low surface gravity and small volume which limits vibrational energy dispersal
\citep{cintala78,cheng02,richardson04}.
Seismic disturbances can destabilize loose material resting on slopes, causing
downhill flows \citep{titley66,lambe79}, crater degradation and crater erasure 
\citep{richardson04,thomas05,richardson05,asphaug08,yamada16}
 and particle size segregation
or sorting \citep{miyamoto07,matsumura14,tancredi15,perera16,maurel17}.
Flat deposits at the bottom of craters on Eros known as ``ponds''  can be explained
with a seismic agitation model \citep{cheng02}, though electrostatic dust levitation may also be required
to account for their extreme flatness and fine grained composition \citep{robinson01,colwell05,richardson05}.
Regions of different crater densities on asteroid 433 Eros are explained by large impacts
that erase craters \citep{thomas05}.
Seismic shaking accounts for slides, slumps, and creep processes
on the Moon \citep{titley66}  and on Eros \citep{veverka01},  particle
size segregation or sorting on Itokawa  \citep{miyamoto07,tancredi15} 
and smoothing of initially rough ejecta
on Vesta, a process called `impact gardening' \citep{schroder14}.

During the contact-and-compression 
phase of a meteor impact, a hemispherical shock wave propagates 
away from the impact site \citep{melosh89}. 
As the shock wave attenuates, it degrades into normal stress (seismic) waves (e.g., \citealt{richardson04,richardson05,jutzi09,jutzi14}).
The seismic pulse, sometimes called `seismic jolt' \citep{nolan92} or `global jolt' \citep{greenberg94,greenberg96},
travels as a pressure wave  through the body.  
Seismic agitation is most severe  
nearest the impact site and along the shortest radial paths through the body
because of radial divergence and attenuation of the seismic pulse
\citep{thomas05,asphaug08}.
After the seismic energy has dispersed through the asteroid, continued
seismic shaking or reverberation of the entire asteroid 
may continue to modify the surface \citep{richardson05}.
Small impacts could excite seismic waves that overlap in time as they attenuate, causing 
 continuous seismic noise known as `seismic hum' \citep{lognonne09}.
The regime that is important for a particular impact and surface modification 
process depends on the frequency spectrum of seismic waves
launched by the impact and the frequency dependent attenuation, scattering and wave speed of the seismic waves 
\citep{richardson05,michel09}.

The size of an impactor that catastrophically disrupts an asteroid exceeds by a few orders
of magnitude the size of one that causes
sufficient seismic shaking to erase craters \citep{richardson05,asphaug08}.
In a rarer event, an asteroid could be hit by a projectile smaller than the disruption threshold
but large enough to  significantly shake  the body.
It is this regime that we consider here.    We explore the nature
of shape changes caused by vibrational oscillations 
excited by a subcatastrophic impact. 
With the imminent arrival in 2018 of the OSIRIS-REx mission
at  Bennu  (Asteroid 101955), we investigate
the possibility that the unusual shape of
Bennu's equatorial ridge  is due to a energetic but subcatastrophic impact.

Asteroids are often modeled as either fractured monoliths or rubble piles. 
In dry granular media on Earth,  pressure waves propagate through particles and from  particle to particle through a network of
contact points, called `force chains'  \citep{cundall79,ouaguenouni97,geng01,clark12}.  
Laboratory experiments find that
the elastic wave speed tends to scale with the classic speed $\sqrt{E/\rho}$ where $E$ is the Young's modulus and $\rho$ is the density,
and is not usually dependent on the particle size, but is weakly dependent on the constraining pressure and porosity \citep{duffy57}.
Thus a continuum elastic material model can approximate the seismic behavior of granular materials 
and has been used to model seismicity in asteroids (e.g., \citealt{murdoch17}).
However, if the  force chains are dependent on gravitational acceleration, terrestrial and lunar granular
materials may not provide good analogs for asteroids which have low surface gravity. 
Rubble pile asteroids may have a small, but finite, level of tensile strength \citep{richardson09} 
due to van der Waals forces between fine particulate material 
\citep{sanchez14,scheeres18}, so both compressive and tensile restoration forces
may be present allowing seismic waves to reflect.   Even without cohesion,
contacts under pressure allow seismic waves to propagate (e.g., \citealt{sanchez11,tancredi12}).  Ballistic contacts
also  allow a pressure pulse to propagate,  as illustrated by the classic toy known as Newton's cradle.  

Unfortunately, little is currently known about how seismic waves are
dispersed, attenuated and scattered in asteroids.  The rapidly attenuated seismic pulse or jolt model  
 \citep{thomas05} is 
consistent with strong attenuation in laboratory granular materials at kHz frequencies \citep{odonovan16}, but
qualitatively differs from
the slowly attenuating seismic reverberation model \citep{cintala78,cheng02,richardson04,richardson05}, that is supported
by measurements of slow seismic attenuation in lunar regolith \citep{dainty74,toksoz74,nakamura76}.
While both seismic jolt and reverberation processes can cause crater erasure and rim degradation, 
size  segregation induced by the Brazil nut effect relies on reverberation 
(e.g., \citealt{miyamoto07,tancredi12,matsumura14,tancredi15,perera16,maurel17}).
 
Despite the poorly constrained seismic wave transport behavior in asteroids,  
a linear elastic material simulation model may  describe 
 the propagation of impact generated seismic waves (e.g., \citealt{murdoch17}).  To model the propagation of 
seismic waves, we use the mass-spring and N-body elastic body model we have developed
to study tidal and spin evolution of viscoelastic bodies 
\citep{quillen16_crust,frouard16,quillen16_haumea,quillen17_pluto}.   
As shown by \citet{kot15}, in the limit of large numbers of randomly distributed mass nodes and an interconnected spring
network comprised of  at least 15 springs per node,
 the mass spring model approximates an isotropic continuum elastic solid.
We use  our mass-spring model  to examine the surface distribution of vibrational energy
excited by an impact.  In this respect we go beyond previous works which have primarily focused
on simulation of rubble piles (e.g., \citealt{walsh12,holsapple13,schwartz13,schwartz14,perera16}) and transmission of seismic pulses 
in planar sheets or spherical bodies (e.g., \citealt{tancredi12,murdoch17}).

In the following section (\S \ref{sec:bennu}), 
we summarize properties of Bennu.  
Normalized units for the problem are discussed in section \S \ref{sec:units}
and we estimate frequencies for its vibrational normal modes (section \S \ref{sec:normalmodes}).
In section \S \ref{sec:seismic} we discuss excitation of seismic waves by an impact.
We identify a regime  where low frequency normal modes are likely to be excited by an impact.
In section \S \ref{sec:sims} we describe our mass/spring model simulations.
We simulate impacts 
 by applying a force pulse to the simulated asteroid surface and 
the strength and duration of the applied force pulse are estimated from scaling relations described
in section \S \ref{sec:seismic}.
Normal modes are identified in the spectrum of the vibrationally excited body.
We examine the pattern of vibrational kinetic energy on the surface
for three different shape models.
In section \S \ref{sec:vibflow} we explore 
how impact excited seismic vibrations could induce granular flows on Bennu's surface.
A summary and discussion follows in \S \ref{sec:sum}.


\begin{figure}
\includegraphics[width=\columnwidth]{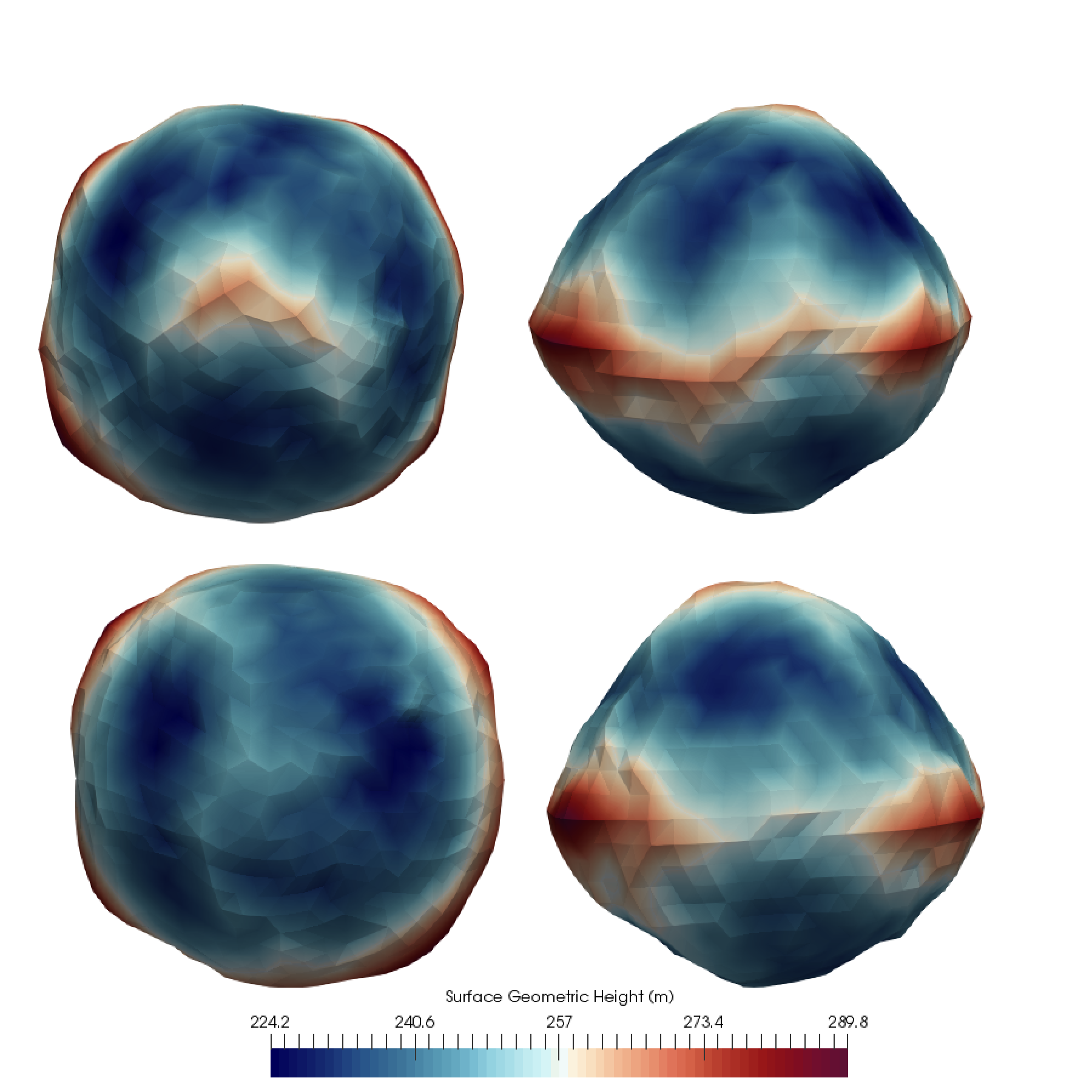}
\caption{The preliminary  Bennu 
shape model by \citet{nolan13} is shown in orthographic (perspective) projections. 
Colors show the geometric height
or radius from geometric center.   The left two panels show polar views and the right
two panels show views with the polar axis on the top.
The polar views seen on the upper and lower left panels show that
the equatorial ridge has a square-like shape.}
\label{fig:shape1}
\end{figure}

\begin{figure}
\includegraphics[width=\columnwidth, trim=20 0 10 10, clip]{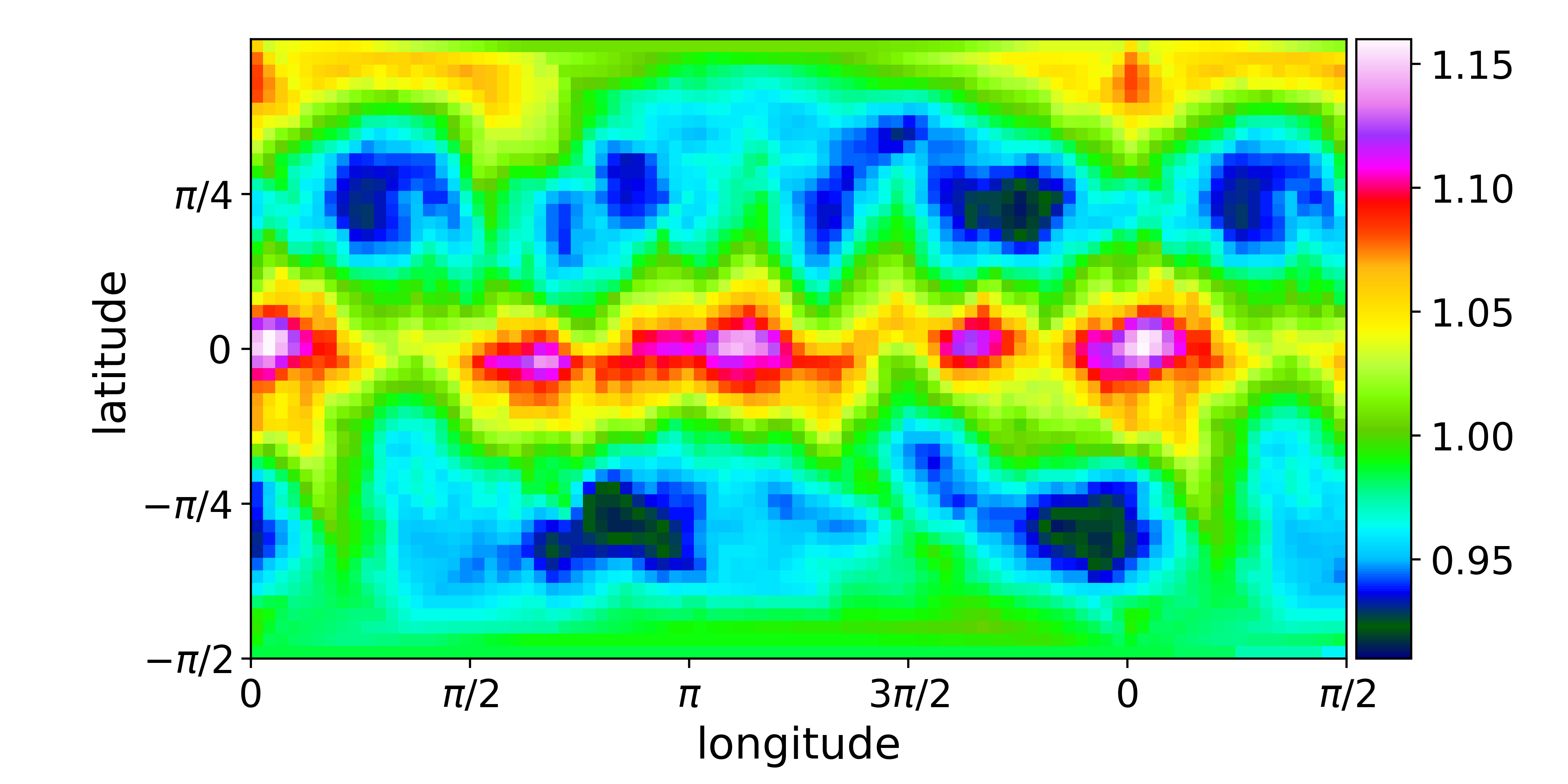}
\caption{Surface geometric height (distance from center) of the Bennu shape model
is shown as a function of latitude and longitude. The colorbar
is in units of the mean equatorial radius.  
The longitude extends past $2 \pi$ so the right side
of the figure repeats the region between 0 and $\pi/2$.  
We do this so that the equatorial peak near 0 is easier to see.
There are four peaks near the equator.  Each peak has an antipodal counterpart.
The equatorial ridge deviates from the equatorial plane, 
going above and below $0^\circ$ latitude by a few degrees.
}
\label{fig:shape2}
\end{figure}

\section{Bennu}
\label{sec:bennu}

The OSIRIS-REx mission, launched in 2016, \citep{lauretta17} to Bennu (Asteroid 101955), 
aims to fire
a jet of high-purity nitrogen gas onto Bennu's surface so as to excite at least 60 g of regolith 
that can be collected and returned to Earth. 
A C-complex asteroid, Bennu is interesting due to its primitive nature.
Spectroscopic measurements are consistent with CM-carbonaceous-chondrite-like material
and Bennu's thermal inertia implies that its surface 
supports a regolith comprised of sub-cm-sized grains \citep{emery14}.
A suite of remote sensing observations during 2018 and 2019 will be used to create
a series of global maps to characterize the geology, mineralogy, surface processes, 
and dynamic state of Bennu.  
These maps will also be used to choose a sample selection site and 
place the returned sample in geological context.
Measurements from the OSIRIS-REx rendezvous
will  be used to test theories for the formation of Bennu's equatorial ridge
\citep{scheeres16}.

Ground-based radar has been used to characterize Bennu's shape, spin state, and surface roughness 
\citep{nolan13}.
Bennu's shape is nearly spherical but like 
1999 KW4  \citep{scheeres06}, Bennu has an equatorial ridge. 
The shape model consists of 1148 vertices, with a spacing of 25 m between vertices and two subdivided regions 
with additional resolution where there are protruding features that could be boulders. 
The shape model was generated by using the vertex locations, estimated Doppler broadening and the  radar scattering function,
to match  the radar images and enforced a uniform mass distribution and principal-axis rotation \citep{nolan13}.
In Figure \ref{fig:shape1} and \ref{fig:shape2} we show the geometric height (distance from geometric
center of body).  Figure \ref{fig:shape1} shows 4 orthographic projections. 
The left two panels show polar views and the right two panels show equatorial views.   
The polar axis is  aligned with the axis of rotation.
We use  latitude $\lambda$ and longitude $\phi$ angles of the body
with respect to the center of mass and spin axis, which we assume is 
aligned with a principal body axis (following \citealt{nolan13}).  
The body spin and higher equatorial elevation reduce the radial acceleration
on the equatorial ridge.  The acceleration is only about 30 $\mu{\rm m/s}^2$
on the equatorial ridge, as compared to 85 $\mu {\rm m/s}^2$ on the poles 
(see Figure 11 by \citealt{scheeres16}).
Vibrational excitations on the equator are most likely to modify the surface  or change
the shape of the asteroid.

The individual radar images by \citet{nolan13} show that the asteroid shape significantly deviates from a sphere and the body's equatorial
ridge has some smaller scale structure.  Individual radar images show significant deviations from a smooth curve that
would be consistent with an ellipsoid, so peaks on
Bennu's equatorial ridge could be real and not due to noise in the radar images or uncertainties in the shape model.   
A polar view  shows that Bennu's equator has a slightly square shape,  (see the 
 the left two plots in figure \ref{fig:shape1}).  
The shape model  exhibits four equatorial peaks,  and each peak
has an antipodal counterpart  on the opposite side of the equatorial ridge. 
The peak  longitudes are approximately  20$^\circ$, 110$^\circ$, 190$^\circ$ and 290$^\circ$.
The four peaks are not exactly separated by $90^\circ$ and the equatorial ridge 
is not exactly equatorial as it deviates by a few degrees in latitude above and below the equatorial plane.
Unfortunately the ground-based shape model may not be sufficiently constrained or unique to know whether Bennu's equatorial ridge
is geometrically suggestive of a periodic process, like Pan's sculpted 
ravioli-shaped equatorial ridge\footnote{\url{https://photojournal.jpl.nasa.gov/catalog/PIA21436}}.
With the imminent arrival of OSIRIS-REx at Bennu, early Dec. 2018, we should soon
have a high resolution imaging of Bennu to improve upon this shape model.

We summarize theories of Bennu's shape evolution following \citet{scheeres16}.
The equatorial ridge may have formed through landslides of surface regolith
flowing down to the equatorial region which, because
of body spin, is a geopotential low \citep{guibout03,scheeres06,walsh08,walsh12,hirabayashi15a}.
Another mechanism for formation of the equatorial ridge involves infall of material
that was fissioned off the parent body due to a binary asteroid interaction
\citep{jacobson11}.
Alternatively, the interior could undergo a plastic deformation that
propagates outwards along the equatorial plane of the body \citep{hirabayashi15b}.
None of these theories propose an explanation for a possible four pointed structure
on the equatorial ridge.   This motivates us to explore impact associated shape changes.


\begin{table*}
\vbox to 85mm
{
	\centering
	\caption{\large Physical Quantities for Asteroid Bennu}
	\label{tab:bennu}
\begin{tabular}{llll} 
\hline
    & & physical & N-body  \\
\hline
Asteroid Mass 	         &	$M$         & $7.8 \times 10^{10}$ kg & 1\\
Mean equatorial radius & $ R_{eq} $  &   246 m  & $\approx 1$  \\
Radius of equivalent volume sphere & $R_{\rm vol}$  & $\approx R_{eq}$ & 1\\
Average Density        & $\rho$     & 1260 kg m$^{-3}$ & ${3}/{(4\pi)}$ \\
Spin Rotation period & $P_{\rm rot}$  & 4.2978 hours & 9.18 $t_{\rm grav} $\\
Gravitational timescale & $t_{\rm grav} $ & 1685 s & 1 \\
Spin angular rotation rate &  $\Omega = 2\pi/P_{\rm rot}$  & $4.06 \times 10^{-4}$ s$^{-1}$ & 0.68 $t_{\rm grav}^{-1}$ \\ 
Energy density & $e_{\rm grav}$  &  110 Pa & 1 \\
Young's modulus  & $E$  & 11.3 MPa & $ 10^5 e_{\rm grav}$ \\
Gravitational speed     & $V_{\rm grav}$ & 0.146 m s$^{-1}$ & 1 \\
P-wave speed     & $V_p$ & 104 m s$^{-1}$ & 712 $V_{\rm grav}$\\
Characteristic frequency       & $f_{\rm char} = V_p/R_{\rm vol} $  & 0.4 Hz & 712 $t_{\rm grav}^{-1}$ \\
\hline
\multicolumn{4}{l}
{\multirow{7}{400pt} 
{\small Notes:    The third column gives quantities in physical units.
The fourth column gives quantities in N-body or gravitational units.
The mass, mean equatorial radius and density,
and spin rotation period
are those by \citet{chesley14}.  The energy density $e_{\rm grav}$ is computed via equation \ref{eqn:eg}
using the asteroid mass and mean equatorial radius.  
The gravitational timescale $t_{\rm grav}$ is computed with equation \ref{eqn:tgrav}.
The gravitational velocity $V_{\rm grav}$ is computed with equation \ref{eqn:Vgrav}.
The characteristic frequency $f_{\rm char} = V_p/R_{\rm vol}$.
The P-wave speed
is that by \citet{cooper74}  for lunar surface regolith and the Young's modulus is estimated
from this, the estimated density of Bennu and for a Poisson ratio $\nu=1/4$.
}}
\end{tabular}}
\end{table*}

\subsection{Sizes,  Units and  Coordinates}
\label{sec:units}

Measurement of the Yarkovsky force gave an estimate for Bennu's bulk density of 
$1260\pm 70$ kg/m$^3$, which yields a
$GM$ value of $5.2\pm 0.6$ m$^3$/s$^2$ (\citealt{chesley14}, total mass $M \approx 7.8 \times 10^{10}$ kg).
Here $G$ is the gravitational constant.
Due to its low density and high porosity ($\sim 40\%$), Bennu is likely to be a rubble-pile \citep{chesley14}.
The ground-based radar observations used to derive the shape model, also characterize its spin state and surface roughness 
\citep{nolan13}.
These observations gave a precise measurement of its rotational period (4.2978 hours).
Non-principal-axis rotation has not been detected.
The escape speed from Bennu is likely under 23 cm/s and lower than this value at lower latitudes 
due to spin and body shape \citep{scheeres16}.
Its mean equatorial radius is estimated at $246 \pm 10$ m \citep{scheeres16} and the radius
of the equivalent volume sphere, $R_{\rm vol}$, (also called the volumetric radius)
is approximately the same value.
We list Bennu's properties (as compiled and measured by \citealt{nolan13,chesley14}) in Table \ref{tab:bennu}.
The sizes and units are summarized in both physical
and N-body or gravitational units.

For our mass-spring simulations  it is convenient to work in units of a gravitational timescale
\begin{align}
t_{\rm grav} &\equiv \sqrt{\frac{R_{\rm vol}^3}{GM}} = \sqrt{ \frac{3}{4 \pi G \rho}} \nonumber \\
&= 1685 {\rm s} \left(\frac{\rho}{1260\ {\rm kg\ m}^3} \right)^\frac{1}{2}. \label{eqn:tgrav}
\end{align}
The inverse of this timescale is the angular rotation rate of an orbit that just barely grazes the surface of the volume equivalent
sphere.
The angular rotation rate corresponding to Bennu's spin rotation period is  
\begin{equation}
\Omega  = 0.68 t_{\rm grav}^{-1} \left( \frac{P_{\rm rot}}{4.2978 \ {\rm hours} }\right)^{-1}. 
\end{equation}
A convenient unit of energy density 
\begin{equation}
e_{\rm grav} = \frac{GM^2}{R_{\rm vol}^4} = 110 \ {\rm kPa} \left( \frac{M}{7.8 \times 10^{10} {\rm kg}} \right)^2
\left(\frac{R_{\rm vol}}{246 \ {\rm m}} \right)^{-4}. \label{eqn:eg}
\end{equation}
Our mass-spring model viscoelastic code typically runs in units of the body mass, $M$,
and the radius of the equivalent volume sphere, $R_{\rm vol}$.
For Bennu the mean equatorial radius is approximately
equal to the radius of the equivalent volume sphere.
In these units the mean asteroid density $\rho = 3/(4\pi)$.
The code  unit of velocity is 
\begin{equation}
V_{\rm grav} = \sqrt{ \frac{GM}{R_{\rm vol}}} =  \frac{R_{\rm vol}}{t_{\rm grav}}. \label{eqn:Vgrav}
\end{equation}
The unit of force $F_{\rm grav} = GM^2/R_{\rm vol}^2$.
Other units can be constructed similarly.  

For an isotropic material
the P-wave,  S-wave and Rayleigh wave velocities are related to the Young's modulus $E$, 
mass density $\rho$,
and Poisson ratio $\nu$ by 
\begin{align}
V_p &= \sqrt{\frac{E}{\rho}} \sqrt{ \frac{(1-\nu)}{ (1 + \nu) (1- 2\nu)}}, \nonumber \\
V_s &= \sqrt{\frac{E}{\rho}} \frac{1}{ \sqrt{2(1+\nu)}}, \nonumber \\
V_{Ra} &\approx V_s \frac{0.862 + 1.14 \nu}{1+\nu}.  \label{eqn:speeds}
\end{align}
where the last relation is approximate (\citealt{freund98}; page 83).
We estimate wave speeds using a Poisson ratio of $\nu=1/4$  because our mass
spring model approximates a continuum elastic solid with this value \citep{kot15}.
With Poisson ratio $\nu=1/4$  the P-wave speed
and ratios between S-wave and Rayleigh wave speeds are
\begin{align}
V_p &= \sqrt{\frac{E}{\rho}} \sqrt{\frac{6}{5}} \approx 1.09 \sqrt{\frac{E}{\rho}}, \nonumber \\
\frac{V_p}{V_s} & \approx 1.73, \nonumber \\
\frac{V_{Ra}}{V_s} & \approx 0.92, \nonumber \\
\frac{ V_{Ra}}{V_p} & \sim 0.53 . \label{eqn:speeds_nu4}
\end{align}
For a homogeneous body, 
it takes a time $t = R_{\rm vol}/V_p$ for a P-wave to travel from surface to core. 
The lowest frequency vibrational normal modes are proportional to the inverse of this travel 
time or a
characteristic frequency $f_{\rm char} \equiv V_p/R_{\rm vol}$.

Surface layers of the moon have a low P-wave velocity of about $V_p \approx 104$ m/s \citep{cooper74} 
suggestive of a fractured and porous material and consistent with lunar regolith.
Taking a Poisson ratio of $\nu = 1/4$ 
and the estimated density of Bennu, the P-wave lunar regolith
velocity corresponds to a Young's modulus of $E=11.3$ MPa.
We will adopt this Young's modulus and this P-wave speed to explore propagation of the impact
excited seismic waves in our model, as did \citet{murdoch17} to study seismicity on 
Didymoon (provisionally designated S/2003 (65803) 1).

We use a right handed Cartesian coordinate system in the body frame  ${\bf r} = (x,y,z)$ and an 
associated spherical coordinate system with  radius, and latitude and longitude angles $(r,\lambda,\phi)$.
The two are related by
 $(x,y,z) = r( \cos \lambda \cos \phi, \cos \lambda \sin \phi, \sin \lambda)$.
We take the z-axis, $\hat {\bf z}$ along the body's maximum moment of inertia, assuming
that this axis is  the spin axis and with North pole at positive $z$.
Rotation is defined with the surface moving toward the East (increasing $\phi$) as seen
from an inertial frame.  
When using the preliminary Bennu shape model  \citep{nolan13},  
we define latitude and longitude with respect to the 
coordinates used by the shape model. These angles are not part of any cartographic standard.


\subsection{Frequencies of the slowest normal modes for a homogeneous sphere}
\label{sec:normalmodes}

Is it possible to excite normal modes in a granular system?
Laboratory studies of pressure waves and pulses in granular media
show weaker attenuation at lower frequencies (e.g., \citealt{odonovan16}), so the lowest frequency waves
should be the last to decay.
Laboratory studies find that
pulse propagation co-exists with long lived slower, multiply scattered coda-like signals \cite{jia04,jia99,hostler05}.
Lower frequency normal mode oscillations are seen in laboratory experiments of granular media \citep{odonovan16}, 
though these are typically at about 1 kHz, corresponding to the lowest frequency modes of a box containing
granular media in a lab, rather than the Hz
scale relevant for Bennu's normal modes (estimated below).  Due to the lower attenuation at low frequencies,
normal mode oscillations may be sustainable in an asteroid.

The first in-depth study of an elastic homogeneous sphere is that by \citet{Lamb_1881}.
Modes are separated into spheroidal $S$ and toroidal $T$ types 
with each mode characterized by the number of surface nodal lines $l$,
internal (or radial) nodal lines $n$ and azimuthal order $m$
(e.g., \citealt{snieder15}).  
We ignore torsional modes because they do not give radial 
displacements,  so they would not be strongly excited by impacts or tidal
perturbations. 
For the non-rotating homogeneous sphere, the mode frequencies are independent of $m$.
In a spinning body,  the normal modes are split and the frequencies 
depend on $m$ \citep{montagner08}.  
However,
as the spin rate of Bennu is small 
(1000 times smaller than the characteristic frequency $f_{\rm char}=V_p/R$),
the frequency splitting in Bennu and other nearly spherical asteroids should be small. 
Following \citet{apple}, we use notation $_nf_l$ for the spherical (S) normal mode frequencies.
With $n=0$ (no radial nodes) the frequencies
\begin{equation}
_0f_l = \frac{\sqrt{l(l+1)} }{2 \pi} \frac{V_{Ra}}{R}  \qquad  {\rm for} \qquad l >0 \label{eqn:smodes}
\end{equation}
(equation 5 by \citealt{apple})
where $V_{Ra}$ is the Rayleigh wave velocity. 
Spherical modes with $n=0$ can be described as 
constructively interfering Rayleigh waves propagating across the surface. 

Evaluating the frequencies of the 
$n=0$ modes (using equation \ref{eqn:smodes}) and using $\frac{V_{Ra}}{V_p} = 0.53$ for 
Poisson ratio $\nu=1/4$ (as discussed in section \ref{sec:units}) we compute for $l=1 $ to 5, 
\begin{align}
_0f_1 &= 0.119  f_{\rm char}, \nonumber \\
_0f_2 &= 0.207 f_{\rm char} ,\nonumber \\
_0f_3 &= 0.292  f_{\rm char}, \nonumber \\ 
_0f_4 &= 0.377  f_{\rm char}, \nonumber \\ 
_0f_5 &= 0.462  f_{\rm char}.
\label{eqn:n0modes}
\end{align}

Spherical modes without surface nodal lines ($l=0$) are often called radial modes
and they can be described in terms of interfering P-waves.
Again following \citet{apple} the
radial modes for a homogeneous sphere are found from the roots of
\begin{equation}
\cot x - \frac{1}{x} + \frac{x}{4} \left(\frac{V_p}{V_s} \right)^2 = 0. \label{eqn:root}
\end{equation}
The radial spherical model frequencies  are 
\begin{equation}
_nf_0 = \frac{x}{2 \pi} \frac{V_p}{R}, \label{eqn:rmodes}
\end{equation}
where $x$ is a root of equation \ref{eqn:root}.
For a homogeneous sphere and Poisson ratio $\nu=1/4$ giving
 $\left(\frac{V_p}{V_s}\right)^2 = 3$ (using  equations \ref{eqn:speeds}),
we find that the first 5 radial frequencies (with $n=0$ to 4) are 
\begin{align}
_0f_0 &= 0.41  f_{\rm char},\nonumber \\
_1f_0 &=0.96  f_{\rm char},  \nonumber \\
_2f_0 &=1.48 f_{\rm char},\nonumber \\
_3f_0 &=1.98  f_{\rm char},\nonumber \\
_4f_0 & =2.49  f_{\rm char}. \label{eqn:l0modes}
\end{align}


\begin{figure*}
\begin{center}
\includegraphics[width=9cm, trim=20 10 80 5, clip]{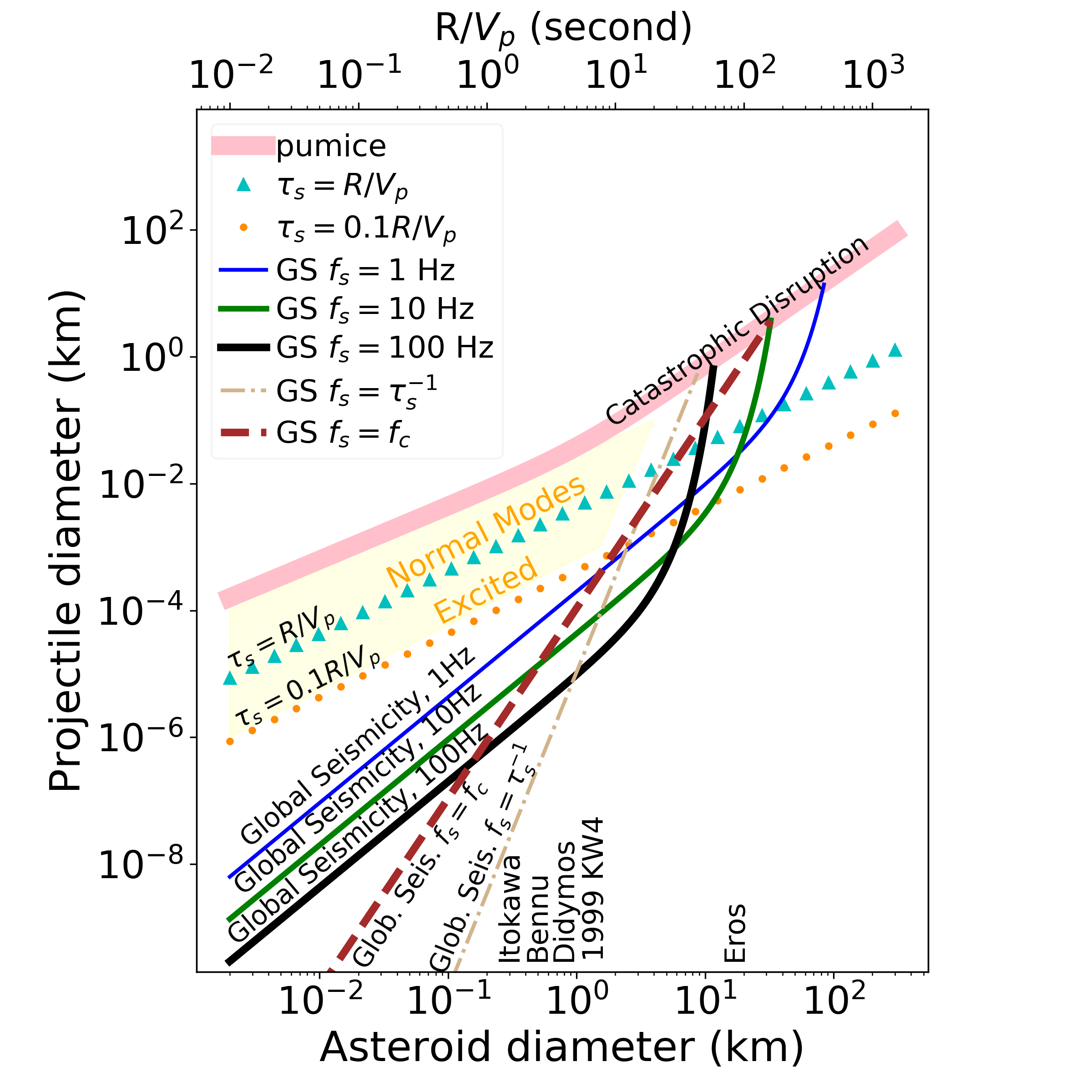}
\end{center}
\caption{Scaling for impacts.  Diameters of projectiles 
capable of catastrophic disruption, global seismic shaking  and excitation
of normal modes.  For this plot we use 
seismic efficiency $\epsilon_s = 10^{-5}$,  seismic amplification  $S = 1$, 
P-wave velocity $V_p = 100 $ m/s, impactor velocity $V_{\rm proj}$ =  5 km/s, and
density for impactor and asteroid of 2.0 g/cc. 
The top pink line shows the diameter in km of a projectile capable of catastrophically disrupting a 
parent body, with diameter shown on the x-axis in km. The disruption threshold is that
by \citet{jutzi10} for pumice using coefficients from their Table 3 and for a 5 km/s impact. 
The green triangles and orange dots  show projectiles giving a source time equal to the characteristic frequency 
$f_{\rm char} = R/V_p$ and one tenth of this.  These 
are computed with equation \ref{eqn:diam_ratio}.
With $\tau_s \sim R/V_p$ we expect excitation of low frequency normal modes.
The top axis shows the time for a P-wave to traverse from the surface to the 
center of the asteroid $R/V_p$.  
Three lower lines (blue, green, black) give minimum impactors capable of causing global seismic shaking (GS)
or seismic waves with accelerations
greater than the surface gravity. These are computed using equation 9 by \citet{richardson05}.
The accelerations depend on the 
seismic frequency, $f_s$, and we have plotted projectile diameters for peak seismic
frequencies $f_s=1$, Hz, 10 Hz and 100Hz.
The dot-dashed tan line and dashed brown line show global seismic shaking
with frequency estimated from the seismic source time (equation \ref{eqn:tan})
and turn over frequency (equation \ref{eqn:brown}).
The yellow region shows the regime where a strong impact lies just below
catastrophic disruption, but above that giving global seismic reverberation.
Impacts in this regime would excite low frequency normal modes.
\label{fig:loglines}}
\end{figure*}

\section{Excitation of Seismic Waves by an Impact}
\label{sec:seismic}

Using scaling arguments (following \citealt{mcgarr69,walker04,lognonne09}), we can describe
the excitation of seismic waves from a meteoroid impact in terms of two ratios, 
one involving the integrated stress of the seismic pulse and the momentum of impactor
and the other involving the energy of seismic waves launched by the impact.   
The seismic impulse $I_s$ is the total momentum from the impact transferred into 
seismic motions,
\begin{equation}
I_s \approx \int_0^{\tau_s} F_s(t) dt \label{eqn:I_s}
\end{equation}
where $\tau_s$ is the seismic source duration and $F_s(t)$ is a time dependent applied force.
The contact-and-compression 
phase of an impact excites a hemispherical shock wave in the ground that propagates 
away from the impact site \citep{melosh89}. 
The shock wave attenuates and degrades into an entirely elastic (seismic) wave. 
The structure of the excited elastic wave is expected to be complex, with multiple pulses
associated with the elastic precursor to the shock wave, an elastic remnant to a plastic wave during
the transition between shock and elastic wave, and
reverberations associated with different
seismic impedances in the target, rock fracture and compactification  
(see section 5.2.6 by \citealt{melosh89}).
However, following \citet{lognonne09},
we can approximate the excitation of the seismic wave with a smooth force function $F_s(t)$ 
applied to the free
asteroid surface and in the direction normal to the surface.
We neglect a possible horizontal stress component that might arise from a grazing collision.

The seismic amplification $S$ is a dimensionless factor describing the ratio of momentum
\begin{equation}
S \equiv \frac{I_s}{m_{\rm proj} V_{\rm proj}},  \label{eqn:S}
\end{equation} 
where $m_{\rm proj}$ is the mass of the  projectile
and $V_{\rm proj}$ is the projectile speed at the moment of impact relative to the asteroid.
The amplification factor for a normal impact has $S \sim 1$, corresponding to a  
perfectly inelastic impact, however hypervelocity impacts with energetic ejecta
can have $S \gtrsim  10 $ \citep{mcgarr69,holsapple04}.

The kinetic energy of the impactor $E_{\rm proj} = \frac{1}{2} m_{\rm proj} V_{\rm proj}^2$ 
and this can be compared to the total radiated seismic energy, $E_s$, giving a
seismic efficiency factor 
\begin{equation}
\epsilon_s \equiv \frac{E_s}{E_{\rm proj}}. 
\end{equation}
The seismic efficiency $\epsilon_s$ is poorly constrained, and
ranges from $\epsilon_s \sim 10^{-2}$ to $10^{-6}$ 
(see experiments and discussions by \citealt{mcgarr69,schultz75,melosh89,richardson05,shishkin07,lognonne09,yasui15,guldemeister17}).

\citet{wolf44} derived an expression for 
 the power radiated in seismic waves into a homogeneous elastic half-space 
by a sinusoidally varying vertical force applied to the surface and that is 
valid for wavelengths larger than the seismic source.
\citet{mcgarr69} used this expression to estimate the seismic
radiated power assuming an applied force function in the form
\begin{equation}
F_s(t) = F_s (1 - \cos (2 \pi t/\tau_s) ) \label{eqn:Fst}
\end{equation}
 that is applied only during $0 < t< \tau_s$
with $\tau_s$, the seismic source duration.
\citet{lognonne09} (see their equation 5) used the expression by 
\citet{mcgarr69} to relate the seismic efficiency
to the seismic source duration giving
\begin{equation}
\epsilon_s  
\sim 8 \pi^2  0.384\ S^2 \frac{m_{\rm proj}}{\rho V_p^3 \tau_s^3},
\label{eqn:eps_s}
\end{equation}
where $\rho$ is the target asteroid density and $V_p$ is the P-wave speed in the target asteroid. 

For a homogeneous spherical asteroid of radius $R$ and mass  $M = \frac{4}{3} \pi \rho R^3$, 
we can rewrite equation \ref{eqn:eps_s} to give 
\begin{align}
\epsilon_s &\approx 127 S^2 \frac{m_{\rm proj}}{M}  \left( \frac{R}{V_p \tau_s} \right)^3 . \label{eqn:eps1}
\end{align}
Using our definition for the characteristic frequency, $f_{\rm char} = V_p/R$, and writing
the mass ratio  $m_{\rm proj}/M$  in terms of a diameter ratio,  equation \ref{eqn:eps1}
is equivalent to
\begin{align}
\frac{D_{\rm proj}}{D}  &\approx 0.2  \left( \frac{\rho}{\rho_{\rm proj}} \right)^\frac{1}{3}
S^{- \frac{2}{3}} \epsilon_s^{\frac{1}{3}}
 \tau_s f_{\rm char} \label{eqn:diam_ratio}
\end{align}
and
\begin{align}
 \tau_s f_{\rm char} &\approx  5 \epsilon_s^{- \frac{1}{3}} S^\frac{2}{3} 
\left( \frac{D_{\rm proj}}{D} \right)
\left( \frac{\rho_{\rm proj}}{\rho} \right)^\frac{1}{3}. \label{eqn:ftau}
\end{align}
Here $\rho_{\rm proj}$ is the projectile density, and $D$ and $D_{\rm proj}$ are asteroid
and projectile diameters.

At a fixed amplification factor $S$ and seismic efficiency $\epsilon_s$, equation \ref{eqn:ftau} implies that 
the seismic source duration times the characteristic frequency is proportional to the ratio of
projectile and asteroid diameters.   In other words, the projectile diameter sets the seismic source
time, with larger projectiles causing longer seismic pulses (see discussion by \citealt{guldemeister17}).
A longer seismic pulse gives a seismic 
spectrum with more power at lower  wave frequencies.  
Impact simulations show larger projectiles producing 
a lower frequency spectrum of seismic waves
(see Figure 10 by \citealt{richardson05}), and this is consistent with 
 source time increasing with projectile size. 

How large a projectile  would excite low frequency
normal modes?  For a homogeneous body, the $n=0$, $l=2$  normal mode has frequency 
$_0f_2 \approx 0.2 f_{\rm char}$ (equation \ref{eqn:n0modes}).   
A seismic source with source time $\tau_s$ should have power
at frequency $f \sim 1/\tau_s$.   So the projectile that gives source time $\tau_s f_{\rm char} \sim 1$
should excite low frequency normal modes. 
Equation \ref{eqn:diam_ratio} applied with  $\tau_s f_{\rm char} \sim 1$
gives us an estimate for the diameter of a projectile capable of exciting seismic waves
at low frequencies, similar to those of the slowest normal modes.
To illustrate the regime for excitation of low frequency normal modes 
by an impact, we plot  equation \ref{eqn:diam_ratio}  in Figure  \ref{fig:loglines} 
for $\tau_s f_{\rm char} =1$ and 0.1 as turquoise triangles and orange circles, respectively, 
with $\epsilon_s = 10^{-5}$, $S=1$ and $\rho_{\rm proj} = \rho$.
For Bennu with a diameter $\sim 500$ m, a projectile with diameter 2m (and with impact angle normal to the surface)
could produce a seismic source
with source timescale that is approximately the inverse of the characteristic frequency and
 similar in size to the periods of its slowest normal modes.

\subsection{Catastrophic disruption and global reverberation thresholds}
\label{sec:glob}

Equation \ref{eqn:diam_ratio}  estimates the size of a projectile that can
excite low frequency normal modes.
How close is this projectile size to a disruption threshold or a threshold for global seismic reverberation?  
It is customary to characterize impacts in terms of a specific energy threshold or 
the kinetic energy in the collision divided by target mass (e.g., \citealt{melosh97,benz99,jutzi10}).
A catastrophic disruption threshold, $Q_D^*$, is
the specific energy required to disperse the target into a set of individual objects, 
the largest one having  half the mass of the original target.
To estimate  the disruption threshold, we use equation 3 by \citet{jutzi10}, 
coefficients for pumice from their Table 3, 
a 5 km/s projectile velocity (typical of asteroid encounters; \citealt{bottke94}) 
and assume that asteroid and projectile have a similar density.
In Figure \ref{fig:loglines}
we plot as a wide pink line the projectile diameter corresponding to this 
disruption threshold as a function of asteroid diameter.  
An impactor of diameter 
$ D_{\rm proj} \sim 10 $m  at a projectile velocity of 5 km/s 
is large enough to catastrophically disrupt Bennu.   
Figure \ref{fig:loglines} illustrates that projectiles capable of exciting normal modes
in an asteroid
are likely to be an order of magnitude smaller than a projectile causing catastrophic disruption.
 
Assuming a single seismic wave frequency and computing accelerations from the wave,
\citet{richardson05} estimated  the diameter of a projectile $D_{\rm proj}$
sufficient to cause seismic vibration across the whole body that is 
above the surface gravitational acceleration $g = GM/R^2$.
The projectile diameter 
\begin{equation}
D_{\rm proj} \approx \left[\frac{G^2 \rho^3 D^5}{9 \epsilon_s \rho_{\rm proj} V_{\rm proj}^2 f_s^2} \right]^\frac{1}{3}
\exp \left( \frac{2 \pi f_s D^2}{K_s \pi^2 Q 3} \right), \label{eqn:Dpr}
\end{equation}
(their equation 15 but corrected with a factor of 1/3 in the exponential).
Here $f_s$ is the frequency  of the seismic waves.
The seismic attenuation coefficient is $Q$ and $K_s$ is a seismic diffusivity.
In Figure \ref{fig:loglines}
we have plotted this function for three frequencies $f_s = 1,10,100$ Hz (as blue, green and black lines)
using $\rho = \rho_{\rm proj}  = 2.0 {\rm g/cc}$, seismic efficiency $\epsilon_s =10^{-5}$,  
projectile velocity $V_{\rm proj} = 5 $ km/s,
attenuation coefficient
$Q = 2000$ and seismic diffusivity $K_s = 0.1 {\rm km}^2 {\rm s}^{-1}$.  
We have adopted the $Q$ and $K_s$ values  used by \citet{richardson05}.  The $Q$ attenuation coefficient is based on
$Q = 3000$ to 5000 from long period seismic instruments on the Moon \citep{dainty74,toksoz74}
and $Q =1600 $ to 2300 from short period lunar seismic instruments \citep{nakamura76}.
The seismic diffusivity is that estimated from lunar seismic observations  \citep{dainty74,toksoz74}.
As expected,
these lines lie below the catastrophic disruption threshold.  These lines are also
below those we estimated for projectiles capable of exciting normal modes.
Projectiles capable of exciting normal modes would cause much larger levels of agitation
than those only just capable of crater erasure.

Equation \ref{eqn:Dpr} is computed assuming  a single
seismic frequency dominates the spectrum.
Previously we estimated the seismic source time. We can use
this timescale to estimate a frequency typical of seismic waves excited by the impact.
Taking seismic frequency $f_s \sim 1/\tau_s$, from the seismic source time,
we use equation \ref{eqn:diam_ratio} for $\tau_s$ to estimate the
frequency of generated seismic waves.  Neglecting seismic attenuation, and
inserting this frequency into equation \ref{eqn:Dpr} gives
\begin{align}
D_{\rm proj}  
&\sim \frac{ 0.7  G^2 \rho^2 S^\frac{4}{3} D^5}{\epsilon_s^{\frac{5}{3}} V_{\rm proj}^2 V_p^2} \label{eqn:tan}
\end{align}
for the projectile diameter capable of causing global seismic reverberation.
This too is plotted in Figure \ref{fig:loglines} as a dot-dashed tan line 
using $\rho = \rho_{\rm proj}  = 2.0 {\rm g/cc}$, seismic efficiency $\epsilon_s =10^{-5}$,  
projectile velocity $V_{\rm proj} = 5 $ km/s, P-wave velocity
$V_p = 100 $ m/s, and seismic amplification factor $S=1$.  The line is below
the constant frequency lines and suggests that smaller impacts can cause significant levels
of agitation on smaller asteroids.  

\subsection{Seismic spectrum corner frequency}
\label{sec:corner}

We lack a model for the frequency spectrum of seismic waves excited by an impact and up to 
this point we have estimated the peak frequency using simple source time model, 
that by \citet{lognonne09} and based on estimates by \citet{mcgarr69}. 
However we can also constrain the seismic spectrum radiated by an impact by considering the size
of the source region.
A seismic source spectrum is limited by the size and duration of
the source.   If an earthquake source is a delta function in time and the fault area infinitesimally small,
then the displacement spectrum is flat (as discussed at the end of section 10.1.4 by \citealt{aki02}).
Here the seismic wave amplitudes are described with material displacements from rest.
Otherwise the source spectrum is weaker at frequencies above a particular frequency,
that we call the corner frequency.   
An instantaneous quake cannot  efficiently radiate seismic waves
with wavelengths smaller than its fault length as the source is coherent in phase.
The corner frequency is due to spatial interference 
 (section 10.1.5 by \citealt{aki02}) and is approximately the
wave speed divided by the fault length.    
Most seismic sources, including
those derived from simulations of asteroid impacts, have power spectra dropping at high frequencies
(see Figure 10 by \citealt{richardson05}).

Crater diameter has previously been used to estimate a source size for impact
radiated seismic waves \citep{meschede11}. 
We can estimate the turn over or corner frequency of the impact radiated
seismic spectrum
using an estimate for the diameter of the crater formed during impact.
Crater sizes on Bennu would be in the strength-scaling regime \citep{holsapple93} with diameter
\begin{equation}
D_{\rm crater} \sim 30 D_{\rm proj}  \label{eqn:30}
\end{equation}
as approximated in equation 32 by  \citet{richardson05}
(also see their Figure 16 comparing this approximation to the regimes discussed by \citealt{holsapple93}).
\citet{housen18} has carried out experiments of impacts into  porous cohesionless materials.
An extrapolation of their highest velocity experiments (by two orders of magnitude
past the smallest dimensionless $\pi_2$ value on their Figure 15)
gives a crater diameter approximately consistent with equation \ref{eqn:30}.

A corner frequency set from the crater diameter and P-wave velocity 
\begin{equation}
f_{\rm corner} \approx \frac{V_p}{D_{\rm crater}} \approx \frac{ V_p} {30 D_{\rm proj}}. \label{eqn:fcorner}
\end{equation}
The displacement spectrum typically drops at frequencies larger than $f_{\rm corner}$ 
with a power law form, and is flat at lower frequencies.
For an impactor with diameter of 1 m and $V_p \sim 100$ m/s,
equation \ref{eqn:fcorner}
gives a corner frequency $f_{\rm corner} \approx 3$ Hz and fairly near the characteristic frequency 
of Bennu that we estimated at 0.4 Hz in Table \ref{tab:bennu}.
In units of the characteristic frequency 
\begin{equation}
\frac{f_{\rm corner} }{f_{\rm char}} \sim  \frac{D}{60 D_{\rm proj}} \label{eqn:fc}
\end{equation}
so an impactor with diameters about 60 times smaller than that of the asteroid
would give a seismic spectrum with  corner frequency in the vicinity of the characteristic frequency and
so within a decade of the slowest normal modes.  
This implies that a significant fraction of the seismic wave power for this impact would
be emitted at low frequencies and possibly at frequencies similar to those of the slowest
normal modes.    

Low frequency seismic waves usually take longer to attenuate than high frequency
waves, so seismic reverberation caused by a large impact
may be long lasting.  The energy in an excited normal mode typically lasts 
an attenuation factor $Q$ times the normal mode period.  Another consequence of
low frequency content in the seismic waves excited by a strong impact is that seismic reverberation could persist. 

Neglecting attenuation and using the corner frequency (equation \ref{eqn:fc}) for the seismic frequency  
($f_s = f_{\rm corner}$)  we can estimate the projectile diameter capable 
of causing global seismic reverberation
again with 
equation \ref{eqn:Dpr}, giving
\begin{equation}
D_{\rm proj} = \frac{G \rho D^3}{V_{\rm proj} V_p}  
\left( \frac{5 }{3 \epsilon_s}  \frac{\rho}{\rho_{\rm proj}} \right)^\frac{1}{2}. \label{eqn:brown}
\end{equation}
This is plotted as a brown dashed line in Figure \ref{fig:loglines}.  The line also lies below
reverberation thresholds estimated with single  seismic frequencies.

\subsection{The regime for subcatastrophic impacts that are capable of exciting normal modes}
\label{sec:regime}

\citet{richardson05} estimated a frequency dependent projectile size
capable of causing global seismic reverberation at the level of surface acceleration.
Using an estimate for the seismic source time and a turn-over or corner frequency
for the spectrum based on crater size
we modified the estimate so that it was independent of frequency.  In both cases
the projectile diameter capable of causing global seismic reverberation lies well
below the catastrophic disruption threshold for small asteroids and  below
an estimate for the size of a projectile capable of exciting low frequency seismic waves.

Figure \ref{fig:loglines} delineates a regime (shown in yellow) where a catastrophic impact can
excite vibrational normal modes and with accelerations above surface gravity acceleration.
The asteroid is not catastrophically disrupted but the surface accelerations caused by the seismic
waves would exceed surface gravity and so  launch material off the surface.  Material could
hop or flow across the surface, but would not necessarily be ejected. 
When the vibrational spectrum is dominated by a few normal modes,
the vibrational energy is not evenly distribution across the surface.
The asteroid surface may slump, reaching a final shape that could be sensitive
to the morphology of the normal modes themselves.

Figure \ref{fig:loglines} gives the regime for an asteroid with a slow P-wave velocity consistent
with a porous material (such as lunar regolith)
and a seismic efficiency of $\epsilon_s = 10^{-5}$.
The catastrophic impact line used here is sensitive to the  
 material of the asteroid.  If the asteroid is softer or comprised of solid ice, then the catastrophic
impact line is higher on the plot.
With a higher seismic efficiency, more energy is excited by the impact and the global seismicity lines
are lower on the plot.  
If the P-wave speed is higher, then we expect the seismic diffusivity $K_s$ to be larger.
This  pushes the attenuation cutoffs for the global seismicity lines to the right on the plot 
and extends the regime
for normal excitation to larger bodies. 
With a somewhat higher seismic efficiency of $\epsilon_s = 10^{-4}$ 
and faster P-wave velocity (3 km/s typical of ice),
the larger icy moon Pan (28 km diameter  and with a  sculpted equatorial ridge), 
may lie in a regime where a 0.3 km diameter icy impactor could excite normal modes.

Figure \ref{fig:loglines} shows a regime for a subcatastrophic impact to
excite vibrational normal modes.   How rare are these impacts on Bennu? 
The mean time between impacts as a function of impactor diameter has been
computed for Mars crossing asteroid 433 Eros by \citet{richardson05} based
on the asteroid size distribution model by \citet{obrien05}.
We can estimate the mean time between impactors by multiplying by the 
ratio of cross sectional areas.
The cross sectional area of Bennu is 
0.19 km$^2$, computed from its mean equatorial radius.
The mean time period between 2 m diameter impacts on 433 Eros 
(with cross sectional area about 360 km$^2$) 
is about 1000 years, 
following Figure 16A by \citet{richardson05}.
The ratio of cross sectional areas is 2000 giving a mean time between  2 m diameter impactors
on Bennu of 2 million  years.
Bennu should experience a shape changing encounter about every few million years.  
The mean time between such impacts is similar to its orbital lifetime as a near-earth object.
Such an encounter would not be unlikely, but neither would it happen  frequently.

\subsection{Force for the seismic pulse}
\label{sec:force}

To carry out simulations of an impact, 
we require a seismic source time and a force amplitude $F_s$.
Using the definitions for amplification factor $S$ (equation \ref{eqn:S}) and $I_s$  (equation \ref{eqn:I_s} and approximating
the integral as $F_s \tau_s$), and using equation \ref{eqn:ftau} for $\tau_s$,
 the force of the seismic impulse in gravitational units
\begin{align}
\frac{F_s}{GM^2/R^2} & \approx 
S   \left(\frac{\rho_{\rm proj}}{\rho} \right)
\left( \frac{D_{\rm proj}}{D} \right)^3
 \left( \frac{V_{\rm proj} V_p }{V_{\rm grav}^2} \right) 
 \left(\frac{ 1}{\tau_s f_{\rm char}} \right).
\end{align}
We insert equation \ref{eqn:diam_ratio} for the diameter ratio 
\begin{align}
\frac{F_s}{GM^2/R^2} & \approx  0.008 S^{-1} \epsilon_s (\tau_s f_{\rm char})^2
   \left( \frac{V_{\rm proj} V_p }{V_{\rm grav}^2}\right) \\
   &\approx 2\  S^{-1}
   \left( \frac{\epsilon_s}{10^{-5} }\right)
   (\tau_s f_{\rm char})^2
 \nonumber \\
    & \times
       \left( \frac{V_{\rm proj}}{5\ {\rm km~s}^{-1}} \right)
       \left( \frac{V_p}{100 {\rm\ m~s}^{-1}} \right)
       \left( \frac{V_{\rm grav}}{\ 0.146 {\rm\ m~s}^{-1}} \right)^{-2}  . \label{eqn:Fs}
\end{align}
In the last line, we have used $V_{\rm grav}$ for Bennu from Table \ref{tab:bennu}.
If we  replace $\tau_s f_{\rm char} $ using equation \ref{eqn:ftau}
we find
\begin{align}
\frac{F_s}{GM^2/R^2} & \approx 0.2 (\epsilon_s S)^\frac{1}{3}
 \left(\frac{D_{\rm proj}}{D} \right)^2
 \left(\frac{\rho_{\rm proj}}{\rho} \right)^\frac{2}{3}
 \left( \frac{V_{\rm proj} V_p }{V_{\rm grav}^2}\right) .  \label{eqn:Fs2}
\end{align}

The source time $\tau_s f_{\rm char}$ and the force of the impulse $F_s$ are not independent, and
both are approximately set by the impactor diameter.  
In section \S  \ref{sec:sims} we will use equation \ref{eqn:ftau}
for the seismic source time and equation \ref{eqn:Fs} for the impulse force  
to model an impact generated seismic pulse with our elastic body simulations code.

It may be convenient to estimate the seismic energy of the impact in gravitational units.
As we computed $F_s$ in gravitational units, the total seismic energy radiated by the impact
\begin{align}
\frac{E_s}{GM^2/R} & \approx \frac{\epsilon_s}{2} 
	\left(\frac{\rho_{\rm proj}}{\rho} \right)
	\left(\frac{D_{\rm proj}}{D} \right)^3 
	\left(\frac{V_{\rm proj}}{V_{\rm grav}} \right)^2  \nonumber \\
   &\approx 4 \times 10^{-3} 
	S^{-2} \epsilon_s^2
	\left(\frac{V_{\rm proj}}{V_{\rm grav}} \right)^2 
	 (\tau_s f_{\rm char})^3 \nonumber \\
	&\approx 4 \times 10^{-4} S^{-2}
	  \left( \frac{\epsilon_s}{10^{-5} }\right)^2
	   \left( \frac{V_{\rm proj}}{5 {\rm\ km~s}^{-1}} \right)^2  
	   \nonumber \\ &\qquad \times
	       \left( \frac{V_{\rm grav}}{\ 0.146 {\rm\ m~s}^{-1}} \right)^{-2}    
	       (\tau_s f_{\rm char})^3. \label{eqn:Es}
\end{align}


\section{Elastic Body Simulations}
\label{sec:sims}

\subsection{Mass-spring models}

To simulate elastic and vibrational response and transmission of seismic waves
we use a mass-spring model \citep{quillen16_crust,frouard16,quillen16_haumea,quillen17_pluto}
that is built on the modular N-body code \texttt{rebound}  \citep{rebound}.  
An elastic solid is approximated as a collection of mass nodes that are
connected by a network of springs.
Springs between mass nodes are damped and
so the spring network approximates the behavior of a Kelvin-Voigt viscoelastic 
solid with Poisson ratio of 1/4 \citep{kot15}.
When a large number of particles is used to resolve the spinning body the mass-spring model behaves
like an isotropic continuum elastic solid \citep{kot15} including its ability to transmit seismic waves.

The mass particles in the resolved spinning body are subjected to three types of forces: 
the gravitational forces acting between every pair of particles in the body, and the elastic and damping spring forces 
acting only between sufficiently close particle pairs. 
Springs have a spring constant $k_s$ and a damping rate parameter $\gamma_s$.  
The number density of springs, spring constants and spring lengths set the shear modulus, $\mu$, whereas
the spring damping rate, $\gamma_s$, allows one to adjust the shear viscosity, $\eta$, 
and viscoelastic relaxation time, $\tau_{\rm relax} = \eta/\mu$. 
For a Poisson ratio of 1/4, the Young's modulus $E = 2(1+ \nu) \mu = 2.5 \mu$.
The equation we used to calculate $E$ from the spring constants of the springs in our code is equation 29
by \citet{frouard16} which was derived by \citet{kot15}.

We work with mass in units of $M$,  the mass of the asteroid,
distances in units of volumetric radius, $R_{\rm vol}$, the radius of a spherical body with the same
volume, time in units of $t_{\rm grav}$  (equation \ref{eqn:tgrav}) and
elastic modulus $E$ in units of $e_{\rm grav}$ (equation \ref{eqn:eg})
which scales with the 
gravitational energy density or central pressure.  
All mass nodes have the same mass
and all springs have the same spring constants.  
The spring constants are chosen so that the Young's modulus of the mass-spring model
is equal to that of Bennu
in gravitational units and assuming a P-wave speed similar to lunar regolith, as listed in
Table \ref{tab:bennu}.
The simulation timestep is chosen so that it is shorter than the time it takes elastic waves to travel
between nodes \citep{frouard16,quillen16_haumea}.

Initial node distribution and spring network are similar to those of the triaxial ellipsoid random spring model
described by \citet{frouard16,quillen16_haumea}, however we are not restricted to a triaxial outer boundary.
We also use other outer boundaries, such
as Bennu's shape model \citep{nolan13}, or an axisymmetric bi-cone model.
For the triaxial ellipsoid the surface obeys $\frac{x^2}{a^2} + \frac{y^2}{b^2} + \frac{z^2}{c^2} = 1$
with $a,b,c$ equal to half the lengths of the principal axes.  An oblate ellipsoid is described
with an axis ratio $c/a<1$ and  $b=a$.
For the axisymmetric bi-cone model, the surface obeys $|z(\varpi)| =  s_{\rm cone} (\varpi_{\rm base} - \varpi)$
with $\varpi = \sqrt{x^2 + y^2}$, $\varpi_{\rm base}$ is the radius at the equator and $s_{\rm cone}$ is
a slope.  The bi-cone shape model consists of two cones with bases glued together at the equator.
Surface sizes are normalized so that their enclosed volume is equivalent to that of a sphere
with radius 1 or with volume of  $ 4 \pi/3$.

Particle positions within a cube containing the body's surface are drawn from a uniform spatial distribution
but only accepted as mass nodes into the spring network if they are within the shape model 
and if they are more distant than $d_I$ from every other previously generated particle.
Once the particle positions have been generated, every pair of particles within distance $d_s$ of each
other are connected with a single spring.  The parameter $d_s$ is the maximum rest
length of any spring.   For the random spring model
we chose $d_s/d_I$ so that the number of springs per node is greater than 15, as recommended
by \citet{kot15}.  All nodes are inter-connected via the spring network. 
Springs are initiated at their rest lengths.  For more discussion on generating particle and spring distributions,
see \citet{quillen16_haumea}.

With all spring constants the same and all nodes the same mass, the spring network
approximates a homogeneous and isotropic elastic body.  The mass-spring model network would be 
capable of simulating materials with varying density or strength by varying the number
of springs per node, the spring constants, 
the masses of the nodes or the number of nodes per unit volume.
We allow the body to rotate by setting the initial node velocities consistent with 
solid body rotation about a principal axis.
The spin rate is chosen to match that of Bennu in gravitational units.
As all forces are applied to pairs of particles and along the vector
connecting each pair,  momentum conservation  is assured.

At the beginning of
the simulation the body is not exactly in equilibrium  because springs are initially set to their rest lengths
but not taking into account self-gravity.  When the simulation begins,
the body vibrates.  As a result
we run the simulation for a time $t_{\rm damp}$ with a higher damping rate $\gamma_{\rm high}$.
After the body has reached equilibrium, we lower the spring damping coefficients and run the simulated impact.

To track deformations on the surface we identify a subset of particles that are near the surface.
A particle is labeled as near the surface if it lies within a distance $d_{\rm surf}$ of 
the  surface model used to generate the initial particle distribution.   
The impact pulse is applied only to these particles. 
Surface points give us an unstructured set of latitudes and longitudes.
By interpolating their positions onto a grid we construct maps 
of physical quantities, such as the radial displacement or radial  velocity component on the surface, and these are shown as Figures below.


\begin{table*}
\centering
	\caption{\large Simulation Parameters}
	\label{tab:sims}
	\begin{tabular}{llll} 
\hline
Common simulation parameters   \\
\hline
Timestep & $dt$          & $2 \times 10^{-5}$ \\
Minimum distance between mass nodes & $d_I$ & 0.09  \\
Ratio of max spring length to $d_I$  & $d_s/d_I$  &  2.5 \\
Spring constant & $k_s$  & 1203  \\
Surface distance  & $d_{\rm surf}$ & 0.15 \\
Young's modulus of spring network &  $E$ &$10^5$  \\ 
Number of nodes &  $N_{\rm nodes}$ & 3780 \\
Number of springs per node & $N_{\rm springs}/N_{\rm nodes}$ & 17 \\
Spin rate & $\Omega$ & 0.7 \\
Seismic source time & $\tau_s$ & 0.003 \\
Seismic force        & $F_s$ & 7.0 \\
Seismic energy      & $E_s$ & 0.004 \\
Impact source angle & $\Theta_i$ & $15^\circ$\\
Impact longitude & $\phi_c$ & $0^\circ$ \\
\hline
Parameters for shape models \\
\hline 
Cone slope & $s_{\rm slope}$  & 0.8 \\
Oblate axis ratio & $c/a$          & 0.8 \\
\hline
Parameters for individual simulations   \\
\hline
Simulation & shape model & Impact latitude ($\lambda_c$)   \\
\hline   
Be0    & Bennu   &  $0^\circ$  \\
Be15  & Bennu   &  $15^\circ$ \\
Be65  & Bennu   &  $65^\circ$ \\
Co0    & Cone     &  $0^\circ$ \\
Co15  &  Cone    &  $15^\circ$ \\  
Co65  &  Cone    &  $65^\circ$ \\  
Ob0    &  Oblate  &  $0^\circ$ \\  
Ob15  &  Oblate  &  $15^\circ$ \\  
Ob65  &  Oblate  &  $65^\circ$ \\  
\hline
\multicolumn{3}{l}{\multirow{5}{350pt} 
{\small Notes:  Units in this table are in gravitational or N-body units, as described in section \ref{sec:units}.
Seismic source time, force and source angle, and energy correspond to a 2\ m radius impactor on Bennu
giving $\tau_s f_{\rm char} = 2$ for seismic efficiency $\epsilon=10^{-5}$, seismic amplification
factor $S=1$, $\rho_{\rm proj} = \rho$,  $V_{\rm proj} = 5$ km/s and quantities for Bennu listed
in Table \ref{tab:bennu}.  $\epsilon_s, S, V_{\rm proj} $ values are consistent with those
used to create Figure \ref{fig:loglines}.
}}
\end{tabular}
\end{table*}

\subsection{Simulated impacts}

We simulate an impact by applying a force impulse 
 to particles near the surface.
The source function of the impact is described with five parameters,
a source time $\tau_s$,
a force amplitude, $F_s$,  an area over which this force is applied that is described with angular
distance $\Theta_i$,
and the  latitude and longitude $\lambda_c, \phi_c $ of the central position defining the impact site. 
The central latitude and longitude
define a direction $\hat {\bf n}$ from the center of mass and in the body's reference
frame.  A surface particle with direction $\hat {\bf n}'$ (from the center of mass)
is perturbed by the impact
if $0 < \hat {\bf n} \cdot \hat {\bf n}' < \cos \Theta_i$  and it lies within angular distance $\Theta_i$
of the impact center.
The force is applied radially. 
The same force is applied equally to each surface particle within the angular patch and
the total force evenly distributed among the particles within the patch.
A force is only applied between $t_0$, the start of the
impact, and $t_0 + \tau_s$.
For the source time function $F_s(t)$ we use  equation \ref{eqn:Fst}, depending
on the source time $\tau_s$ and the force amplitude $F_s$.

To relate simulation parameters to the scaling relations in section \ref{sec:seismic}
we need asteroid radius, mass, density and P-wave velocity.
We chose a seismic efficiency $\epsilon_s$,  seismic amplification factor  $S$,
 projectile velocity $V_{\rm proj}$ and density ratio $\rho_{\rm proj}/\rho$.
We choose an impactor diameter, then compute source time $\tau_s f_{\rm char}$ using equation 
\ref{eqn:ftau}, force amplitude $F_s$ using equation \ref{eqn:Fs2}, seismic energy using
equation \ref{eqn:Es} 
and seismic source angular size $\Theta_i = D_{\rm crater}/D$
following equation \ref{eqn:30}.  We convert the source time to gravitational units 
by dividing $\tau_s f_{\rm char}$ by $f_{\rm char}$ in gravitational units.
We run a series of simulations for an 
impact with a projectile radius of 2m on asteroid Bennu.  We compute
seismic source time and angular size and force amplitude
with seismic efficiency $\epsilon_s = 10^{-5}$, 
seismic amplification
factor $S=1$, $\rho_{\rm proj} = \rho$,  projectile velocity $V_{\rm proj} = 5$ km/s and physical
quantities for Bennu listed
in Table \ref{tab:bennu}.    

To explore the sensitivity of the seismic response to an equatorial ridge we use the Bennu shape
model \citep{nolan13} and a bi-cone model consisting two cones glued together
at the equator and we compare these to an oblate model with similar axis
ratios but lacking a peaked equatorial ridge.  The Bennu shape model  exhibits four peaks
on its equatorial ridge, but the other two shape models are axisymmetric.
For each shape model, we run equatorial impacts, and impacts at latitudes of $15^\circ$ and $65^\circ$.
Parameters for our simulations are listed Table \ref{tab:sims}.

\FloatBarrier

\begin{figure*}
\begin{center}
\includegraphics[trim=0 70 0 40, clip, height=6.0in]{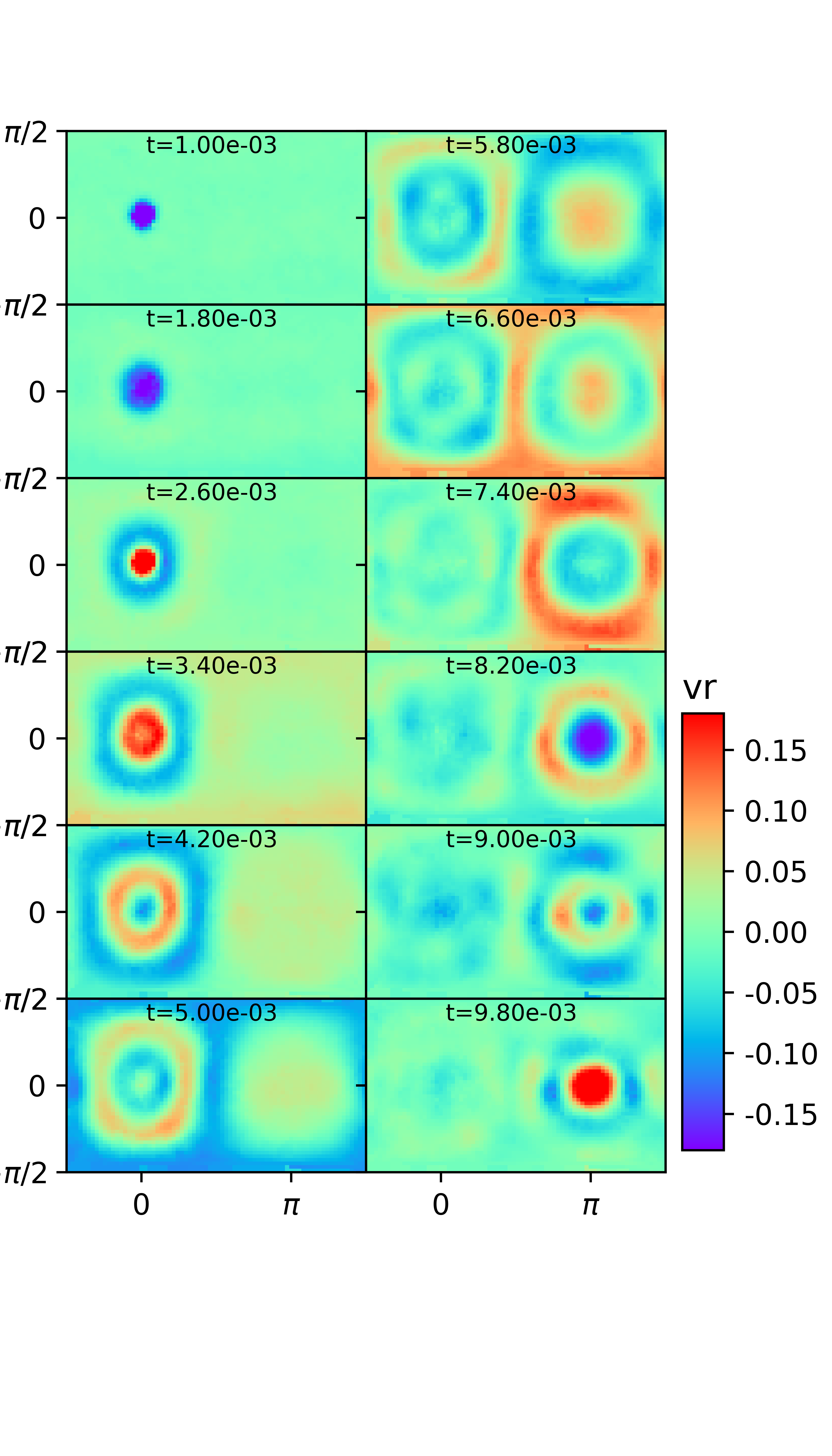}
\end{center}
\caption{The radial velocity component on the surface as a function of time after an equatorial impact impulse
as seen in the Be0 mass spring model simulation.  This simulation uses the Bennu shape model.
The impact occurred at a longitude of $0^\circ$ and on the equator.
Each panel shows a different time with
time after impact written  in black on the top of each panel.
Surface motions are gridded onto 
cylindrical projections with  $x$ axis the
longitude and $y$ axis the latitude.  
The color bar shows the radial velocity  in gravitational units.
}
\label{fig:wave_cyl}
\end{figure*}

\begin{figure*}
\begin{center}
\includegraphics[trim=0 70 0 20, clip, height=8.0in]{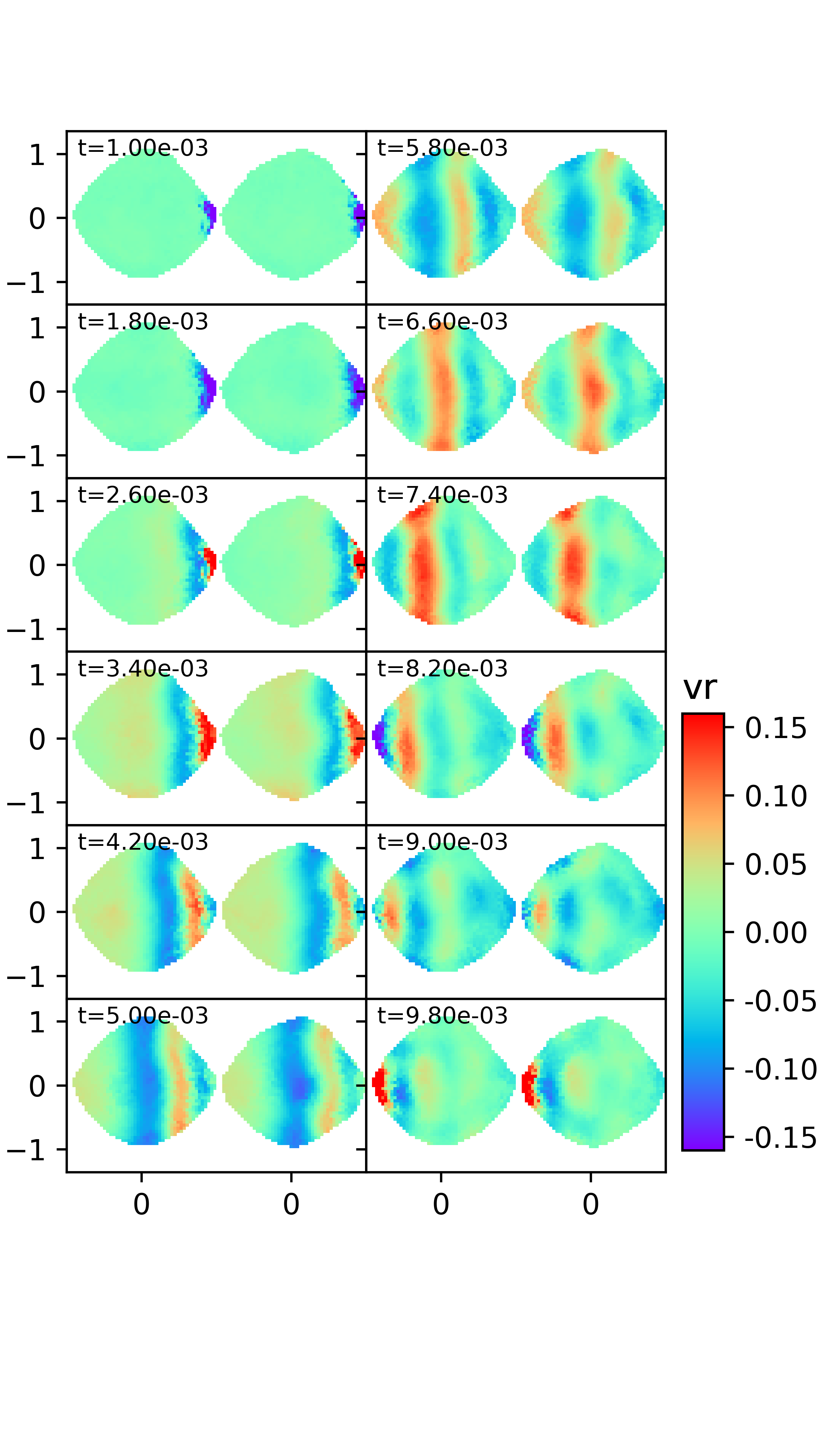}
\end{center}
\caption{Same as Figure \ref{fig:wave_cyl} except the radial velocity
component is shown using a distant perspective (orthographic projection). Both sides
of the body are shown in each panel.  The equator is at $y=0$ and is oriented horizontally.}
\label{fig:wave_orth}
\end{figure*}

\subsection{Excitation of seismic waves by a simulated impact}

The mass-spring model approximates a homogeneous and isotropic elastic body
in which elastic waves can propagate \citep{kot15}, however,  we have 
not previously used it  to simulate seismic waves.
Tracking the motion of surface nodes only, we display the radial component of velocity
as a function of time after impact for the Be0 simulation
 in Figure \ref{fig:wave_cyl} using a cylindrical projection and in Figure \ref{fig:wave_orth}
using front and back orthographic projections.
These figures illustrate that a strong seismic pulse is excited by the impact 
that travels across the body surface.  The impact occurs on the equator and
at longitude $0^\circ$ and the simulated body is the Bennu shape model.
The impulse  is applied inward and so causes a negative radial velocity which appears blue
in Figures \ref{fig:wave_cyl} and \ref{fig:wave_orth}.  The second panel on the top left of
\ref{fig:wave_cyl} shows a blue ring propagating ahead of a red rebound moving outward
away from the impact site.
Seismic focusing (e.g., \citealt{schultz75,meschede11}) is seen as the pulse becomes strong
at the impulse's antipode.

We estimate the time it takes the pulse to travel across the surface of the body.
The pulse takes $\Delta t  \approx 0.004$ to travel across 90 degrees in longitude
or twice this $\sim 0.008$ to travel all the way to the other side of the body (to the impact's antipode) 
where the wave is focused.   
Studies of antipodal focusing of seismic waves from an impact  
find that  the main contribution to peak antipodal displacements  
comes from the constructive interference of low frequency Rayleigh waves \citep{schultz75,meschede11}. 
We estimate the Rayleigh wave speed from
the Young's modulus, P-wave speed and with Poisson ratio $\nu=1/4$, giving
$V_{Ra} \approx 0.53 \times 712 = 378$ in gravitational units, following
our estimates in section \S \ref{sec:units} and parameters listed  Table \ref{tab:bennu}.
To travel along the surface to the antipode,
we estimate  a time $t  = \pi/V_{Ra} \approx  \pi/378 =  0.0083$.  
The pulse travel time across the surface in the simulations is approximately consistent
with the expected Rayleigh wave speed.

\subsection{Spectrum of seismic waves excited by a simulated impact}
\label{sec:spectrum}

\begin{figure}
\includegraphics[width=\columnwidth]{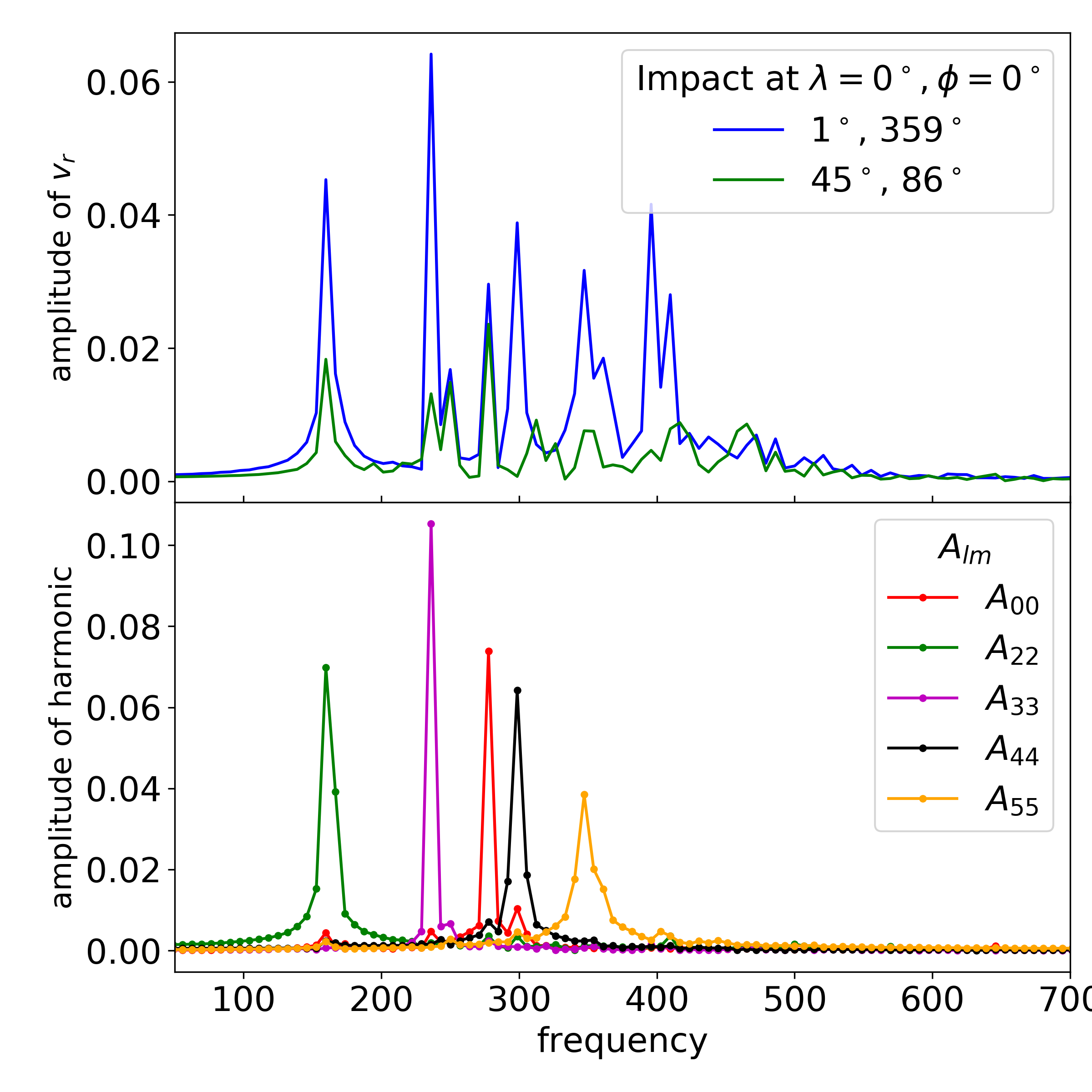}
\caption{The top panels shows the frequency spectrum of the radial velocity of two individual surface particles
from the same simulation shown in Figures \ref{fig:wave_cyl} and \ref{fig:wave_orth} of an  equatorial
impact on the Bennu shape model.
The latitude and longitude of the two particles are shown in the key in the same panel.
In the bottom panel we show the spectrum of different spherical harmonic amplitudes.
These are measured using  particles near the surface and with equations \ref{eqn:Alm} and \ref{eqn:Afft}. 
Peaks in the spectra
can be identified with the  shapes of normal modes.        
The strongest normal modes excited by
this simulated impact are the football mode $l=2$, the $l=3$ triangular mode and the radial
$l=0$ mode.  
The seismic source duration was long enough to excite low frequency normal modes.
}
\label{fig:spec}
\end{figure}

We examine the frequency spectrum of individual particle motions in the Be0 simulation.
Using a fast Fourier transform, we compute the spectrum
$\tilde v_r(f) = \int v_r(t) e^{i 2 \pi f t} dt $ from the particle's radial velocity.
We carry out the integral within a time window $\Delta t = 0.15$ long and beginning
just after the force pulse has ended.
In Figure \ref{fig:spec} top panel, we  plot  
the amplitude $|\tilde v_r(f)|$ as a function of frequency for two surface particles,
one near the equator and at longitude 
$\phi \approx 0^\circ$ and the other at a latitude of $\lambda \approx 45^\circ$
and longitude near $90^\circ$.
The spectra in Figure \ref{fig:spec} (top panel) are not flat
but dominated by a series of peaks.    A particle could
be located at a node and so experience no vibration at an excited normal mode.
This is why we plot  the spectra of
two  particles at different locations so we are less likely to miss a strong frequency of vibration.
Each peak in the spectra is associated with the normal modes defined by specific $l,n$ indices.
Normal modes with the same $l,n$ but with different $m$ values should have similar frequencies
as the body is nearly spherical.

Using spherical harmonics we identify which mode corresponds to each frequency peak
and then discuss how well our mass spring model reproduces the normal mode
frequencies  
that we predicted for a homogeneous elastic sphere.  

On the surface of a nearly spherical body, the motions of a vibrational normal mode can be described
with spherical harmonics,
\begin{equation}
Y_l^m(\theta,\phi) = \sqrt{ \frac{2l+1}{4 \pi}\frac{(l-m)!}{(l+m)!} } P_l^m(\cos \theta) e^{i m \phi} 
\end{equation}
with spherical coordinate angles  $\theta,\phi$.  The azimuthal angle
$\phi \in [0,2\pi]$ is equivalent to longitude and the inclination or colatitude angle
$\theta \in  [0,\pi]$  (with convention $\theta \equiv  \pi/2 - \lambda$). 
The functions $P_l^m(x)$ are  Legendre polynomials.
It is convenient to work with a real rather than complex set of functions,  
\begin{equation}
Y_{lm}(\theta,\phi) = \left\{ \begin{array}{c} 
           Y_l^m(\theta,\phi) \\ \sqrt{2}(-1)^{m}{\rm Im} [Y_l^{|m|} (\theta,\phi) ] \\ \sqrt{2}(-1)^{m}{\rm Re}[ Y_l^m (\theta,\phi)]
\end{array} \right. \qquad  {\rm for}
 \qquad
  \begin{array}{c} m=0 \\ m<0 \\ m>0 \end{array} .
\end{equation}
The spherical harmonics are a complete set of orthonormal functions; 
\begin{equation}
\int_0^{2 \pi} d\phi \int_0^\pi \sin(\theta) d \theta \ Y_{lm}(\theta,\phi) Y_{l'm'} (\theta,\phi) 
=  \delta_{mm'}\delta_{ll'}.
\end{equation}

Using surface particles only we compute amplitudes of radial motions $v_r(\theta, \phi, t) $ with the  integral
\begin{equation}
A_{lm} (t) = \int_0^{2 \pi} d\phi \int_0^{\pi} d \theta \ Y_{lm}(\theta, \phi) v_r(\theta,\phi,t) \label{eqn:Alm}
\end{equation}
over the sphere.
This gives us
the amplitude as a function of time for each particular spherical harmonic with index $l,m$.
We take the Fourier transform of $A_{lm}(t)$ 
with 
\begin{equation}
\tilde A(f) = \int dt A_{lm}(t) e^{2 \pi i f t} . \label{eqn:Afft}
\end{equation}
The amplitude for 5 harmonics $|\tilde A_{lm}(f)|$
are plotted in Figure \ref{fig:spec} for the same simulation as shown in Figures \ref{fig:wave_cyl}
and \ref{fig:wave_orth} but using simulation outputs extending to $t=0.15$ after impact. 
We don't show the $l=1$ spherical harmonic as it would only show motion of the body's center of mass. 
We compute the spectrum for $l=m$ spherical harmonics 
 to make sure that $l$ nodes and anti-nodes are present
on the equator, as our impact for this simulation took place on the equator.
The Fourier transforms 
for each spherical harmonic amplitude $|\tilde A_{lm}|$ peaks at a single frequency.
This frequency should be equivalent to that of normal modes with 
 indices $l,n$  and with  associated normal mode frequency $_nf_l$.
We expect that the $l$ of the spherical harmonic is equal to the $l$ index of the normal 
mode.

\begin{table*}
\vbox to 80mm{
	\centering
	\caption{\large Peak Frequencies}
	\label{tab:normalmodes}
\begin{tabular}{llll} 
\hline
Spherical harmonic    &  Peak of $|\tilde A_{lm}|$   & Related mode  & Expected frequency  \\
\hline
$ Y_{00}$   &	278         &  $_0f_0 $  & 292 \\
$ Y_{22}$   &	160         &  $_0f_2 $  & 148   \\
$Y_{33}$   &	236         &  $_0f_3 $  & 208 \\
$ Y_{44}$   &	299         &  $_0f_4 $ & 268  \\
$ Y_{55}$   &	347         &  $_0f_5$ & 329  \\
\hline
\multicolumn{4}{l}
{\multirow{9}{350pt} 
{\small Notes:  The spherical harmonics are listed in the first column.
The peak frequencies (in N-body or gravitational units)
of the amplitudes for each spherical harmonic are measured for
  the Be0 simulation shown in Figures \ref{fig:wave_cyl}, \ref{fig:wave_orth}, and \ref{fig:spec}  using the
  spectra shown in the bottom panel of Figure \ref{fig:spec}.  The simulation has an equatorial impact
  on the Bennu shape model. 
  The measured frequencies are listed  
in the second column.   The relevant normal mode frequency is shown in the third column
in the form $_nf_l$.   The expected frequency of this mode 
for the same simulation is listed in the fourth column.  These frequencies
are computed following section \S \ref{sec:normalmodes} (equations \ref{eqn:n0modes}
and \ref{eqn:l0modes})  
for a homogenous isotropic elastic sphere with a Poisson ratio of $\nu=1/4$ and P-wave speed of 712
in N-body units.  The predicted and measured normal mode frequencies are similar implying
that the mass-spring model simulations are behaving as expected.
}}
\end{tabular}
}
\end{table*}

In Table \ref{tab:normalmodes} we show the peak frequencies for
the spherical harmonic amplitudes measured  from the spectra in the bottom panel of 
Figure \ref{fig:spec}.  These are in N-body units (in units of $t_{\rm grav}^{-1}$).  In the same table we use
the formulas given in section \S \ref{sec:normalmodes} (equations \ref{eqn:n0modes}
and \ref{eqn:l0modes})  
to compute the expected frequencies of the normal modes, assuming 
a homogenous elastic sphere with a Poisson ratio of $\nu=1/4$ and P-wave speed of 712
in N-body units, matching that we estimated for the simulated elastic modulus.

A comparison between predicted and measured normal mode frequencies in 
Table \ref{tab:normalmodes} implies 
 that the simulated Rayleigh wave speed  is about 10\% higher than we predicted
for the simulation, as the $l>0$ modes have somewhat higher frequencies than we expect.
The radial mode $_0f_0$ has frequency slower than we expected, so perhaps the P-wave speed
is somewhat slower.  Our simple elastic model  exhibits normal modes
and with frequencies near those we predict for a homogenous elastic sphere. 

The spectrum shown in Figure \ref{fig:spec} shows that 
the strongest modes excited by
the simulated impact are the football mode with $l=2$, the  triangular mode with $l=3$ and the radial
$l=0$ mode.   We chose
the impulse source time, force amplitude and area so as to mimic an impact that we
suspected would be in a regime capable of exciting slow normal modes.
We have verified that predominantly slow normal modes were excited in the simulation.
This was expected as we chose the impactor and its associated
seismic source function to be in a regime
consistent with excitation of the slowest normal modes.

\begin{figure*}
\begin{center}
$\begin{array}{lll}
\includegraphics[width=2.0in,trim=40 10 0 0, clip]{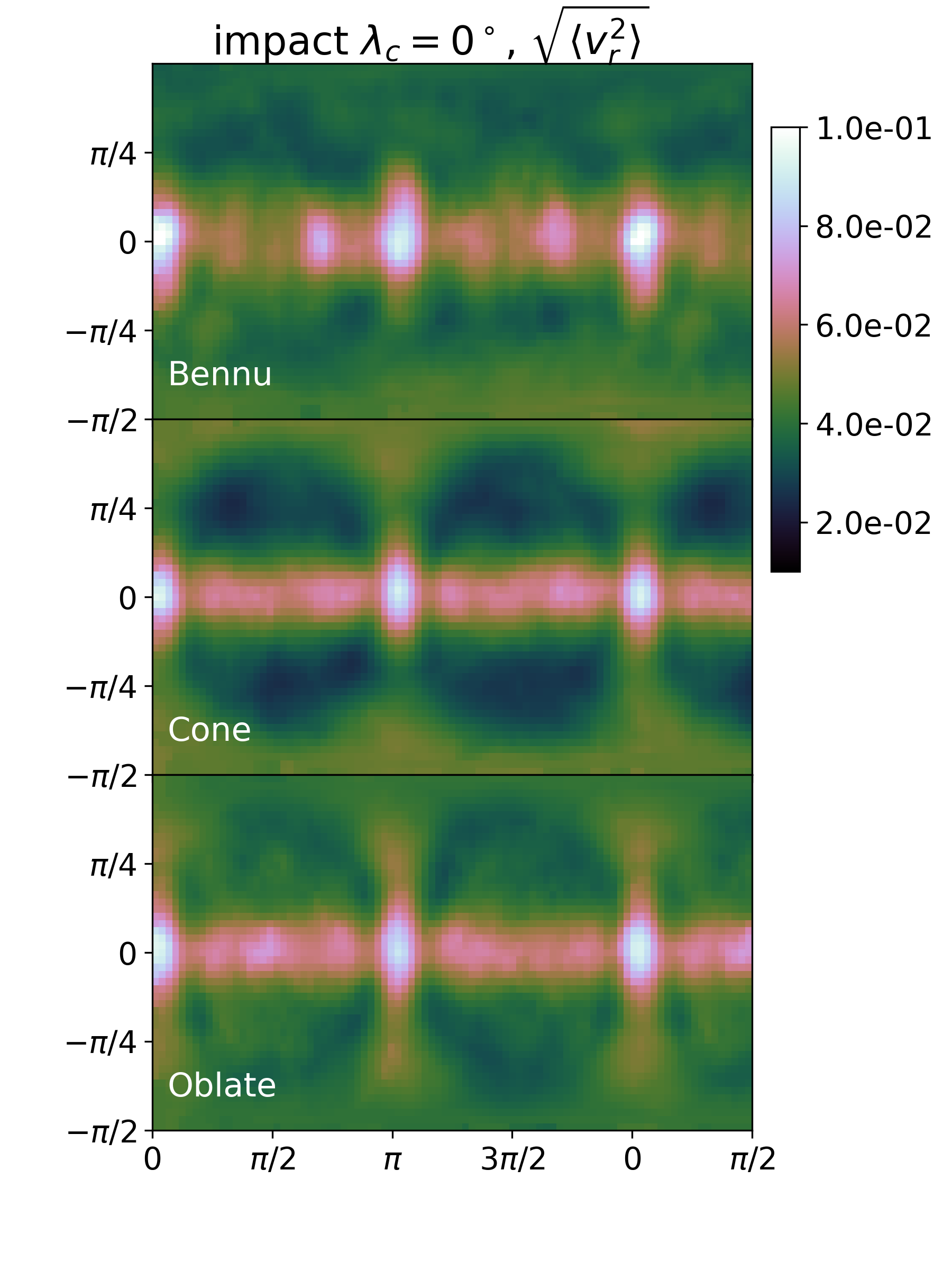} &
\includegraphics[width=2.0in,trim=40 10 0 0, clip]{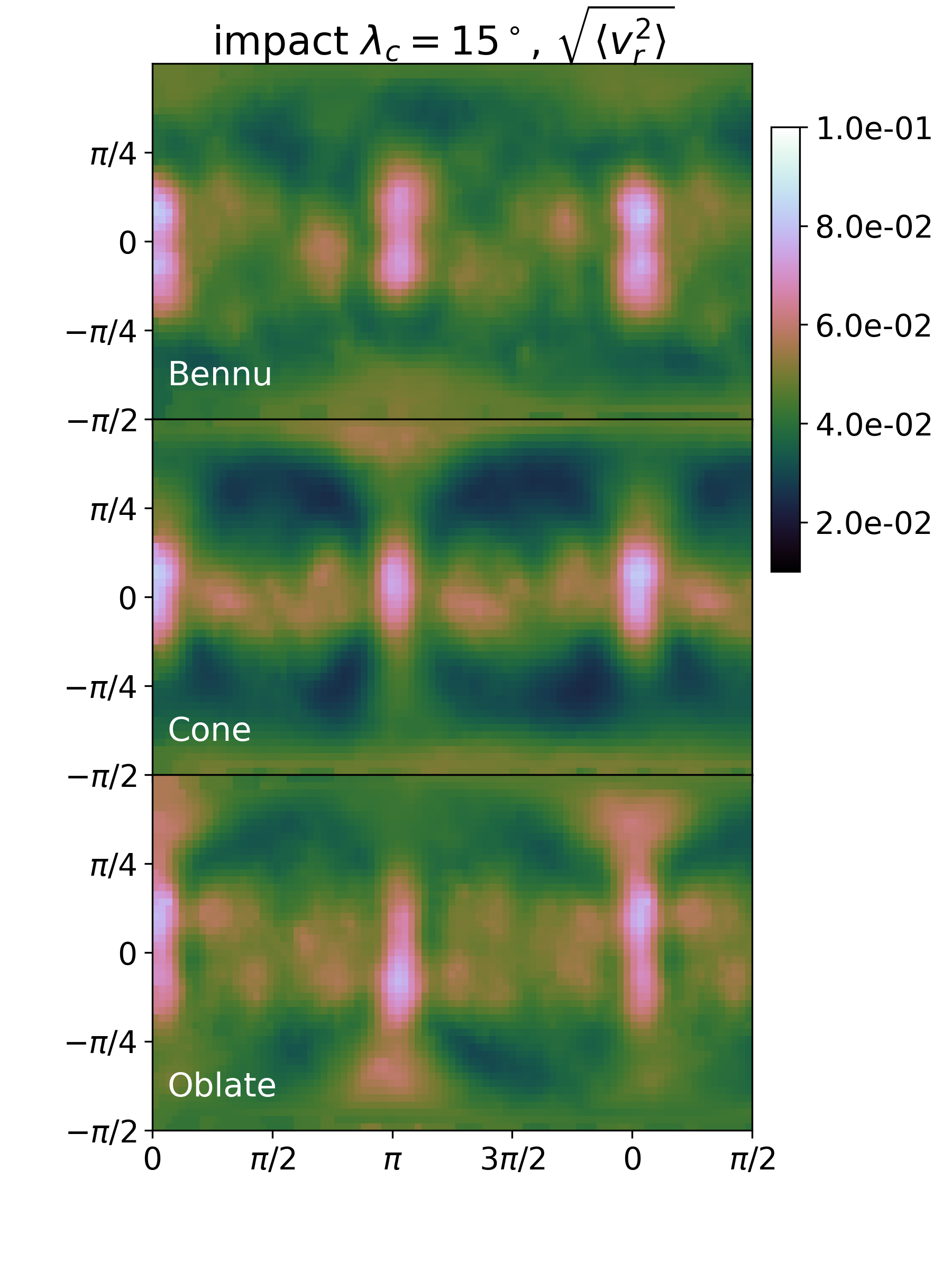} &
\includegraphics[width=2.0in,trim=40 10 0 0, clip]{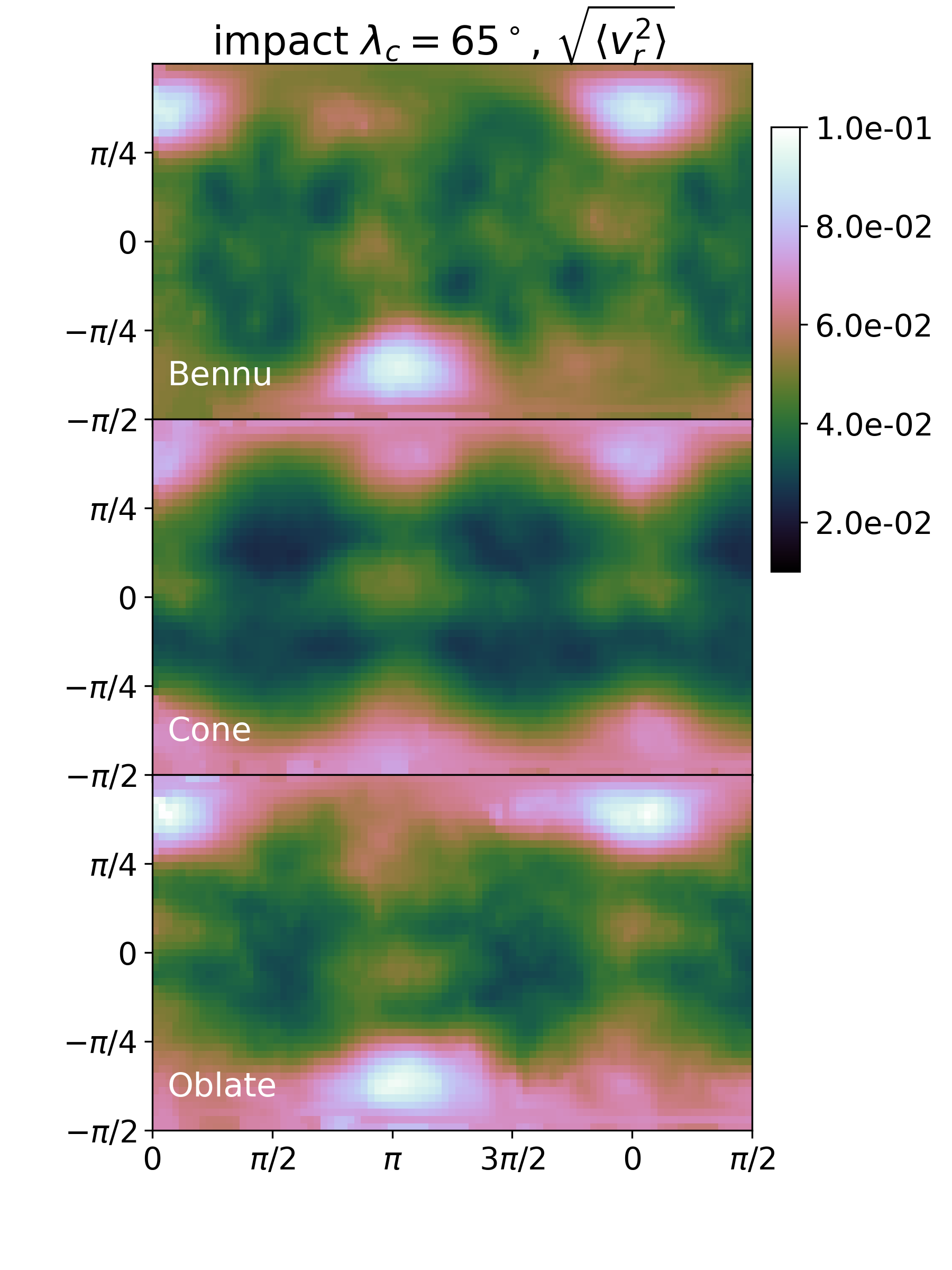}
\end{array}$
\end{center}
\caption{We show the root mean square of radial motions $\sqrt{\langle v_r^2\rangle}$
for three sets of simulations averaged over a time of $\Delta t = 0.09$ or 12 football normal mode 
oscillation periods. 
Axes for individual panels are longitude and latitude, however the longitude on the right
hand side extends past $2 \pi$ so that structure near the impact longitude can be more clearly seen.
In top, middle and bottom panels we show impacts on Bennu, Cone and Oblate shape models.
 The left column shows
equatorial impacts, the middle column shows impacts at a latitude of $15^\circ$ 
and the right column shows impacts at $15^\circ$.  
Impacts were at a longitude of $\phi=0^\circ$.
The simulations are those listed in Table \ref{tab:sims}. 
The Bennu shape model shows peaks for low latitude impacts
at $\phi \sim \pm \pi/2$ because the surface itself has four equatorial
peaks.
}
\label{fig:vr2ave}
\end{figure*}

\begin{figure}
\includegraphics[width=\columnwidth]{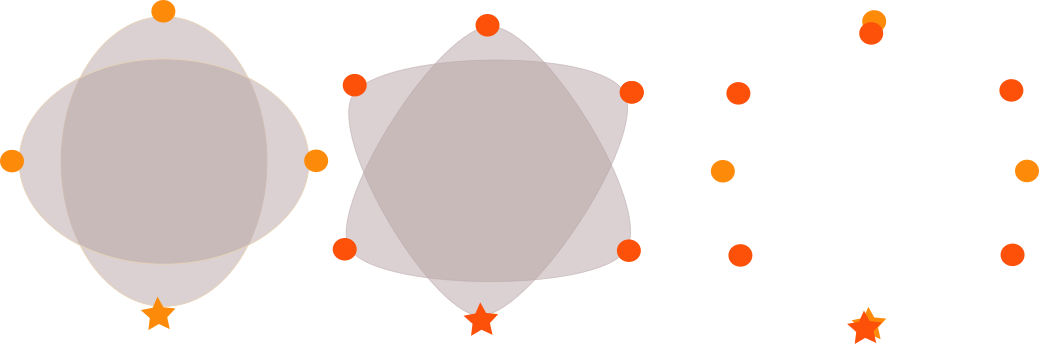}
\caption{We show a cartoon illustration of an impact hitting the equator of a body.  
The point of impact is represented by a star.
On the left, the football ($l=2$) mode
is excited giving four antinodes, shown as orange filled circles.   In the middle panel, the triangular mode ($l=3$)
is excited giving six antinodes (red circles).  On the right panel we show the location of both sets of antinodes.
The antinodes only coincide at impact and its antipodal point.   
}
\label{fig:modes}
\end{figure}

\subsection{Patterns of surface motions}

Because only a few normal modes are strongly excited in our simulations, vibrational motions 
on the surface are not evenly distributed.
From  particles near the surface we compute the root mean square of radial velocity variations
$\sqrt{\langle v_r^2 \rangle}$ by taking the average of $v_r^2$ for each particle
from a series of simulation outputs.  
The patterns of radial velocity variations for nine simulations, using Bennu, Cone and Oblate
shape models and for impacts at latitudes of 0, 15 and 65$^\circ$
are shown in Figure \ref{fig:vr2ave}. 
The parameters for the simulations are listed in Table \ref{tab:sims}.    
The $x$ axes are longitude and the $y$ axes are latitude. 
The longitude range exceeds $2 \pi$ so that
the pattern of vibrations near longitude $\phi=0$, where the impact took place,
 is easier to see.    The portion of the surface on the far right
of these figures is the same as that on the far left.  
To compute the averages for $\langle v_r^2 \rangle$
we used simulation outputs spaced by 5 time steps or $10^{-4}$ and extending over a time period
$\Delta t = 0.09$.  The  time window
is about 12 football mode ($l=2$) periods.  
The football mode is the slowest mode seen in our spectra. 
These figures show the vibrational energy distribution that might be experienced by
the asteroid surface if the seismic waves are damped slowly.  
We note that the spring network only approximates a continuum elastic model, so  on long
time periods,  coupling between modes may mimic seismic wave scattering. 

The color bars in  Figure \ref{fig:vr2ave}
show the size of the radial velocities in N-body units.    
The sizes for $\sqrt{\langle v_r^2 \rangle} \sim 0.05$  are not large, but they when integrated
over the volume are approximately 
 consistent with the total simulated seismic energy $E_s = 0.004$ that
we estimated for our pulse length and force amplitude using parameters from our simulations
(listed in Table \ref{tab:sims} and using equations \ref{eqn:Fs2} and \ref{eqn:Es}).

We estimate the associated
accelerations  by multiplying the velocity $\sqrt{\langle v_r^2 \rangle}$
by an angular oscillation frequency  $\omega = 2 \pi f$.
The frequency of the modes (in Table \ref{tab:normalmodes}) range from $f = 150$ to 300
in units of $t_{\rm grav}^{-1}$.  The 
 accelerations at the peaks are about 60 in N-body units,  allowing
seismic surface motions to launch material off the surface for short periods of time.
Material is not ejected as the initial velocity of lofted material should not be larger than the maximum $v_r$ of the surface.
The acceleration divided by the surface gravitational acceleration is sometimes called 
the acceleration parameter, $\Gamma$.
We should not be surprised by the size of the acceleration parameter, 
as our simulated impact was chosen to be above
the global seismic reverberation threshold with $\Gamma \sim 1$.
Dividing $\sqrt{\langle v_r^2 \rangle}$ by an angular frequency typical of the
oscillation gives us the size of seismic displacements.  
The displacements are small, $\sim 10^{-4}$ in units of body radius, corresponding
to 2.5 cm for Bennu's radius.

Figure \ref{fig:vr2ave}
shows a similarity between surface shape and the distribution of vibrational energy.
The equatorial ridges in the Bennu model and bi-cone model simulations show higher levels of vibrational
energy than mid latitudes, particularly for the low latitude impacts.  
The Bennu shape model  shows four peaks in the vibrational energy distribution.  We
attribute this to the four peaks on its equatorial ridge.  We know this is not due to the angular phases
of the normal modes because the axisymmetric shape models don't show the same four peaks.

Although our spectra (discussed in section \ref{sec:spectrum})
show strong excitation of both football ($l=2$) and triangular 
$l=3$ spherical modes, we see 
the strongest vibrational motions at the impact sight and its antipodal point.
Interference between $l=3$ and $l=2$ modes on the equator can reduce the strength
of antinodes in the football modes that are $\pm 90^\circ$ from the impact point, see Figure \ref{fig:modes}
for an illustration.
Also, due to excitation of different $m$ harmonics
when excited at a single point,  the football mode has power along the entire plane perpendicular
to the impact point.     This is seen in Figure \ref{fig:vr2ave} as a vertical bar
at a longitude of $\phi \sim \pi$.

At an impact latitude of $15^\circ$,  the morphology
of vibrational energy is  different than for an equatorial impact.
Again peaks are seen at impact site and its antipode but these now lie above and below the equator.
Associated surface slumping toward the equator would be lopsided.  
Impacts at higher latitudes cause little excitation near the equator.

Our vibrational energy maps of Figure \ref{fig:vr2ave} show
that vibration tends to be more vigorous along the equator where surface elevation
is highest.   This is similar to a process known as `topographic amplification' (e.g., \citealt{lee09}). 
We would predict, based on the vibration energy  maps,
 that a near equatorial impact 
would preferentially cause surface
slumping toward the equator and to a larger extent at impact site and its antipodal point.
We do not see a single football mode excited that gives vibration preferentially at
four equatorial peaks
as are seen on Bennu's equatorial ridge.  The Bennu shape mode does show increased vibration
at four peaks but that is because these locations are elevated and the normal modes
themselves show more vibration at these points.  Without elevation variations along  the equator,
as is true for our bi-cone and oblate models, four equatorial peaks are not seen in the vibration maps.
If seismic waves are long-lasting, an impact is unlikely to explain the formation of
the four peaks currently seen on Bennu's equatorial ridge, from an initially axisymmetric ridge,
unless the $l=3$ mode damps faster than the football mode.

If the simulated body has a harder core would the change in the normal mode frequencies
 allow four peaks to be seen?   
The $n=0$ modes can be described as constructively interfering Rayleigh waves propagating 
across the surface.  The $l=2$ mode is comprised of a longer wavelength Rayleigh wave
that penetrates deeper
than the $l=3$ mode. With a harder core, we expect a faster $l=2$ mode.  This would reduce
the ratio of the $l=2$ and $l=3$ mode frequencies compared to that
of a homogeneous body.  With closer
frequencies, both modes must simultaneously be excited. 
So an impact on a model with a hard core would not preferentially excite the football $l=2$ mode.
And with closer frequencies, their damping rates would be similar.


\begin{figure*}
\begin{center}
$\begin{array}{lll}
\includegraphics[width=2.0in,trim=40 10 0 0, clip]{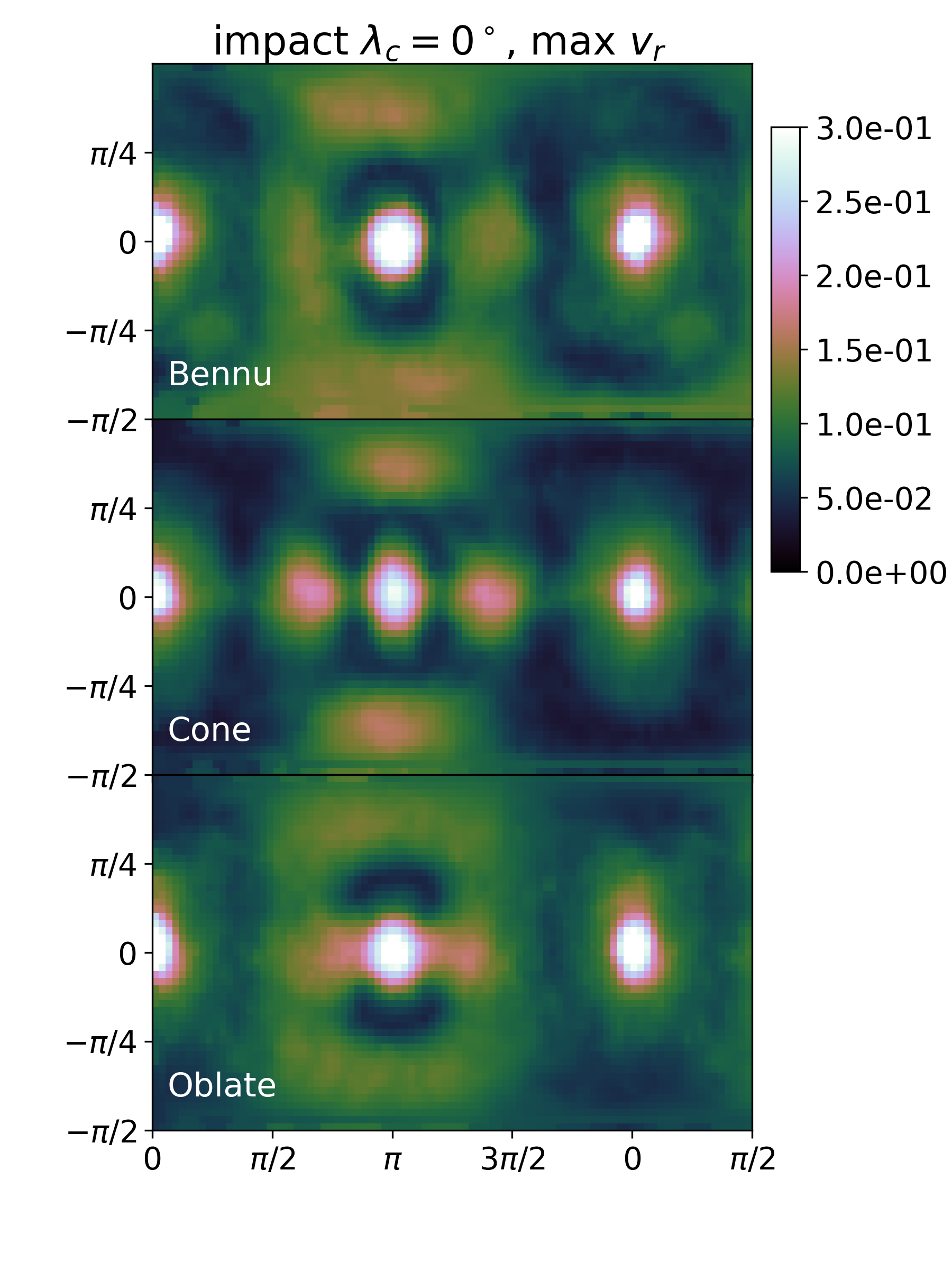}&
\includegraphics[width=2.0in,trim=40 10 0 0, clip]{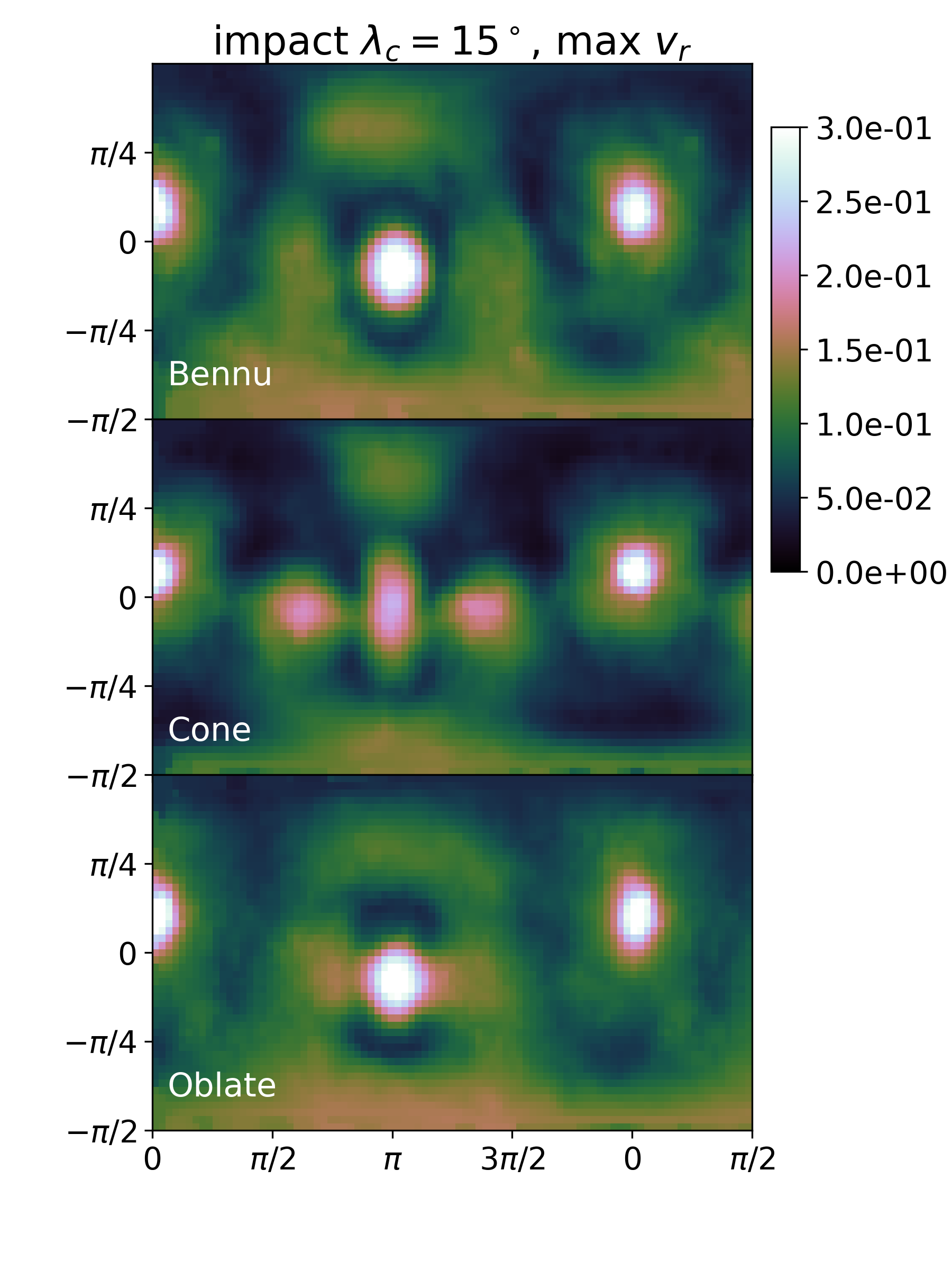}&
\includegraphics[width=2.0in,trim=40 10 0 0, clip]{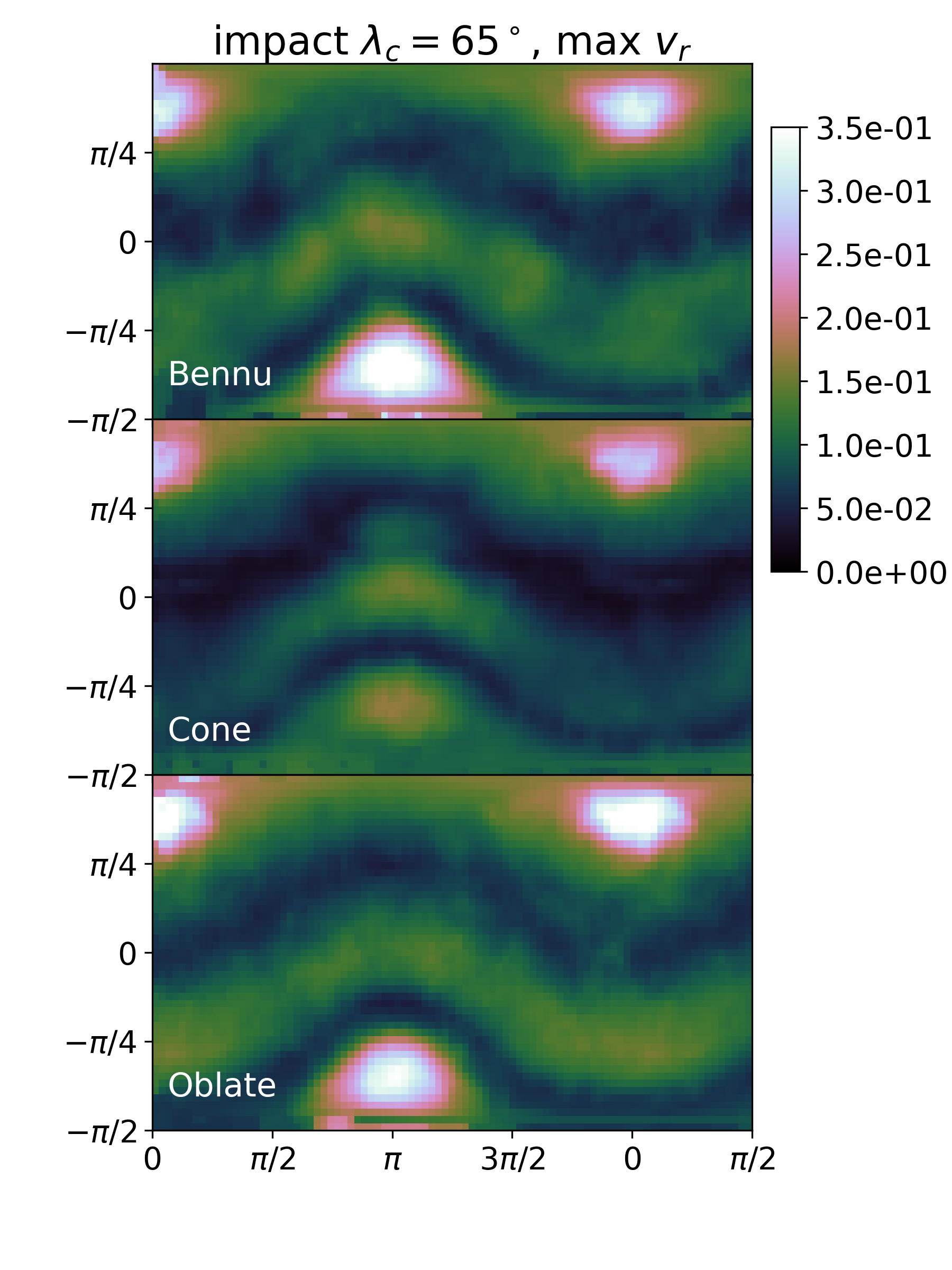}
\end{array}$
\end{center}
\caption{
Similar to Figure \ref{fig:vr2ave} except 
we show the maximum values of the radial velocity $v_r$  during a time $\Delta t = 0.01$, approximately
the time it takes the surface Rayleigh wave to focus on the impact antipode.
Four equatorial peaks are seen in the bi-cone shape model for low latitude impacts.
}
\label{fig:max}
\end{figure*}

\begin{figure*}
\begin{center}
\includegraphics[trim=0 70 0 20, clip, height=8.0in]{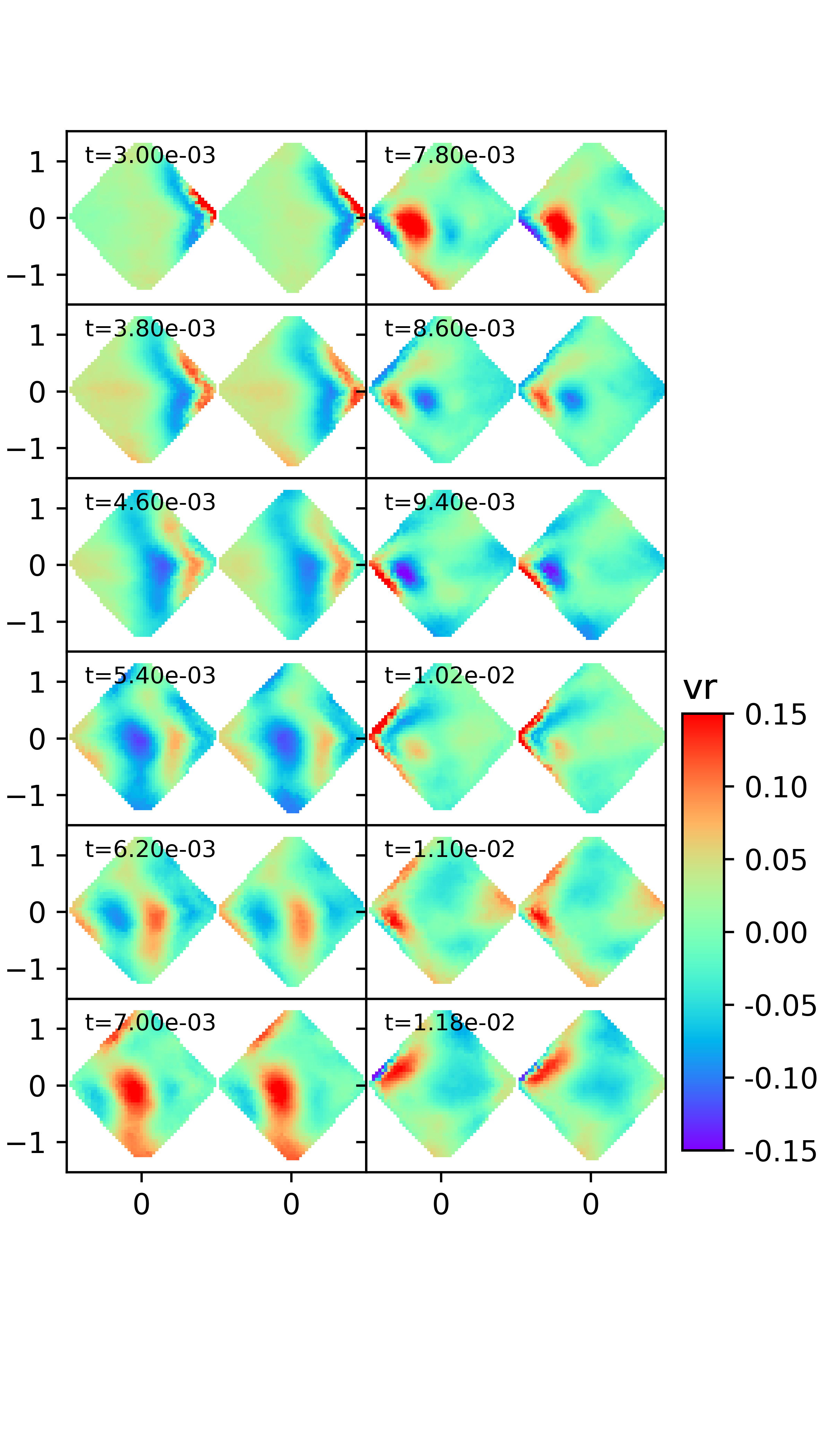}
\end{center}
\caption{Same as Figure \ref{fig:wave_orth} except showing the C15 simulation with a Cone shape model
and a $15^\circ$ latitude impact site.}
    \label{fig:wave_orth_cone}
\end{figure*}

If the seismic waves are rapidly damped, would we see different structure in the vibrational motions?
In Figure \ref{fig:max} we plot the maximum (and positive) value of $v_r$ spanning the window of time
$\Delta t = 0.01$,
from just after impulse has ended to just after the seismic pulse is focused at the antipode.
Four or five equatorial peaks are seen for low latitude impacts.   Four peaks are easiest to 
see in the bi-cone shape model simulations.
We attribute the structure on the equatorial ridge of the bi-cone model to the non-spherical
shape of the body, allowing moderate wavelength seismic 
waves to focus and constructively interfere on the equator.
Figure \ref{fig:wave_orth_cone} shows the wave front 
and is similar to Figure \ref{fig:wave_orth} but shows the C15 simulation using
a bi-cone shape model and with a $15^\circ$ latitude impact. 
The wave front curvature is most easily seen in the left panels of Figure \ref{fig:wave_orth_cone}
and the resulting equatorial focusing on the equator and about $90^\circ$ 
from the impact sight is seen on the top right panel.

We notice that the equatorial peaks near $\pm \pi/2$ seen in the equatorial impacts in Figure \ref{fig:max}
are not $180^\circ$ apart.   The two weaker equatorial peaks on Bennu seem to be nearly
$180^\circ$ apart (see Figure \ref{fig:shape2}).    Focusing of the seismic
wave would be sensitive to body composition and shape, so a more complex  model
might produce peaks in a maximum $v_r$ map with locations 
 similar to  on Bennu's equatorial ridge.

In summary, if seismic reverberations are long lasting, vibrational energy on the surface
is primarily seen at impact point and its antipode.  Regions of higher surface elevation 
(such as the equatorial ridge) show
more vibration. We expect slumping toward the equator from impact site and its antipode.
If seismic reverberations are quickly damped, then  motions
are highest in regions where the impact excited pulse is focused.    In addition
to the impact antipode, focus points could
occur on the equatorial ridge for low-latitude impacts and near $90^\circ$ from the impact point.
 

\section{Vibration induced granular flows on the surface of Bennu}
\label{sec:vibflow}

Above we have shown that 
surface vibrations caused by a large impact reflect the structure of vibrational normal modes,
with some regions on the surface experiencing more shaking than others.
We have also seen that the seismic pulse on a non-spherical body is not uniform in strength
as it traverses the surface.
In this section we explore how these two types of motions might induce granular flows on Bennu's surface. 

Spectral measurements and Bennu's estimated thermal inertia imply that the surface of Bennu
supports regolith that is comprised of grains with typical sizes between 1 mm and 1 cm \citep{emery14}.
The equatorial ridge is redder than the poles, suggesting that the ridge material
contains smaller grains that preferentially migrates to the geo-potential low \citep{binzel15}. 
Notably \citep{tardivel18} predict the opposite trend, that deformation during spin-up  would 
 generate a rocky equator and sandy tropics. 
We lack constraints on the depth of Bennu's regolith layer, however the asteroid porosity and near
spherical shape (implying an absence of embedded monoliths) could be consistent with thick layers of regolith, subsurface rubble or porous and cracked rock. 
The depth of  flow induced by vibration is likely to be smaller than the lateral dimensions of our problem
so we restrict our discussion to surface flows (e.g., \citealt{aradian02}).
This assumption neglects 
 rearrangements deep inside the asteroid that could be caused by the passage of a seismic pulse. 

Granular flows are complex, exhibiting a rich phenomenology even in the absence
of vibration (e.g., \citealt{aranson02,forterre08}).
As the vibrational modes have periods similar to a Hz, if shape changes are caused by
impacts, they must occur during a short time
(a few hundred periods is only a few minutes).
Significant mass flows, those large enough to change the body shape,
are not possible in a short time unless the thickness of the flowing layer
significantly exceeds the typical grain size.    
The granular flow must be in a dense 
and liquid-like state, similar to flows on inclined planes,
rather than a gaseous or solid-like state where jamming and avalanches can take place. 
For discussion on granular flow regimes  see \citet{campbell90,forterre08}. 

Bennu's surfaces exhibits downhill slopes (see Figures by \citealt{scheeres16}), 
so its surface material is not currently fluid-like.
Bennu's surface slopes are below the critical angle of repose of the granular
medium on its surface.  
However, vigorous vibration  can cause fluid-like behavior in
a granular medium, or vibro-fluidization  (e.g., \citealt{savage88}),  lowering
the critical angle of the surface.
We adopt  a depth-averaged description for a surface flow, using variables similar to those used for 
Saint-Venant equations for shallow water waves
that have been modified for granular flows \citep{savage89,aradian02,forterre08}.
The advantage of this approach is that we don't need to model the velocity profile
in the flowing layer. 
We assume that there is a well defined interface between a flowing layer and
static underlying material.
We describe the flow with a thickness for the flowing layer, $h$, and ${\bf u}$ a depth averaged velocity
in the flowing layer, 
giving mass flux ${\bf Q} = \rho h{\bf u}$ (see section 4 by \citealt{forterre08} or \citealt{aradian02}).  
Conservation of mass is described with 
\begin{equation}
\frac{\partial h }{\partial t}  + {\boldsymbol \nabla} 
\cdot \left(h {\bf u} \right) = \frac{\partial H}{\partial t}. \label{eqn:com}
\end{equation}
We have assumed an incompressible medium.  
The divergence is restricted to gradients on the surface.
With $ \frac{\partial H}{\partial t}=0$,
there is no exchange between the surface flow and the underlying
static base.  If there is mass exchange, the static base increases or decreases
in height at a rate given by $ \frac{\partial H}{\partial t}$.

At each location  on the surface the mass flux  ${\bf Q}$ should depend on 
the surface gravitational acceleration, $g_{\rm geo} = |{\bf a}_{\rm geo}|$,  
the slope of the surface $\beta$, the amplitude of seismic vibrations 
characterized by the mean $\langle v_r^2 \rangle$ or the maximum $v_r$ during the seismic pulse,  
and the depth of the flowing layer $h$.
Here ${\bf a}_{\rm geo}$ is the direction of acceleration on the surface due to gravity and body spin.
We assume that the flow velocity ${\bf u}$ is in the downhill direction,
that given by the gradient of the geopotential on the surface. 
The flow rate $u$ and mass flux $\rho h {\bf u}$ 
should increase with increasing surface slope, vibration velocity, and 
decreasing surface gravity.

We explore two types of models, a hopping block model with flowing layer
depth $h$ independent of position on the surface
and a fluidized layer model with $h$ set by the level of vibrations.
The hopping block model is similar to the hopping and sliding block model (slipping with friction)  used  by \citet{richardson05}
to estimate surface flows caused by seismic reverberation.  
The average flow velocity $u$ varies with position on the surface for both models.

In section \S \ref{sec:slope} we compute the surface slope, as flow rates should depend on it.
In section \S \ref{sec:block}
we consider a hopping block model with a depth $h$ for the
flowing layer that is independent of surface position.  The model is computed for
flow caused by a seismic pulse (dependent on max $v_r$) and for
prolonged seismic reverberation (and dependent on $\langle v_r^2 \rangle$).
In section \S \ref{sec:vibro}  we 
use the notion of granular temperature to estimate a depth $h$ for vibro-fluidization
and  an empirical flow rule for granular flows on inclined planes to estimate 
a depth-averaged velocity ${\bf u}$ in the flowing layer.  
 In section \S \ref{sec:height}
these three models are used to estimate the distribution, rate and extent 
that flowing granular material is accumulated on the surface due to
an energetic impact. 

\subsection{Surface acceleration and slope}
\label{sec:slope}

To estimate surface flows we need to compute
 the vertical acceleration at the surface  for
an aspherical and spinning body.
The geopotential is defined as 
\begin{equation}
\Phi_{\rm geo}({\bf r}) \equiv -\frac{1}{2} \Omega^2 (x^2 + y^2) - U({\bf r})
\end{equation}
(following \citealt{scheeres16} for $z$-axis, spin axis and a body principal axis aligned)
where ${\bf r} = (x,y,z)$ are coordinates in the body frame with respect to the center of mass, 
$\Omega$ is the body's spin rate and
$U({\bf r})$ is the gravitational potential energy per unit mass; 
$U({\bf r}) = \int d^3 {\bf r}'  \rho({\bf r}') |{\bf r} - {\bf r}'|^{-\frac{1}{2}}$.
We compute a local gravitational acceleration vector at a surface position ${\bf r}_s$
\begin{align}
{\bf a}_{\rm geo}({\bf r}_s) &=  \left. {\boldsymbol \nabla} \Phi_{\rm geo}({\bf r}) \right|_{{\bf r_s}}
\label{eqn:ageo_vec}
\end{align}
where the gradient is computed in three-dimensions and then evaluated at a point on the surface ${\bf r}_s$.

A surface slope is the relative orientation between the surface normal vector and the local gravitational 
acceleration vector.  
We describe the surface slope with an angle $\beta_{\rm slope}$, with
\begin{equation}
\sin \beta_{\rm slope} ({\bf r}_s) = \frac{  {\bf a}_{\rm geo} ({\bf r}_s) 
\cdot \hat {\bf n}({\bf r}_s) }{| {\bf a}_{\rm geo}  ({\bf r}_s)   |} \label{eqn:beta}
\end{equation}
and with $\hat {\bf n} ({\bf r}_s) $  a unit vector normal to the surface.  
The slope angle $\beta_{\rm slope}=0$ 
if surface normal and gravitational acceleration vectors are aligned.

\subsection{Hopping model}
\label{sec:block}

At acceleration parameter greater than 1 (with accelerations due to vibration exceeding the surface
acceleration), particles on the surface are pushed off the surface into free fall.   
A particle that is launched from the surface with vertical velocity $v$ is in flight
for a time $\Delta t = 2v/g_{\rm geo}$.  With a surface slope $\beta$, the particle
travels downhill horizontally a distance 
$\Delta x \sim v \beta \Delta t \sim  2 \beta v^2/g_{\rm geo} $
or at an average horizontal speed $u \sim \Delta x/\Delta t \sim v \beta$.
Here the notion of horizontal and vertical are with respect to the gravitational acceleration vector.

For long timescale seismic reverberation
we estimate the vertical velocity $v$ using the root mean square of radial
velocity variations $\sqrt{\langle v_r^2 \rangle}$, giving an average horizontal velocity
\begin{equation}
u \sim \beta \sqrt{\langle v_r^2 \rangle}.  \label{eqn:u_hop}
\end{equation}
We assume flow is downhill, which we denote with a vector on the surface
$\hat {\boldsymbol \beta}$.
With a constant depth $h$ of flowing material equation \ref{eqn:com} gives
a rate the surface elevation changes
\begin{equation}
\frac{1}{h} \frac{\partial H({\bf r}_s)}{\partial t} \approx {\boldsymbol \nabla} \cdot 
\left(\beta \hat {\boldsymbol \beta} \sqrt{\langle v_r^2  \rangle}  \right) \label{eqn:dHdt_hop}
\end{equation}
at a position ${\bf r}_s$ on the surface, with quantities on the right
hand side 
dependent on surface position ${\bf r}_s$ and with the gradient operator applied on the surface.

For a seismic surface pulse (and assuming that the seismic waves damp quickly)
we estimate the distance  traveled on the surface $\Delta x$ from the maximum
outward velocity 
$v_r$ and with a single hop (and no slip), giving
\begin{equation}
\Delta x \sim \frac{2 \beta |{\rm max} v_r|^2}{g_{\rm geo}}.  \label{eqn:x_hop}
\end{equation}
With a constant depth $h$ of flowing material, equation \ref{eqn:com} for conservation
of mass gives a total height change
following a seismic pulse
\begin{equation}
\frac{\Delta H({\bf r}_s)}{h}   \approx {\boldsymbol \nabla} \cdot 
\left( \frac{2 \beta \hat {\boldsymbol \beta} |{\rm max} v_r |^2}{g_{\rm geo}} \right).
\label{eqn:deltaH_hop}
\end{equation}
and with quantities on the right hand side dependent on surface position ${\bf r}_s$.


\subsection{Vibro-fluidization flow model}
\label{sec:vibro}

Kinetic theories for granular flow are often dependent on the notion of 
a granular temperature \citep{ogawa80,jenkins83,lun84,ding90,warr95,goldhirsch08} that 
 is estimated from the square of velocity fluctuations in the grains.
In a vibrating granular bed,  the granular temperature is proportional to the root mean
square of  the velocity of vibrations  
 \citep{warr95,kumaran98}.  
Following scaling by \citet{kumaran98} for media with weak dissipation (little friction), 
we assume the granular temperature
\begin{equation}
T_{\rm granular} ({\bf r}_s) \approx \langle v_r^2 ({\bf r}_s) \rangle,  \label{eqn:Tg}
\end{equation}
where $ \langle v_r^2 ({\bf r}_s) \rangle$ represents a time integrated average
of the velocity fluctuations caused by seismic vibrations.
The granular temperature gives an estimate for a pressure associated with vibrational kinetic energy,
$P_{\rm kinetic} = \rho T_{\rm granular}$.
We compare the kinetic pressure 
to hydrostatic pressure $\rho g_{\rm geo} h$  at a depth $h$
where $g_{\rm geo} \approx |{\bf a}_{\rm geo}|$ 
is the acceleration near the surface due to gravity and body spin.
Equating hydrostatic pressure at depth $h$ to that associated with vibrational motions gives 
an approximate depth for the base of a vibro-fluidized layer,
\begin{equation}
h({\bf r}_s) \approx \frac{ \langle v_r^2 ({\bf r}_s) \rangle}{g_{\rm geo} ({\bf r}_s)}. \label{eqn:h_fluid}
\end{equation}
This estimate for depth assumes that a significant fraction of the vibrational energy 
goes into random grain motions.  If the grain motions are correlated
 then the fluidized  layer would be deeper
than estimated by equation \ref{eqn:h_fluid}.

What size does equation \ref{eqn:h_fluid} give for  our impact simulations? 
Figure \ref{fig:vr2ave} shows peaks with 
$\langle v_r^2 \rangle \sim  10^{-2}$
in N-body units.  In our N-body units, acceleration for a non-rotating spherical body is 1.
At the equator the surface gravity is about 1/2 of that without rotation, so equation \ref{eqn:h_fluid}
 gives $h \sim 0.02$ in units of radius.  For Bennu this corresponds to a depth of 5 m.
The ratio of fluidization depth to 
grain size (using grains of size 0.3 cm) is $\sim 170$, exceeding many experiments
but large enough that a shallow water wave analogy is appropriate. 

If the depth of the fluidized layer $h$ is set by the vibrational kinetic energy, 
there is exchange between flowing material and the
underlying medium.  
Ignoring  slow variations in the time averaged vibrational energy, equation \ref{eqn:com}
becomes
\begin{equation}
\frac{\partial H ({\bf r}_s)}{\partial t} = {\boldsymbol \nabla} 
	\cdot \left[  h({\bf r}_s) {\bf u}({\bf r}_s) \right] . \label{eqn:dHdt}
\end{equation}
With a relation for the flow velocity ${\bf u}$, this equation gives us an
 estimate for how granular flow on the surface 
increases or decreases the local surface height.


When described in terms of a continuum model,
dense flows exhibit shear-rate dependent stress
giving the flow a viscous-like behavior \citep{bagnold54,aradian02,aranson02,forterre08,holsapple13}.
Gravitational acceleration 
gives a component of the pressure gradient that is 
parallel to the surface and depends on the surface slope.
A description for the vertical velocity profile of a flow and an effective flow viscosity 
(which could be depend on the shear as in the Bagnold description) gives a relation between
averaged flow velocity and surface slope. 
Experiments and numerical simulations of steady granular flows on inclined
planes \citep{pouliquen99,silbert03,midi04,deboeuf06} support an empirical flow rule 
relating the flow rate to the surface slope,
\begin{equation}
\frac{ u} { \sqrt{g_{\rm geo} h}}  \sim  f(\beta). \label{eqn:u_scale}
\end{equation}
The quantity on the left $u/\sqrt{g_{\rm geo} h}$ can be identified as a Froude number making
the function $f()$ dimensionless.
With equation \ref{eqn:h_fluid} for $h$, $u \propto \sqrt{g_{\rm geo} h} = \sqrt{\langle v_r^2  \rangle}$
and is dependent on the root mean square of vibrational velocity
and consistent with the estimate we derived for hopping in equation \ref{eqn:u_hop}.
The function  $f(\beta)$ depends on slope and grain properties such as packing fraction
and the critical angle of repose of the granular medium, $\beta_c$.
Unfortunately the empirical scaling laws for steady flow on inclined planes neglect vibration.
Without vibration and with slope $\beta<\beta_c$ below the critical angle of repose,
there would be no flow.
For our vibro-fluidized layer we modify the scaling relation of equation \ref{eqn:u_scale}
that is successful at describing granular flows on inclined planes. 
We desire a function $f()$ that
gives a more flow at higher slopes and no flow when the slope is zero.
A linear function  does this
\begin{equation}
f(\beta) =  K \beta \label{eqn:ffun}
\end{equation}
with $K$ a unitless scaling factor.

Combining equations \ref{eqn:h_fluid}, \ref{eqn:dHdt}, \ref{eqn:u_scale}, and \ref{eqn:ffun}
we  estimate
 the height increase or decrease due to impact excited vibrations
\begin{align}
\frac{\partial H({\bf r}_s)}{\partial t} \sim  {\boldsymbol \nabla}  \cdot
\left[ K \beta \hat {\boldsymbol \beta}  \frac{ \langle v_r^2 \rangle^\frac{3}{2} }{ g_{\rm geo} } 
\right] \label{eqn:dH1}
\end{align}
with the gradient restricted to the surface and quantities on the right
hand side  (excepting $K$) that are functions of surface position ${\bf r}_s$.
Despite the  uncertainties in estimating the flow of vibrated
granular materials this equation shows expected trends.
There is more flow where there is more vibration, such as at antinodes.
There is no flow predicted where the surface slope is zero and there is more flow
where   surface acceleration is lower  at the equator.
While the direction of flow is  toward the equator, height changes
can occur non-uniformly in azimuth because the vibrational energy depends on longitude.

\begin{figure*}
\begin{center}
$\begin{array}{lll}
\includegraphics[width=2.0in,trim=40 10 0 0, clip]{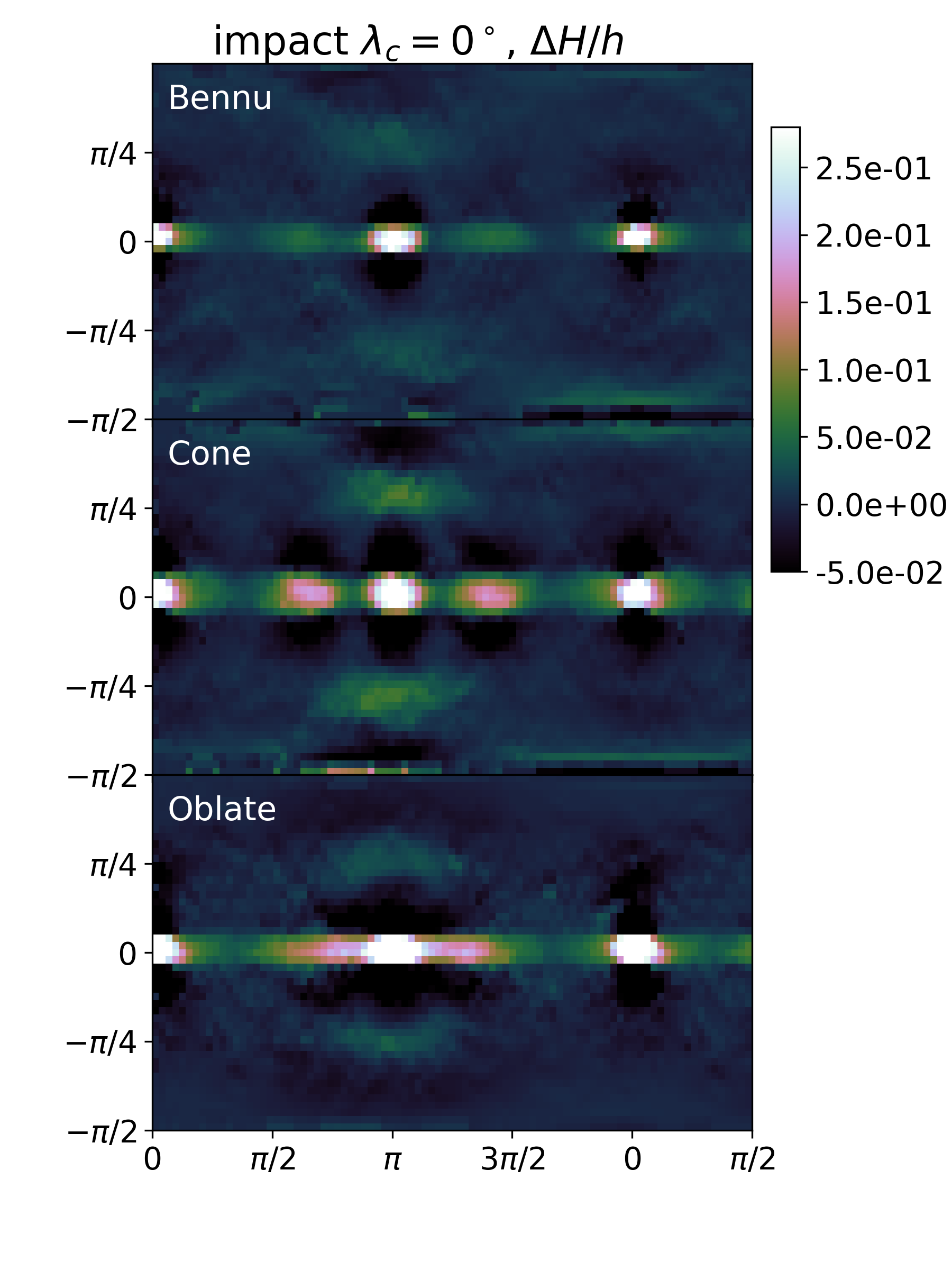}&
\includegraphics[width=2.0in,trim=40 10 0 0, clip]{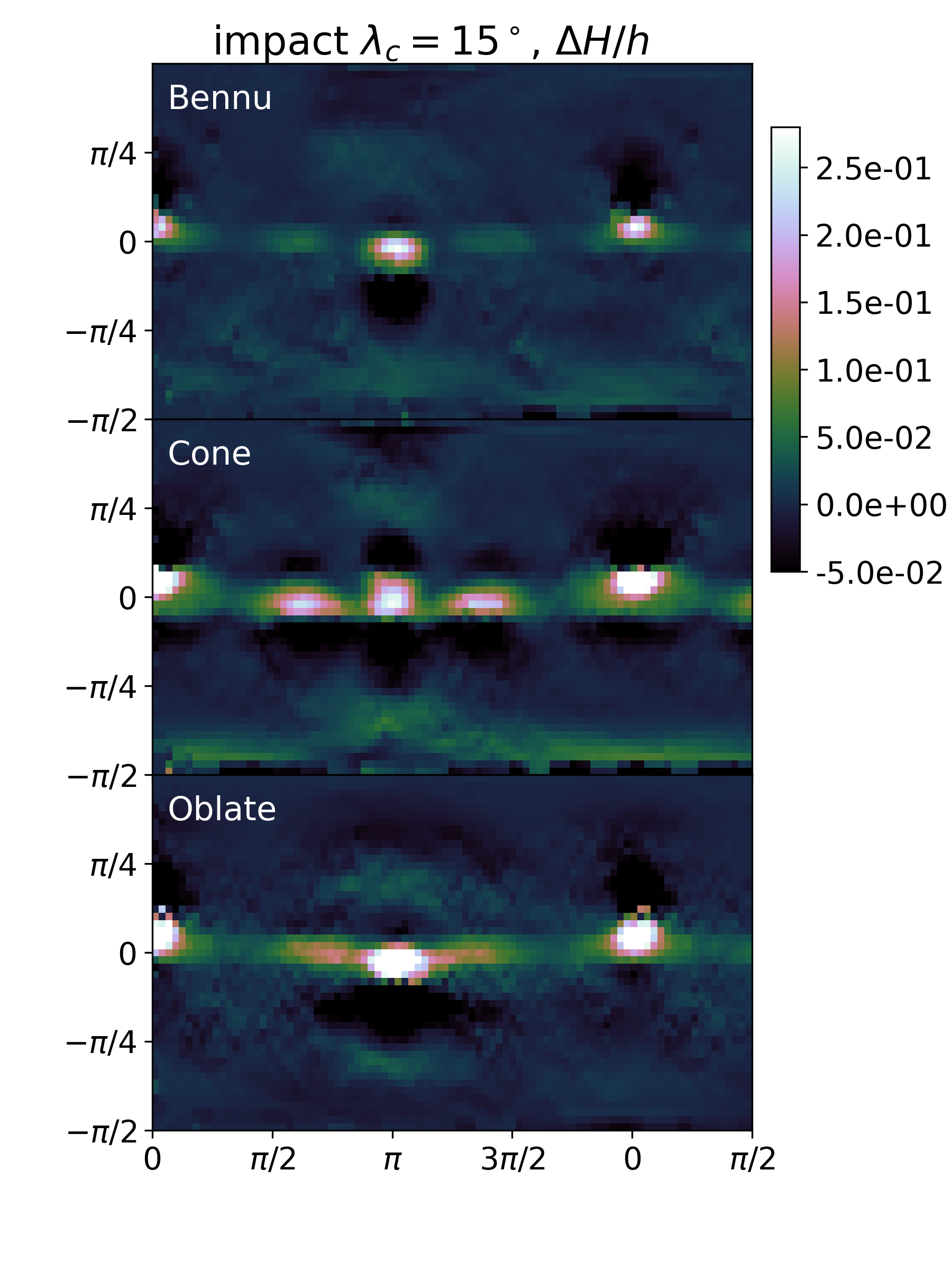}&
\includegraphics[width=2.0in,trim=40 10 0 0, clip]{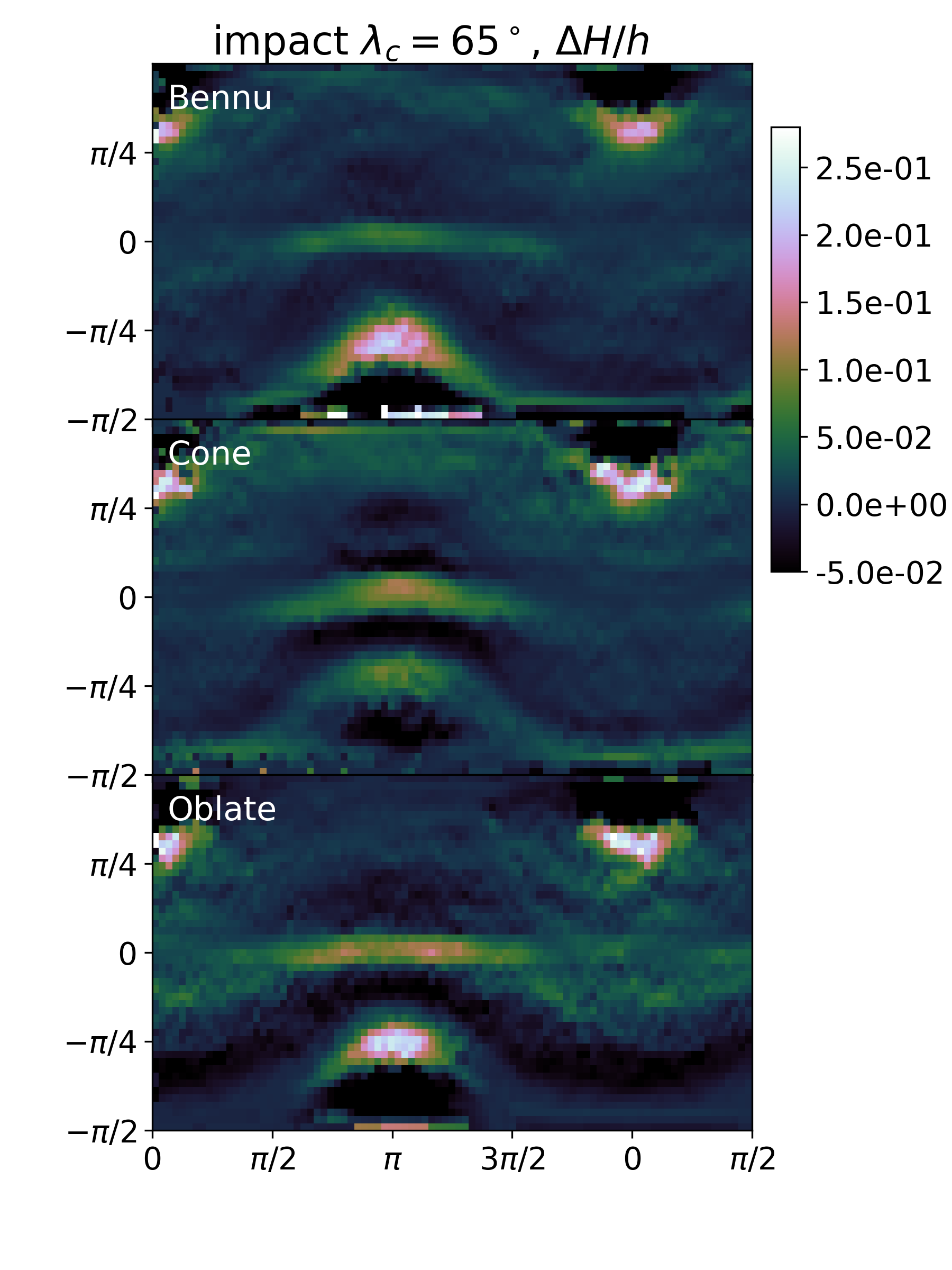}
\end{array}$
\end{center}
\caption{
Estimates for the change in surface height divided by flow layer depth, $\Delta H/h$,
for a seismic jolt  one hop model caused by energetic impacts.  The height changes are estimated
using the maximum positive $v_r$ shown in Figure \ref{fig:max} from the initial seismic pulse
and equation \ref{eqn:deltaH_hop} for the height change divided by depth of flowing layer.
As before, we show nine simulations, for three different impact latitudes (left to right) and three 
different shape models (top to bottom).
}
\label{fig:grad_short}
\end{figure*}

\begin{figure*}
\begin{center}
$\begin{array}{lll}
\includegraphics[width=2.1in,trim=40 10 0 0, clip]{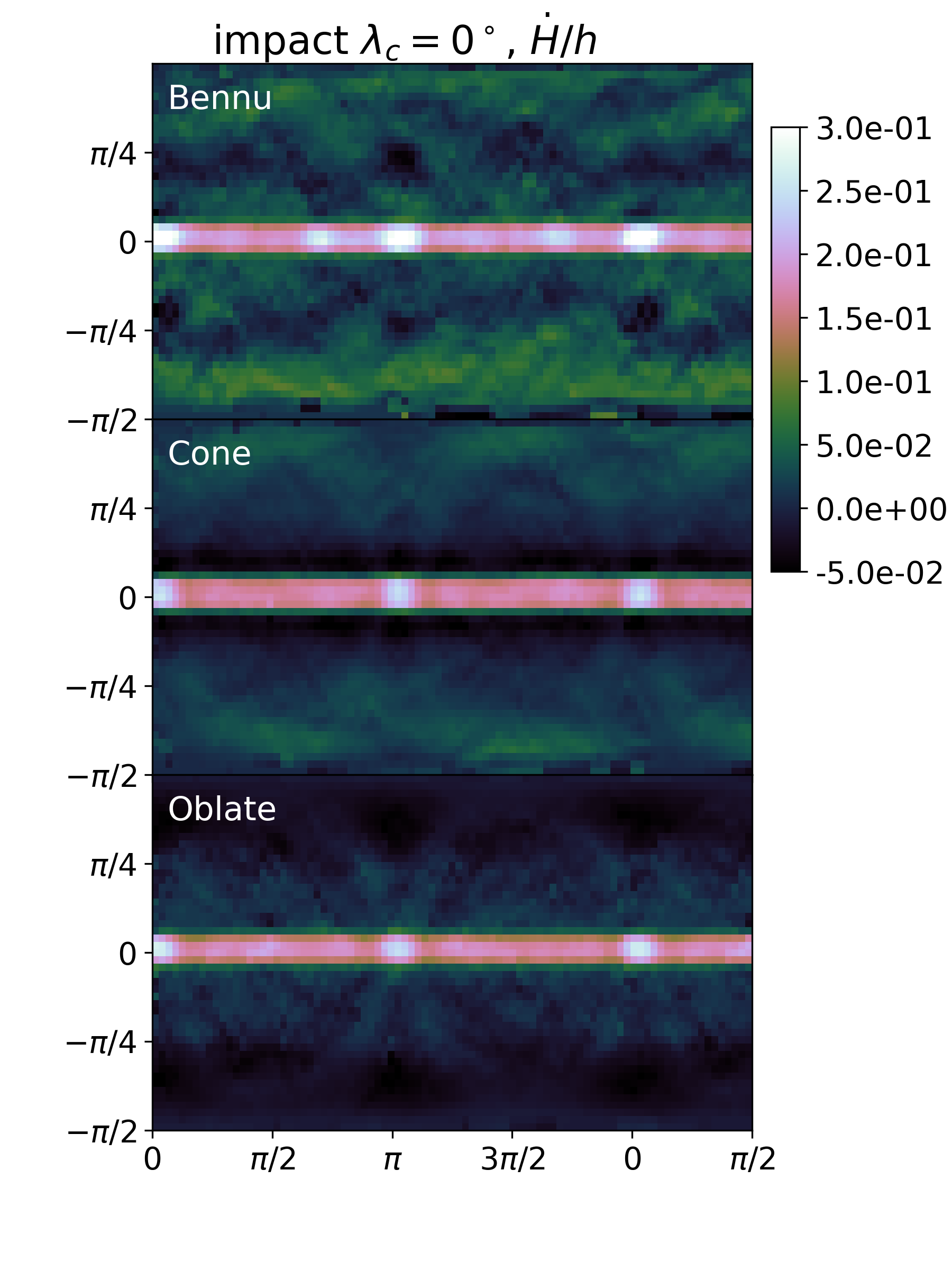}&
\includegraphics[width=2.1in,trim=40 10 0 0, clip]{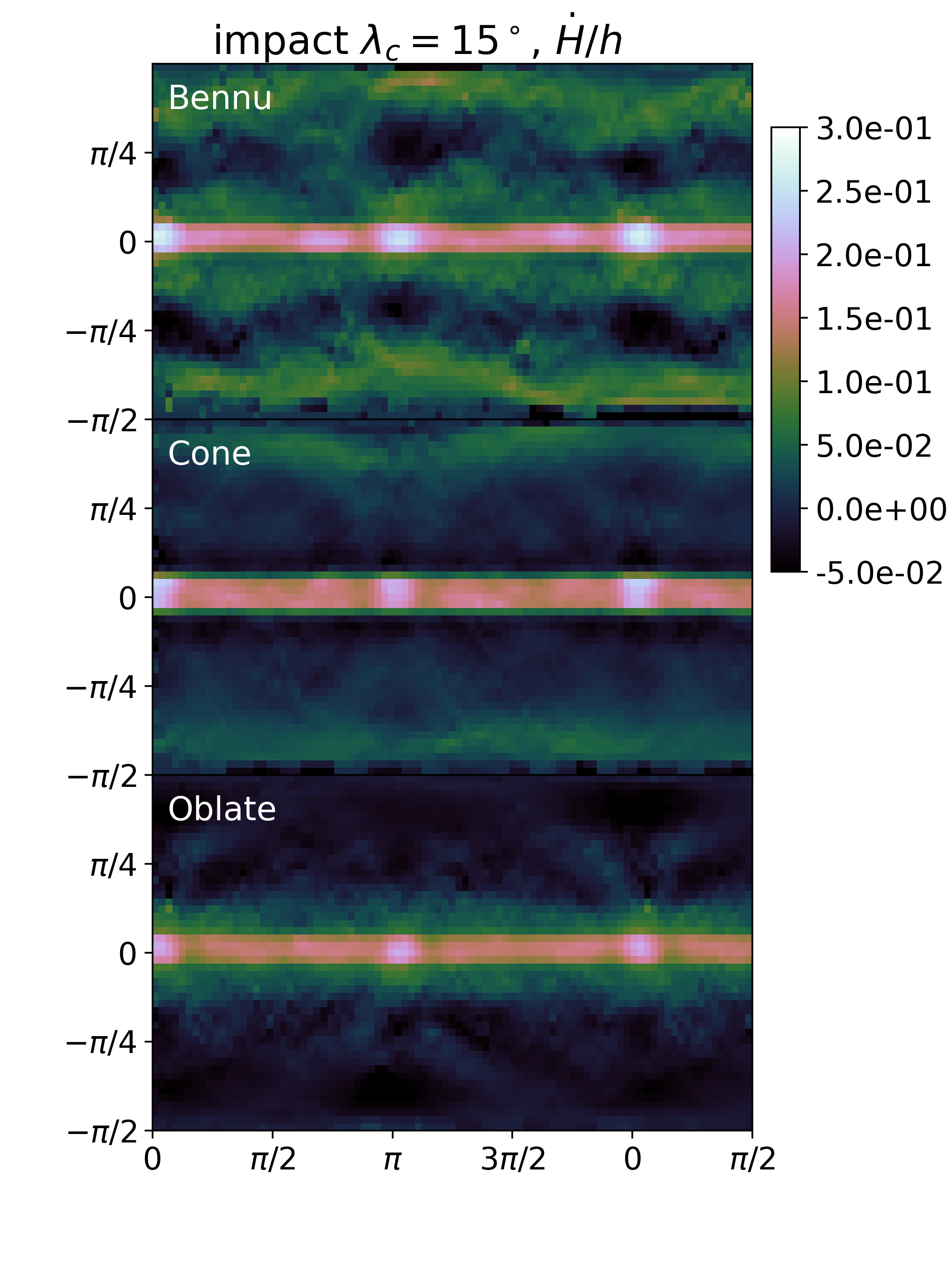}&
\includegraphics[width=2.1in,trim=40 10 0 0, clip]{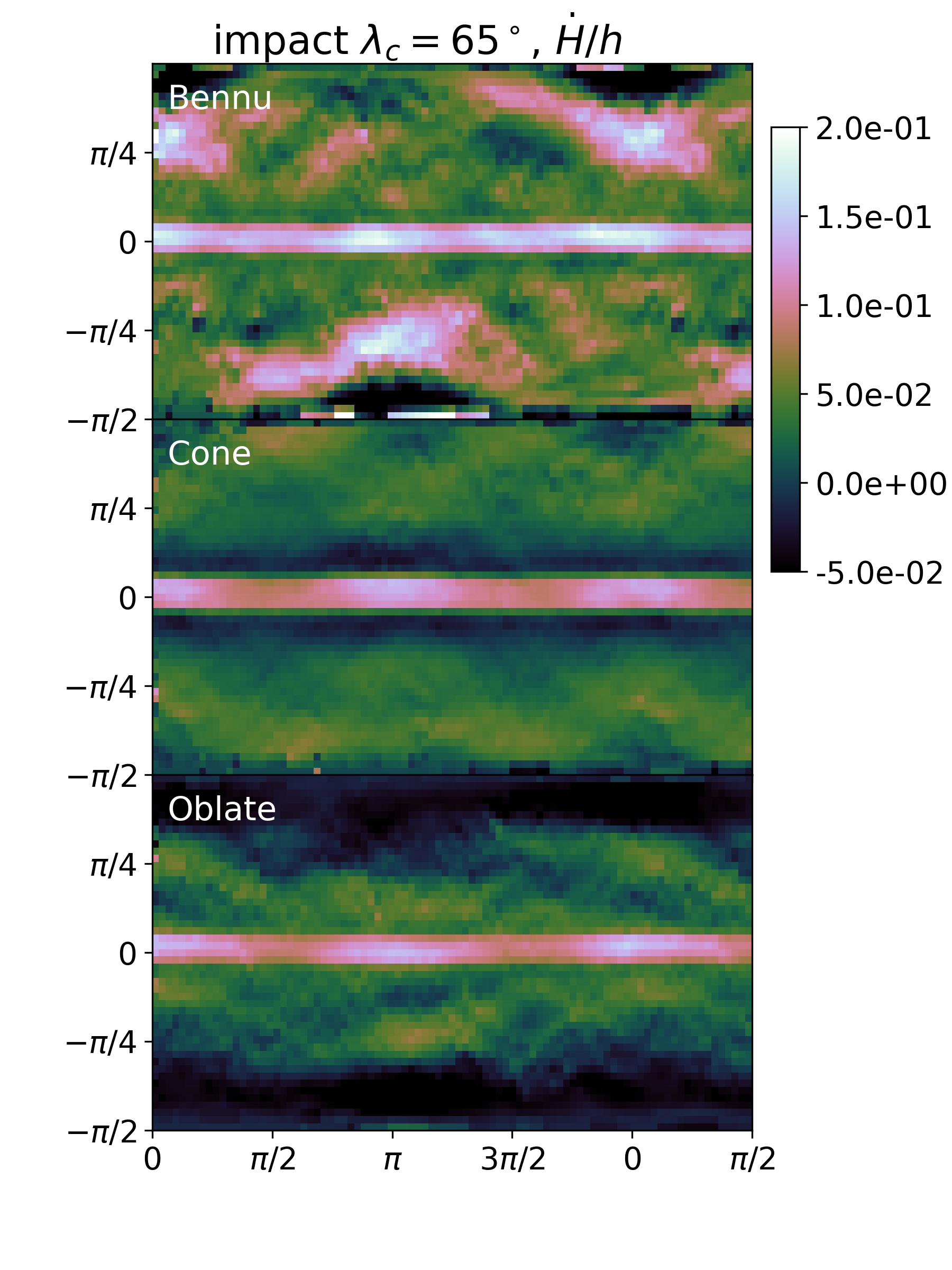}
\end{array}$
\end{center}
\caption{
Estimates for the rate of change of surface height divided by flow layer depth, 
$\frac{1}{h} \frac{\partial H}{\partial t}$ long lasting seismic reverberations and
a constant depth flow model.  The height changes are estimated using
$\langle v_r^2 \rangle$ shown in Figure \ref{fig:vr2ave} and equation \ref{eqn:dHdt_hop} for
our nine simulations.
}
\label{fig:grad_ch}
\end{figure*}

\begin{figure*}
\begin{center}
$\begin{array}{lll}
\includegraphics[width=2.1in,trim=40 10 0 0, clip]{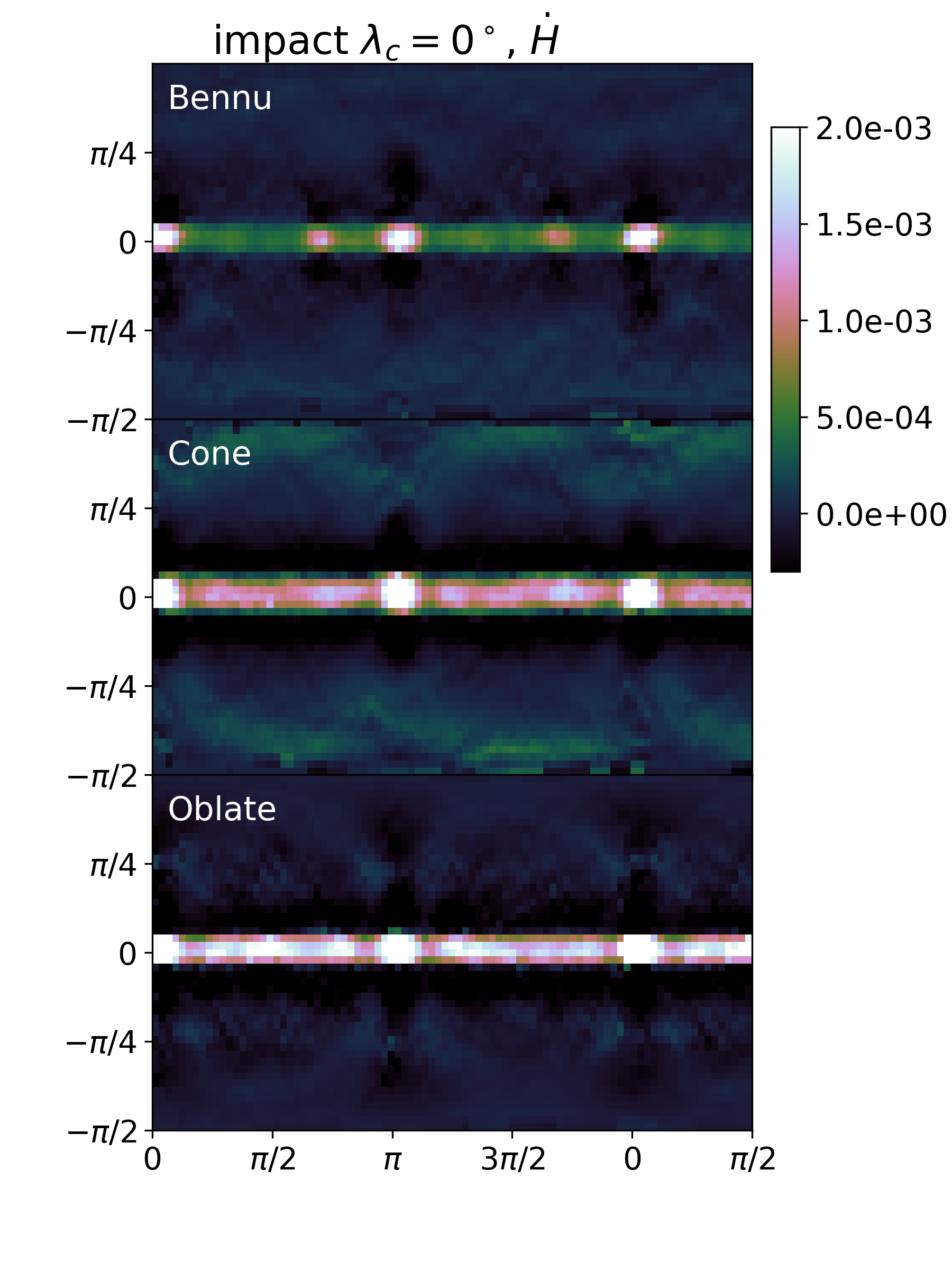}&
\includegraphics[width=2.1in,trim=40 10 0 0, clip]{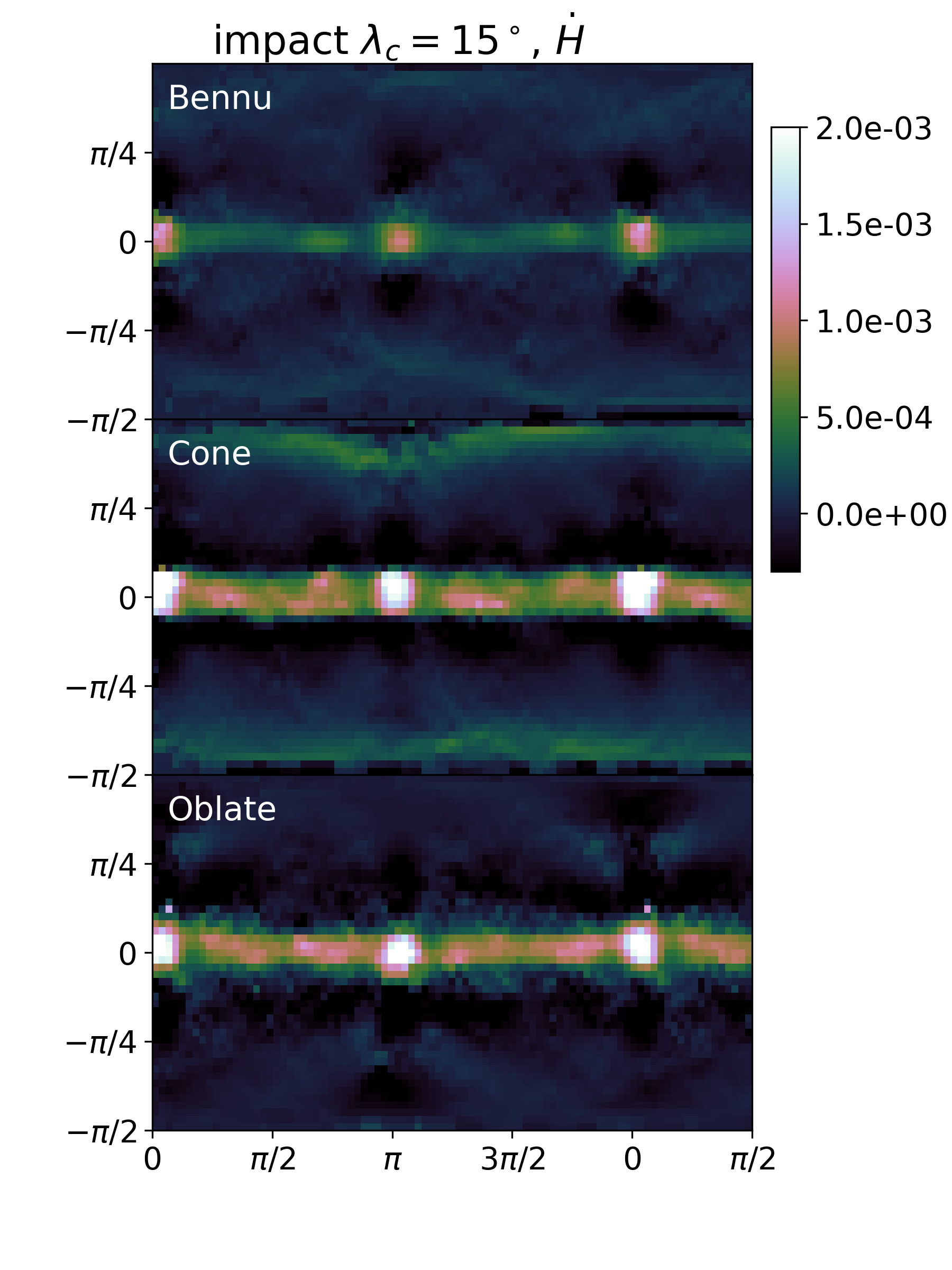}&
\includegraphics[width=2.1in,trim=40 10 0 0, clip]{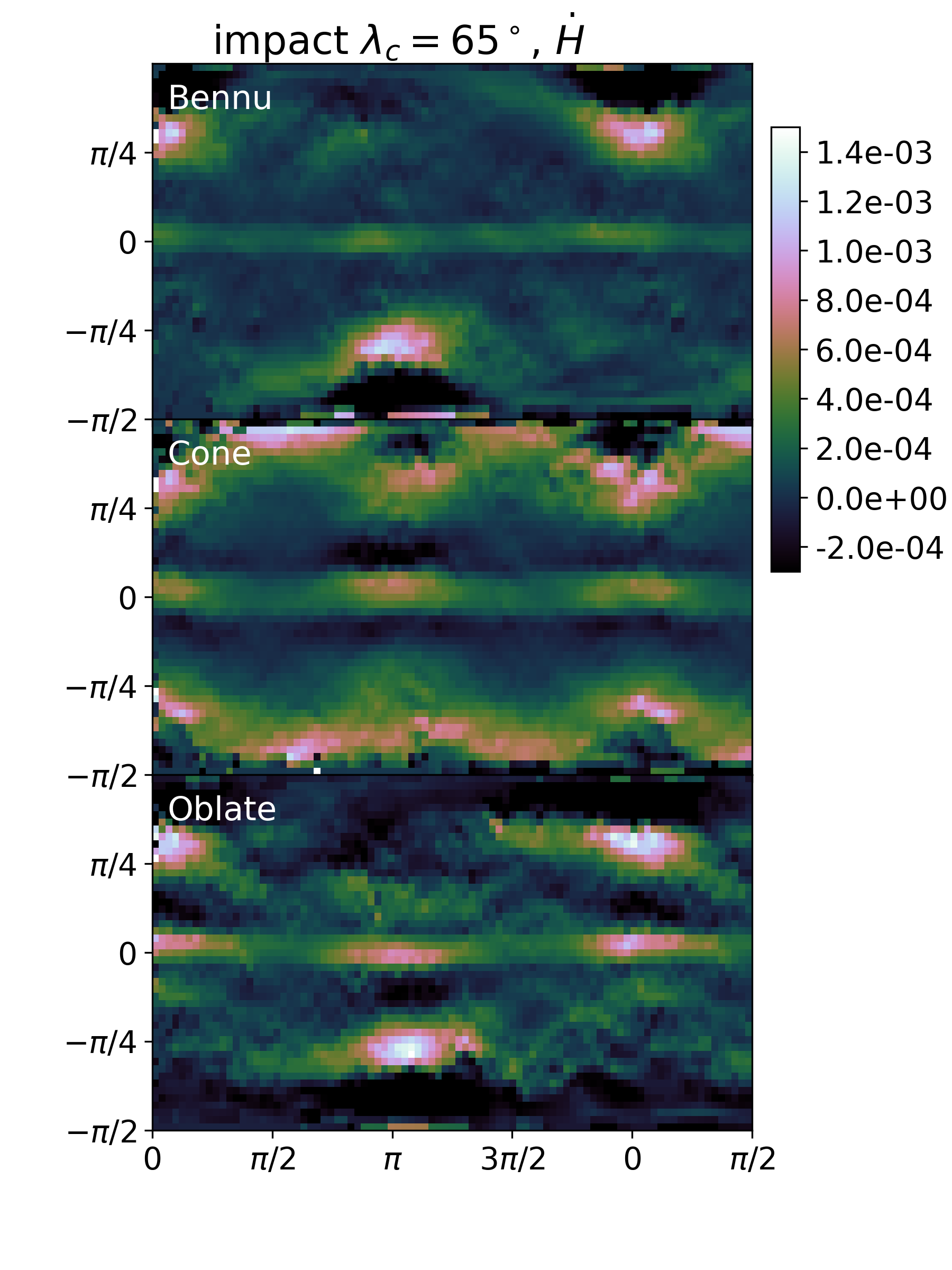}
\end{array}$
\end{center}
\caption{
Estimates for the rate of change of surface height, 
$ \frac{\partial H}{\partial t}$ long lasting seismic reverberations and
a vibro-fluidization flow model.  The height changes are estimated using
$\langle v_r^2 \rangle$ shown in Figure \ref{fig:vr2ave} and equation \ref{eqn:dH1} with $K=1$ for
our nine simulations.
}
\label{fig:grad_long}
\end{figure*}

\subsection{Surface height changes}
\label{sec:height}

We have three estimates for surface height variations, a seismic jolt one hop model, and two seismic reverberation models.     
The seismic jolt  one hop model
 uses the maximum positive radial velocity $v_r$ shown in Figure \ref{fig:max} 
 and equation \ref{eqn:deltaH_hop} for the change in surface height
to produce Figure \ref{fig:grad_short}, showing $\Delta H/h$ for our nine simulations.
The seismic pulse exhibited by our simulations primarily give strong equatorial flows, however
four peaks are seen on the equator in bi-cone and oblate models with near equatorial impacts.
The color bar scale shows $\Delta H/h$ so the size of the variations in height are less than 1/4
of the depth of the flowing layer $h$.   Figure \ref{fig:shape2} shows that the peak to peak elevation
differences on Bennu are $\Delta H \sim 1/10$  of its radius.  
To match these height variations we would require a flow depth
$h$ to be similar to 0.4 times the radius of Bennu itself.  This implies that focusing
of the initial seismic could not account for Bennu's 4 equatorial peaks.

The first of our seismic reverberation models assumes a constant depth for the flowing layer
and estimates the flow rate using a typical hop speed.
For this model 
 we use
$\langle v_r^2 \rangle$ shown in Figure \ref{fig:vr2ave} and equation \ref{eqn:dHdt_hop} 
to produce Figure \ref{fig:grad_ch}, showing $\frac{1}{h} \frac{\partial H}{\partial t}$ for our nine
simulations.  These figures show accumulation at the equatorial ridge, impact point and its
antipode.  To estimate the height change $\Delta H$ we must multiply by the total length
of time of seismic reverberation.  A value of  2000 times the frequency of the $l=2$ normal mode
gives a time $\Delta t = 14$ in gravitational units.  Here 2000 is a $Q$ value for the ratio
of energy lost per cycle. Taking a maximum value
of $\frac{1}{h} \frac{\partial H}{\partial t} \approx 0.3$ we estimate a total height change
of $\Delta H \approx 4 h$.  If the depth of flowing material is a few meters then the equatorial
ridge could increase by $\sim 10$ meters.

The second of our seismic reverberation models assumes a fluidization depth estimated with a granular temperature  
and a flow rate
with scaling motived by studies of inclined plane granular flow. For prolonged seismic reverberations,  
we use $\langle v_r^2 \rangle$ shown in Figure \ref{fig:vr2ave} and equation \ref{eqn:dH1} with $K=1$
to produce Figure \ref{fig:grad_long}, showing $\frac{\partial H}{\partial t}$ for our nine
simulations.  To estimate the height change $\Delta H$ we must multiply by the total length
of time of seismic reverberation.   Taking a maximum value of $2 \times 10^{-3}$ for
$ \frac{\partial H}{\partial t} $ times a time 14 (corresponding to a $Q$ of 2000 times
the period of the $l=2$ mode)
we estimate $\Delta H \approx 0.03$ in units of body radius.   This corresponds to 7 meters on Bennu.
The scaling factor $K$ can exceed 1 and we may have underestimated the fluidization depth, 
so larger elevation gains might be possible.

In summary, a seismic jolt and hopping surface
model seems unlikely to account for formation of Bennu's four equatorial peaks. 
Prolonged seismic reverberation could cause an equatorial ridge 
or other  high elevation regions to increase in height by a few meters.
Impact point and its antipode could also increase in height by a few meters.

\section{Discussion and Conclusions}
\label{sec:sum}

Impacts that are above global seismicity thresholds for resurfacing
and cause crater erasure likely happened on asteroids such as Bennu.  
These subcatastrophic impacts form large craters and should
radiate seismic waves efficiently at low frequencies, thus they could
 couple to the normal modes of whole body.  
Low frequency seismic waves would damp less quickly than high frequency ones.
If the attenuation rate is low,
once low frequency normal modes are excited, the body may oscillate for many oscillation periods.
But as the normal mode motions are not uniformly distributed on the body surface,
the pattern of seismic reverberation
depends on the structure of the normal modes themselves.

Because of the large seismic source size (because a large crater is formed)
the initial surface impulse (or seismic jolt) contains power at long wavelengths and so can be focused
by the shape of the body.   Whether the seismic response decays rapidly
or slowly giving a long period of seismic reverberation,  the
distribution of excited seismic motions would not be uniform across the surface.


We identify a regime in small asteroids less than a kilometer diameter where
rare and subcatastrophic are above
that giving global seismic reverberation.   
These impacts could excite low frequency normal modes.  
As the associated accelerations are well above
that of surface gravity, the surface of the body could preferentially be modified 
 where the excited normal modes 
or the initial seismic pulse is strongest.  For the asteroid Bennu,  
a few meter diameter impactor could be in this regime.
 
We simulate excitation of seismic waves by an impact using a mass-spring elastic N-body model.
The simulations let us explore the distribution of vibrational excitation on the surface
for different asteroid shape models.
The seismic source is modeled as a force applied to surface particles in a small
region that has
amplitude and time length estimated via scaling for a 4m  diameter impactor on Bennu.
We explore impacts on three shape models, the Bennu shape model, a bi-cone model
with an equatorial ridge and an oblate model.
Spectra of particle motions show that low frequency normal modes are excited, including
 the $l=2$ football-shaped spherical mode and the triangular-shaped $l=3$ normal mode.    
With a few thousand particles the
mass-spring model shows normal modes 
at frequencies within 10\% of those predicted for a homogeneous isotropic elastic sphere.

We use our simulations of impacts to display the root mean square of 
radial motions on the surface and the maximum positive radial speed caused by the initial
seismic pulse.      
If seismic reverberations are prolonged, vibrational energy on the surface
is primarily seen at impact point and its antipode.  Regions of higher surface elevation 
(such as the equatorial ridge) show more vibration.
Interference between
excited $l=2$ and $l=3$ normal modes prevents points on the equator other
than the impact point and its antipode from experiencing strong vibration.
Bennu's four equatorial would not be formed via seismic reverberation
from an initially axisymmetric equatorial ridge.

If seismic reverberations are quickly damped, then motions
are highest in regions where the impact excited pulse is focused.   
In addition to the impact antipode, focus points could
occur on the equatorial ridge for low-latitude impacts and near $90^\circ$ from the impact point.
Four equatorial peaks could be formed due to surface slumping but each one need not have an 
antipodal counter part, as is true for the Bennu shape mode by \citet{nolan13}.

We explored three simple flow models.  We estimated flow distance using a single hop 
and using a constant depth for a mobile  layer.  The hop distance
was estimated from the highest positive radial velocity seen in the simulations on the surface.  The estimated
surface height variations were  small and rule out a seismic impulse as
a mechanism for forming four peaks on Bennu's equatorial ridge.   Our second flow
model assumes prolonged seismic reverberation, adopts a velocity based on 
the acceleration parameter (and is equivalent to the average velocity during a hop)
and assumes a constant depth for the mobile layer.
Our third model also assumes prolonged seismic reverberation,  but allows the depth
of the mobile layer to depend on the level of surface vibration.  With a long timescale for the decay
of seismic energy (seismic $Q \sim 2000$),  an energetic impact could cause
shape changes of a few meters on an asteroid like Bennu.  
Primarily we see an increase in the height of an equatorial ridge, the point of impact and
its antipode.  If impact point is near but not on the equator, slumping toward the equator would
be lopsided. In other words, flows toward the equator would preferentially happen from northern latitudes near
the impact point and southern latitudes at its antipode, for an impact at a northern latitude. 
Four equatorial peaks are not formed due to interference between the excited $l=2$ (football shaped)
and $l=3$ (triangular shaped) normal modes.  This rules out an impact related prolonged seismic reverberation 
mechanism for the formation of Bennu's 4 equatorial peaks.

Bennu's 4 equatorial peaks are still a puzzle.   Perhaps a tidal excitation would better excite
the football-shaped mode without exciting an $l=3$ triangular mode, though a nearby encounter
with a larger body would be required  (e.g., see \citealt{press77,yu14,quillen16_crust}).
Alternatively perhaps an excited $l=3$ triangular mode would decay more quickly than
the football-shaped mode letting the body surface slump at four rather than two equatorial peaks. 
The four peaks may reflect a `ring-down' in vibrational energy, similar to that of a bell.
An abrupt earthquake along an internal fault could preferentially excite  the football shaped normal mode.
High resolution imaging of imaging that will be obtained by OSIRIS-REx will improve the shape model,
confirming or refuting the existence of these peaks.

Here we adopted a wave speed typical of lunar regolith (as have some other asteroid seismic studies; \citealt{murdoch17}) 
and with Young's modulus of 11 MPa.
If there are large fractured blocks inside the asteroid, then the wave speed could be higher
or the normal modes of the asteroid could be influenced by the normal mode spectrum of the blocks
(e.g., \citealt{richardson05}).   Cohesion models for asteroid estimate a strength about $10^2$--$10^4$ times
weaker than 11 MPa \citep{sanchez14,scheeres18}.
Even a cohesionless material should  transmit a compression wave, 
but the wave  rebound or reflection from surfaces
would be reduced because the medium would deform under tension or loft particles off the surface 
rather than transmit a pulse. 
The difference emphasizes that we have used an elastic model, comprised of masses and springs,
to explore the behavior of a body that could be a conglomerate.    

Seismic attenuation, dispersion and scattering properties are not known for asteroids, and granular flow
problems are complex even without high levels of subsurface vibration.    It is
not easy to predict whether granular material on a vibrating Chladni plate 
collects at anti-nodes or at nodes (e.g., \citealt{chladni,dorrestijn07}).
We have neglected seismic attenuation
and scattering,  and our flow models are simplistic as they do not depend on  
nature of friction between particles and the particle size distribution (e.g., \citealt{richardson05}).
We have neglected the details of crater formation, crater depth, re-accumulation
of crater ejecta  and have used a  simple model for the seismic pulse generated by the impact.
The amplitude of surface motions may depend on  topography and sub-surface composition.
(e.g., \citealt{kawase96,lee09}) and the medium itself may be transformed or deformed by the passage of
a pressure wave.
Future studies of the strong impact regime considered here
could improve upon our numerical techniques, and adopt more realistic 
asteroid composition, impact and flow models.  

\section*{Acknowledgements}  

A.C. Quillen thanks L'Observatoire de la C\^ote d'Azur for their warm welcome, 
and hospitality March and April 2018.
A. C. Quillen is grateful for generous support from the Simons Foundation and her work
 is in part supported by NASA grant 80NSSC17K0771.
This work is supported by the National Natural Science Foundation of China (grant No: 11673072, 11761131008
to Y.H. Zhao and Y.Y. Chen).
P. S\'anchez is supported in part by NASA grant NNX14AL16G from the Near-Earth Object Observation Program 
and from grants from NASA's SSERVI Institute. 
S.R.S. acknowledges support from the Complex Systems and Space, Environment, 
Risk and Resilience Academies of the Initiative d?EXcellence ?Joint, Excellent, and Dynamic Initiative? (IDEX JEDI) 
of the Universit\'e C\^ote d?Azur.
A.C. Quillen is grateful to the Leibniz Institut f\"ur Astrophysik Potsdam for their
warm welcome, support and hospitality July 2017 and April -- May 2018.
A.C. Quillen thanks Mt. Stromlo Observatory
for their warm welcome and hospitality Nov 2017-- Feb 2018.

We thank Patrick Michel, Mark Wieczorek, Marco Delbo,  Cindy Ebinger,
Jianyang Li, and Sinying Yang for helpful discussions.

\vskip 2 truein

{\bf Bibliography}

\bibliographystyle{elsarticle-harv} 
\bibliography{bennu_refs}

\begin{thebibliography}{114}
\expandafter\ifx\csname natexlab\endcsname\relax\def\natexlab#1{#1}\fi
\providecommand{\url}[1]{\texttt{#1}}
\providecommand{\href}[2]{#2}
\providecommand{\path}[1]{#1}
\providecommand{\DOIprefix}{doi:}
\providecommand{\ArXivprefix}{arXiv:}
\providecommand{\URLprefix}{URL: }
\providecommand{\Pubmedprefix}{pmid:}
\providecommand{\doi}[1]{\href{http://dx.doi.org/#1}{\path{#1}}}
\providecommand{\Pubmed}[1]{\href{pmid:#1}{\path{#1}}}
\providecommand{\bibinfo}[2]{#2}
\ifx\xfnm\relax \def\xfnm[#1]{\unskip,\space#1}\fi
\bibitem[{Aki and Richards(2002)}]{aki02}
\bibinfo{author}{Aki, K.}, \bibinfo{author}{Richards, P.},
  \bibinfo{year}{2002}.
\newblock \bibinfo{title}{Quantitative Seismology, 2nd edition}.
\newblock \bibinfo{publisher}{University Science Books, Sausalito, CA}.
\bibitem[{Aradian et~al.(2002)Aradian, Rapha\"el and de~Gennes}]{aradian02}
\bibinfo{author}{Aradian, A.}, \bibinfo{author}{Rapha\"el, E.},
  \bibinfo{author}{de~Gennes, P.G.}, \bibinfo{year}{2002}.
\newblock \bibinfo{title}{Surface flows of granular materials: a short
  introduction to some recent models}.
\newblock \bibinfo{journal}{Compte Rendue Physique} \bibinfo{volume}{3},
  \bibinfo{pages}{187--196}.
\bibitem[{Aranson and Tsimring(2002)}]{aranson02}
\bibinfo{author}{Aranson, I.S.}, \bibinfo{author}{Tsimring, L.S.},
  \bibinfo{year}{2002}.
\newblock \bibinfo{title}{Continuum theory of partially fluidized granular
  flows}.
\newblock \bibinfo{journal}{Physics Review Letters} \bibinfo{volume}{65},
  \bibinfo{pages}{061303}.
\bibitem[{Asphaug(2008)}]{asphaug08}
\bibinfo{author}{Asphaug, E.}, \bibinfo{year}{2008}.
\newblock \bibinfo{title}{Critical crater diameter and asteroid impact
  seismology}.
\newblock \bibinfo{journal}{Meteoritics \& Planetary Science}
  \bibinfo{volume}{43}, \bibinfo{pages}{1075--1084}.
\bibitem[{Bagnold(1954)}]{bagnold54}
\bibinfo{author}{Bagnold, R.A.}, \bibinfo{year}{1954}.
\newblock \bibinfo{title}{Experiments on a gravity-free dispersion of large
  solid particles in a newtonian fluid under shear}.
\newblock \bibinfo{journal}{Proceedings of the Royal Society of London, Series
  A} \bibinfo{volume}{225}, \bibinfo{pages}{49--63}.
\bibitem[{Benz and Asphaug(1999)}]{benz99}
\bibinfo{author}{Benz, W.}, \bibinfo{author}{Asphaug, E.},
  \bibinfo{year}{1999}.
\newblock \bibinfo{title}{Catastrophic disruptions revisited}.
\newblock \bibinfo{journal}{Icarus} \bibinfo{volume}{142},
  \bibinfo{pages}{5--20}.
\bibitem[{Binzel et~al.(2015)Binzel, DeMeo, Burt, Cloutis, Rozitis, Burbine,
  Campins, Clark, Emery, Hergenrother, Howell, Lauretta, Nolan, Mansfield,
  Pietrasz, Polishook and Scheeres}]{binzel15}
\bibinfo{author}{Binzel, R.P.}, \bibinfo{author}{DeMeo, F.E.},
  \bibinfo{author}{Burt, B.J.}, \bibinfo{author}{Cloutis, E.A.},
  \bibinfo{author}{Rozitis, B.}, \bibinfo{author}{Burbine, T.H.},
  \bibinfo{author}{Campins, H.}, \bibinfo{author}{Clark, B.E.},
  \bibinfo{author}{Emery, J.P.}, \bibinfo{author}{Hergenrother, C.W.},
  \bibinfo{author}{Howell, E.S.}, \bibinfo{author}{Lauretta, D.S.},
  \bibinfo{author}{Nolan, M.C.}, \bibinfo{author}{Mansfield, M.},
  \bibinfo{author}{Pietrasz, V.}, \bibinfo{author}{Polishook, D.},
  \bibinfo{author}{Scheeres, D.J.}, \bibinfo{year}{2015}.
\newblock \bibinfo{title}{Spectral slope variations for osiris-rex target
  asteroid (101955) bennu: Possible evidence for a fine-grained regolith
  equatorial ridge}.
\newblock \bibinfo{journal}{Icarus} \bibinfo{volume}{256}, \bibinfo{pages}{22
  -- 29}.
\newblock \URLprefix
  \url{http://www.sciencedirect.com/science/article/pii/S0019103515001463},
  \DOIprefix\doi{https://doi.org/10.1016/j.icarus.2015.04.011}.
\bibitem[{Bottke et~al.(1994)Bottke, Nolan, Greenberg and Kolvoord}]{bottke94}
\bibinfo{author}{Bottke, W.F.}, \bibinfo{author}{Nolan, M.},
  \bibinfo{author}{Greenberg, R.}, \bibinfo{author}{Kolvoord, R.},
  \bibinfo{year}{1994}.
\newblock \bibinfo{title}{Velocity distributions among colliding asteroids}.
\newblock \bibinfo{journal}{Icarus} \bibinfo{volume}{107},
  \bibinfo{pages}{255--268}.
\bibitem[{Campbell()}]{campbell90}
\bibinfo{author}{Campbell, C.S.}, .
\newblock \bibinfo{title}{Rapid granular flows}.
\newblock \bibinfo{journal}{Annual Review of Fluid Mechanics}
  \bibinfo{volume}{22}, \bibinfo{pages}{57--92}.
\bibitem[{Cheng et~al.(2002)Cheng, Izenberg, Chapman and Zuber}]{cheng02}
\bibinfo{author}{Cheng, A.F.}, \bibinfo{author}{Izenberg, N.},
  \bibinfo{author}{Chapman, C.}, \bibinfo{author}{Zuber, M.},
  \bibinfo{year}{2002}.
\newblock \bibinfo{title}{Ponded deposits on asteroid 433 eros}.
\newblock \bibinfo{journal}{Meteoritics \& Planetary Science}
  \bibinfo{volume}{37}, \bibinfo{pages}{1095--1105}.
\bibitem[{Chesley et~al.(2014)Chesley, Farnocchia, Nolan, Vokrouhlick{\'y},
  Chodas, Milani, Spoto, Rozitis, Benner, Bottke, Busch, Emery, Howell,
  Lauretta, Margot and Taylor}]{chesley14}
\bibinfo{author}{Chesley, S.R.}, \bibinfo{author}{Farnocchia, D.},
  \bibinfo{author}{Nolan, M.C.}, \bibinfo{author}{Vokrouhlick{\'y}, D.},
  \bibinfo{author}{Chodas, P.W.}, \bibinfo{author}{Milani, A.},
  \bibinfo{author}{Spoto, F.}, \bibinfo{author}{Rozitis, B.},
  \bibinfo{author}{Benner, L.A.}, \bibinfo{author}{Bottke, W.F.},
  \bibinfo{author}{Busch, M.W.}, \bibinfo{author}{Emery, J.P.},
  \bibinfo{author}{Howell, E.S.}, \bibinfo{author}{Lauretta, D.S.},
  \bibinfo{author}{Margot, J.L.}, \bibinfo{author}{Taylor, P.A.},
  \bibinfo{year}{2014}.
\newblock \bibinfo{title}{Orbit and bulk density of the osiris-rex target
  asteroid (101955) bennu}.
\newblock \bibinfo{journal}{Icarus} \bibinfo{volume}{235}, \bibinfo{pages}{5 --
  22}.
\newblock \URLprefix
  \url{http://www.sciencedirect.com/science/article/pii/S0019103514001067},
  \DOIprefix\doi{https://doi.org/10.1016/j.icarus.2014.02.020}.
\bibitem[{Chladni(1787)}]{chladni}
\bibinfo{author}{Chladni, E.F.F.}, \bibinfo{year}{1787}.
\newblock \bibinfo{title}{Entdeckungen \"uber die Theorie des Klanges}.
\newblock \bibinfo{publisher}{Breitkopf und H\"artel, Leipzig}.
\bibitem[{{Cintala} et~al.(1978){Cintala}, {Head} and {Veverka}}]{cintala78}
\bibinfo{author}{{Cintala}, M.J.}, \bibinfo{author}{{Head}, J.W.},
  \bibinfo{author}{{Veverka}, J.}, \bibinfo{year}{1978}.
\newblock \bibinfo{title}{{Characteristics of the cratering process on small
  satellites and asteroids}}, in: \bibinfo{booktitle}{Lunar and Planetary
  Science Conference Proceedings}, pp. \bibinfo{pages}{3803--3830}.
\bibitem[{Clark et~al.(2012)Clark, Kondic and Behringer}]{clark12}
\bibinfo{author}{Clark, A.H.}, \bibinfo{author}{Kondic, L.},
  \bibinfo{author}{Behringer, R.P.}, \bibinfo{year}{2012}.
\newblock \bibinfo{title}{Particle scale dynamics in granular impact}.
\newblock \bibinfo{journal}{Physics Review Letters} \bibinfo{volume}{109},
  \bibinfo{pages}{238302}.
\bibitem[{Colwell et~al.(2005)Colwell, Gulbis, Hor{\'a}nyi and
  Robertson}]{colwell05}
\bibinfo{author}{Colwell, J.E.}, \bibinfo{author}{Gulbis, A.A.},
  \bibinfo{author}{Hor{\'a}nyi, M.}, \bibinfo{author}{Robertson, S.},
  \bibinfo{year}{2005}.
\newblock \bibinfo{title}{Dust transport in photoelectron layers and the
  formation of dust ponds on eros}.
\newblock \bibinfo{journal}{Icarus} \bibinfo{volume}{175}, \bibinfo{pages}{159
  -- 169}.
\newblock \URLprefix
  \url{http://www.sciencedirect.com/science/article/pii/S0019103504003847},
  \DOIprefix\doi{https://doi.org/10.1016/j.icarus.2004.11.001}.
\bibitem[{Cooper et~al.(1974)Cooper, Kovach and Watkins}]{cooper74}
\bibinfo{author}{Cooper, M.R.}, \bibinfo{author}{Kovach, R.L.},
  \bibinfo{author}{Watkins, J.S.}, \bibinfo{year}{1974}.
\newblock \bibinfo{title}{Lunar near-surface structure}.
\newblock \bibinfo{journal}{Reviews of Geophysics and Space Sciences}
  \bibinfo{volume}{12}, \bibinfo{pages}{291--308}.
\bibitem[{Cundall and Strack(1979)}]{cundall79}
\bibinfo{author}{Cundall, P.A.}, \bibinfo{author}{Strack, O.D.L.},
  \bibinfo{year}{1979}.
\newblock \bibinfo{title}{A discrete numerical model for granular assemblies}.
\newblock \bibinfo{journal}{Geotechnique} \bibinfo{volume}{29},
  \bibinfo{pages}{47--65}.
\bibitem[{Dainty et~al.(1974)Dainty, Toks\"oz, Anderson, Pines, Nakamura and
  Latham}]{dainty74}
\bibinfo{author}{Dainty, A.}, \bibinfo{author}{Toks\"oz, M.N.},
  \bibinfo{author}{Anderson, K.}, \bibinfo{author}{Pines, P.},
  \bibinfo{author}{Nakamura, Y.}, \bibinfo{author}{Latham, G.},
  \bibinfo{year}{1974}.
\newblock \bibinfo{title}{Seismic scattering and shallow structure of the moon
  in oceanus procellarum}.
\newblock \bibinfo{journal}{Moon} \bibinfo{volume}{9}, \bibinfo{pages}{11--29}.
\bibitem[{Deboeuf et~al.(2006)Deboeuf, Lajeunesse, Dauchot and
  Andreotti}]{deboeuf06}
\bibinfo{author}{Deboeuf, S.}, \bibinfo{author}{Lajeunesse, E.},
  \bibinfo{author}{Dauchot, O.}, \bibinfo{author}{Andreotti, B.},
  \bibinfo{year}{2006}.
\newblock \bibinfo{title}{Flow rule, self-channelization and levees in
  unconfined granular flows}.
\newblock \bibinfo{journal}{Physics Review Letters} \bibinfo{volume}{97},
  \bibinfo{pages}{158303}.
\bibitem[{Ding and Gidaspow(1990)}]{ding90}
\bibinfo{author}{Ding, J.}, \bibinfo{author}{Gidaspow, D.},
  \bibinfo{year}{1990}.
\newblock \bibinfo{title}{A bubbling fluidization model using kinetic theory of
  granular flow}.
\newblock \bibinfo{journal}{AIChE Journal} \bibinfo{volume}{36},
  \bibinfo{pages}{523--538}.
\newblock \URLprefix \url{http://dx.doi.org/10.1002/aic.690360404},
  \DOIprefix\doi{10.1002/aic.690360404}.
\bibitem[{Dorrestijn et~al.(2007)Dorrestijn, Bietsch, Acikalin, Raman, Hegner,
  Meyer and Gerber}]{dorrestijn07}
\bibinfo{author}{Dorrestijn, M.}, \bibinfo{author}{Bietsch, A.},
  \bibinfo{author}{Acikalin, T.}, \bibinfo{author}{Raman, A.},
  \bibinfo{author}{Hegner, M.}, \bibinfo{author}{Meyer, E.},
  \bibinfo{author}{Gerber, C.}, \bibinfo{year}{2007}.
\newblock \bibinfo{title}{Chladni figures revisited based on nanomechanics}.
\newblock \bibinfo{journal}{Physics Review Letters} \bibinfo{volume}{98},
  \bibinfo{pages}{026102}.
\bibitem[{Duffy and Mindlin(1957)}]{duffy57}
\bibinfo{author}{Duffy, J.}, \bibinfo{author}{Mindlin, R.},
  \bibinfo{year}{1957}.
\newblock \bibinfo{title}{Stress strain relations and vibrations of granular
  media}.
\newblock \bibinfo{journal}{Journal of Applied Mechanics} \bibinfo{volume}{24},
  \bibinfo{pages}{585--593}.
\bibitem[{Emery et~al.(2014)Emery, Fern{\'a}ndez, Kelley, (n{\`e}e Crane),
  Hergenrother, Lauretta, Drake, Campins and Ziffer}]{emery14}
\bibinfo{author}{Emery, J.}, \bibinfo{author}{Fern{\'a}ndez, Y.},
  \bibinfo{author}{Kelley, M.}, \bibinfo{author}{(n{\`e}e Crane), K.W.},
  \bibinfo{author}{Hergenrother, C.}, \bibinfo{author}{Lauretta, D.},
  \bibinfo{author}{Drake, M.}, \bibinfo{author}{Campins, H.},
  \bibinfo{author}{Ziffer, J.}, \bibinfo{year}{2014}.
\newblock \bibinfo{title}{Thermal infrared observations and thermophysical
  characterization of osiris-rex target asteroid (101955) bennu}.
\newblock \bibinfo{journal}{Icarus} \bibinfo{volume}{234}, \bibinfo{pages}{17
  -- 35}.
\newblock \URLprefix
  \url{http://www.sciencedirect.com/science/article/pii/S0019103514000827},
  \DOIprefix\doi{https://doi.org/10.1016/j.icarus.2014.02.005}.
\bibitem[{Forterre and Pouliquen(2008)}]{forterre08}
\bibinfo{author}{Forterre, Y.}, \bibinfo{author}{Pouliquen, O.},
  \bibinfo{year}{2008}.
\newblock \bibinfo{title}{Flows of dense granular media}.
\newblock \bibinfo{journal}{Annual Review of Fluid Mechanics}
  \bibinfo{volume}{40}, \bibinfo{pages}{1--24}.
\bibitem[{Freund(1998)}]{freund98}
\bibinfo{author}{Freund, L.B.}, \bibinfo{year}{1998}.
\newblock \bibinfo{title}{Dynamic Fracture Mechanics}.
\newblock \bibinfo{publisher}{Cambridge University Press, Cambridge, England}.
\bibitem[{Frouard et~al.(2016)Frouard, Quillen, Efroimsky and
  Giannella}]{frouard16}
\bibinfo{author}{Frouard, J.}, \bibinfo{author}{Quillen, A.C.},
  \bibinfo{author}{Efroimsky, M.}, \bibinfo{author}{Giannella, D.},
  \bibinfo{year}{2016}.
\newblock \bibinfo{title}{Numerical simulation of tidal evolution of a
  viscoelastic body modelled with a mass-spring network}.
\newblock \bibinfo{journal}{Monthly Notices of the Royal Astronomical Society}
  \bibinfo{volume}{458}, \bibinfo{pages}{2890--2901}.
\bibitem[{{GDR MiDi}(2004)}]{midi04}
\bibinfo{author}{{GDR MiDi}}, \bibinfo{year}{2004}.
\newblock \bibinfo{title}{On dense granular flows}.
\newblock \bibinfo{journal}{The European Physical Journal E}
  \bibinfo{volume}{14}, \bibinfo{pages}{341--365}.
\newblock \URLprefix \url{https://doi.org/10.1140/epje/i2003-10153-0},
  \DOIprefix\doi{10.1140/epje/i2003-10153-0}.
\bibitem[{Geng et~al.(2001)Geng, Howell, Longhi, Behringer, Reydellet, Vanel,
  Cl\'ement and Luding}]{geng01}
\bibinfo{author}{Geng, J.}, \bibinfo{author}{Howell, D.},
  \bibinfo{author}{Longhi, E.}, \bibinfo{author}{Behringer, R.},
  \bibinfo{author}{Reydellet, G.}, \bibinfo{author}{Vanel, L.},
  \bibinfo{author}{Cl\'ement, E.}, \bibinfo{author}{Luding, S.},
  \bibinfo{year}{2001}.
\newblock \bibinfo{title}{Footprints in sand: The response of a granular
  material to local perturbations}.
\newblock \bibinfo{journal}{Physics Review Letters} \bibinfo{volume}{87},
  \bibinfo{pages}{035506}.
\bibitem[{Goldhirsch(2008)}]{goldhirsch08}
\bibinfo{author}{Goldhirsch, I.}, \bibinfo{year}{2008}.
\newblock \bibinfo{title}{Introduction to granular temperature}.
\newblock \bibinfo{journal}{Powder Technolology} \bibinfo{volume}{182},
  \bibinfo{pages}{130--136}.
\bibitem[{Greenberg et~al.(1996)Greenberg, Bottke, M., Geissler, Petit, Durda,
  Asphaug and Head}]{greenberg96}
\bibinfo{author}{Greenberg, R.}, \bibinfo{author}{Bottke, W.},
  \bibinfo{author}{M., N.}, \bibinfo{author}{Geissler, P.},
  \bibinfo{author}{Petit, J.}, \bibinfo{author}{Durda, D.},
  \bibinfo{author}{Asphaug, E.}, \bibinfo{author}{Head, J.},
  \bibinfo{year}{1996}.
\newblock \bibinfo{title}{Collisional and dynamical history of ida}.
\newblock \bibinfo{journal}{Icarus} \bibinfo{volume}{120},
  \bibinfo{pages}{106--118}.
\bibitem[{Greenberg et~al.(1994)Greenberg, Nolan, Bottke, Kolvoord and
  Veverka}]{greenberg94}
\bibinfo{author}{Greenberg, R.}, \bibinfo{author}{Nolan, M.},
  \bibinfo{author}{Bottke, W.}, \bibinfo{author}{Kolvoord, R.},
  \bibinfo{author}{Veverka, J.}, \bibinfo{year}{1994}.
\newblock \bibinfo{title}{Collisional history of gaspra}.
\newblock \bibinfo{journal}{Icarus} \bibinfo{volume}{107},
  \bibinfo{pages}{84--97}.
\bibitem[{Guibout and Scheeres(2003)}]{guibout03}
\bibinfo{author}{Guibout, V.}, \bibinfo{author}{Scheeres, D.},
  \bibinfo{year}{2003}.
\newblock \bibinfo{title}{Stability of surface motion on a rotating ellipsoid}.
\newblock \bibinfo{journal}{Celestial Mechanics and Dynamical Astronomy}
  \bibinfo{volume}{87}, \bibinfo{pages}{263--290}.
\bibitem[{G\"uldemeister and W\"unnemann(2017)}]{guldemeister17}
\bibinfo{author}{G\"uldemeister, N.}, \bibinfo{author}{W\"unnemann, K.},
  \bibinfo{year}{2017}.
\newblock \bibinfo{title}{Quantitative analysis of impact-induced seismic
  signals by numerical modeling}.
\newblock \bibinfo{journal}{Icarus} \bibinfo{volume}{296},
  \bibinfo{pages}{15--27}.
\bibitem[{Hirabayashi et~al.(2015)Hirabayashi, S\'anchez and
  Scheeres}]{hirabayashi15a}
\bibinfo{author}{Hirabayashi, H.}, \bibinfo{author}{S\'anchez, D.P.},
  \bibinfo{author}{Scheeres, D.J.}, \bibinfo{year}{2015}.
\newblock \bibinfo{title}{Internal structure of asteroids having surface
  shedding due to rotational instability}.
\newblock \bibinfo{journal}{Astrophysical Journal} \bibinfo{volume}{808},
  \bibinfo{pages}{63--85}.
\bibitem[{Hirabayashi and Scheeres(2015)}]{hirabayashi15b}
\bibinfo{author}{Hirabayashi, M.}, \bibinfo{author}{Scheeres, D.J.},
  \bibinfo{year}{2015}.
\newblock \bibinfo{title}{Stress and failure analysis of rapidly rotating
  asteroid (29075) 1950 da}.
\newblock \bibinfo{journal}{Astrophysical Journal Letters}
  \bibinfo{volume}{798}, \bibinfo{pages}{L8--13}.
\bibitem[{Hitchman et~al.(2016.)Hitchman, van Wijka and Davidson}]{apple}
\bibinfo{author}{Hitchman, S.}, \bibinfo{author}{van Wijka, K.},
  \bibinfo{author}{Davidson, Z.}, \bibinfo{year}{2016.}
\newblock \bibinfo{title}{Monitoring attenuation and the elastic properties of
  an apple with laser ultrasound.}
\newblock \bibinfo{journal}{Postharvest Biology and Technology}
  \bibinfo{volume}{121}, \bibinfo{pages}{71--77}.
\bibitem[{Holsapple(1993)}]{holsapple93}
\bibinfo{author}{Holsapple, K.A.}, \bibinfo{year}{1993}.
\newblock \bibinfo{title}{The scaling of impact processes in planetary
  sciences}.
\newblock \bibinfo{journal}{Annual Review of Earth and Planetary Sciences}
  \bibinfo{volume}{21}, \bibinfo{pages}{333--373}.
\bibitem[{Holsapple(2004)}]{holsapple04}
\bibinfo{author}{Holsapple, K.A.}, \bibinfo{year}{2004}.
\newblock \bibinfo{title}{Equilibrium figures of spinning bodies with
  self-gravity}.
\newblock \bibinfo{journal}{Icarus} \bibinfo{volume}{172},
  \bibinfo{pages}{272--303}.
\bibitem[{Holsapple(2013)}]{holsapple13}
\bibinfo{author}{Holsapple, K.A.}, \bibinfo{year}{2013}.
\newblock \bibinfo{title}{Modeling granular material flows: The angle of
  repose, fluidization and the cliff collapse problem}.
\newblock \bibinfo{journal}{Planetary and Space Science} \bibinfo{volume}{82},
  \bibinfo{pages}{11--26}.
\bibitem[{Hostler(2005)}]{hostler05}
\bibinfo{author}{Hostler, S.R.}, \bibinfo{year}{2005}.
\newblock \bibinfo{title}{Pressure wave propagation in a shaken granular bed}.
\newblock \bibinfo{journal}{Physical Review E} \bibinfo{volume}{72},
  \bibinfo{pages}{031304}.
\newblock \DOIprefix\doi{10.1103/PhysRevE.72.031304}.
\bibitem[{Housen et~al.(2018)Housen, Sweet and Holsapple}]{housen18}
\bibinfo{author}{Housen, K.}, \bibinfo{author}{Sweet, W.},
  \bibinfo{author}{Holsapple, K.A.}, \bibinfo{year}{2018}.
\newblock \bibinfo{title}{Impacts into porous asteroids.}
\newblock \bibinfo{journal}{Icarus} \bibinfo{volume}{300},
  \bibinfo{pages}{72--96}.
\bibitem[{Jacobson and Scheeres(2011)}]{jacobson11}
\bibinfo{author}{Jacobson, S.A.}, \bibinfo{author}{Scheeres, D.},
  \bibinfo{year}{2011}.
\newblock \bibinfo{title}{Dynamics of rotationally fissioned asteroids: Source
  of observed small asteroid systems}.
\newblock \bibinfo{journal}{Icarus} \bibinfo{volume}{214},
  \bibinfo{pages}{161--178}.
\bibitem[{Jenkins and Savage(1983)}]{jenkins83}
\bibinfo{author}{Jenkins, J.T.}, \bibinfo{author}{Savage, S.B.},
  \bibinfo{year}{1983}.
\newblock \bibinfo{title}{A theory for the rapid flow of identical, smooth,
  nearly elastic, spherical-particles}.
\newblock \bibinfo{journal}{Journal of Fluid Mechanics} \bibinfo{volume}{130},
  \bibinfo{pages}{187--202}.
\bibitem[{Jia(2004)}]{jia04}
\bibinfo{author}{Jia, X.}, \bibinfo{year}{2004}.
\newblock \bibinfo{title}{Codalike multiple scattering of elastic waves in
  dense granular media}.
\newblock \bibinfo{journal}{Physical Review Letters} \bibinfo{volume}{93},
  \bibinfo{pages}{154303}.
\newblock \DOIprefix\doi{10.1103/PhysRevLett.93.154303}.
\bibitem[{Jia et~al.(1999)Jia, Caroli and Velicky}]{jia99}
\bibinfo{author}{Jia, X.}, \bibinfo{author}{Caroli, C.},
  \bibinfo{author}{Velicky, B.}, \bibinfo{year}{1999}.
\newblock \bibinfo{title}{Ultrasound propagation in externally stressed
  granular media}.
\newblock \bibinfo{journal}{Physics Review Letters} \bibinfo{volume}{82},
  \bibinfo{pages}{1863}.
\bibitem[{Jutzi and Michel(2014)}]{jutzi14}
\bibinfo{author}{Jutzi, M.}, \bibinfo{author}{Michel, P.},
  \bibinfo{year}{2014}.
\newblock \bibinfo{title}{Hypervelocity impacts on asteroids and momentum
  transfer i. numerical simulations using porous targets}.
\newblock \bibinfo{journal}{Icarus} \bibinfo{volume}{229},
  \bibinfo{pages}{247--253}.
\bibitem[{Jutzi et~al.(2010)Jutzi, Michel, Benz and Richardson}]{jutzi10}
\bibinfo{author}{Jutzi, M.}, \bibinfo{author}{Michel, P.},
  \bibinfo{author}{Benz, W.}, \bibinfo{author}{Richardson, D.C.},
  \bibinfo{year}{2010}.
\newblock \bibinfo{title}{Fragment properties at the catastrophic disruption
  threshold: The effect of the parent body's internal structure.}
\newblock \bibinfo{journal}{Icarus} \bibinfo{volume}{207},
  \bibinfo{pages}{54--65}.
\bibitem[{Jutzi et~al.(2009)Jutzi, Michel, Hiraoka, Nakamura and
  Benz}]{jutzi09}
\bibinfo{author}{Jutzi, M.}, \bibinfo{author}{Michel, P.},
  \bibinfo{author}{Hiraoka, K.}, \bibinfo{author}{Nakamura, A.M.},
  \bibinfo{author}{Benz, W.}, \bibinfo{year}{2009}.
\newblock \bibinfo{title}{Numerical simulations of impacts involving porous
  bodies. ii. comparison with laboratory experiments}.
\newblock \bibinfo{journal}{Icarus} \bibinfo{volume}{201},
  \bibinfo{pages}{802--813}.
\bibitem[{Kawase(1996)}]{kawase96}
\bibinfo{author}{Kawase, H.}, \bibinfo{year}{1996}.
\newblock \bibinfo{title}{The cause of the damage belt in kobe: "the basin-edge
  effect", constructive interference of the direct s-wave with the
  basin-induced diffracted/rayleigh waves}.
\newblock \bibinfo{journal}{Seismological Research Letters}
  \bibinfo{volume}{67}, \bibinfo{pages}{25--34}.
\bibitem[{Kot et~al.(2015)Kot, Nagahashi and Szymczak}]{kot15}
\bibinfo{author}{Kot, M.}, \bibinfo{author}{Nagahashi, H.},
  \bibinfo{author}{Szymczak, P.}, \bibinfo{year}{2015}.
\newblock \bibinfo{title}{Elastic moduli of simple mass spring models}.
\newblock \bibinfo{journal}{The Visual Computer: International Journal of
  Computer Graphics} \bibinfo{volume}{31}, \bibinfo{pages}{1339--1350}.
\bibitem[{Kumaran(1998)}]{kumaran98}
\bibinfo{author}{Kumaran, V.}, \bibinfo{year}{1998}.
\newblock \bibinfo{title}{Temperature of a granular material `fluidized' by
  external vibrations}.
\newblock \bibinfo{journal}{Physics Review E} \bibinfo{volume}{57},
  \bibinfo{pages}{5660--5664}.
\bibitem[{Lamb(1881)}]{Lamb_1881}
\bibinfo{author}{Lamb, H.}, \bibinfo{year}{1881}.
\newblock \bibinfo{title}{On the vibrations of an elastic sphere}.
\newblock \bibinfo{journal}{Proceedings of the London Mathematical Society}
  \bibinfo{volume}{s1-13}, \bibinfo{pages}{189--212}.
\newblock \URLprefix \url{http://dx.doi.org/10.1112/plms/s1-13.1.189},
  \DOIprefix\doi{10.1112/plms/s1-13.1.189}.
\bibitem[{Lambe and Whitman(1979)}]{lambe79}
\bibinfo{author}{Lambe, T.W.}, \bibinfo{author}{Whitman, R.V.},
  \bibinfo{year}{1979}.
\newblock \bibinfo{title}{Soil Mechanics (Series in Soil Engineering)}.
\newblock \bibinfo{publisher}{Wiley, New York.}
\bibitem[{Lauretta et~al.(2017)Lauretta, Balram-Knutson, Beshore, Boynton,
  Drouet~d'Aubigny, DellaGiustina, Enos, Golish, Hergenrother, Howell, Bennett,
  Morton, Nolan, Rizk, Roper, Bartels, Bos, Dworkin, Highsmith, Lorenz, Lim,
  Mink, Moreau, Nuth, Reuter, Simon, Bierhaus, Bryan, Ballouz, Barnouin,
  Binzel, Bottke, Hamilton, Walsh, Chesley, Christensen, Clark, Connolly,
  Crombie, Daly, Emery, McCoy, McMahon, Scheeres, Messenger,
  Nakamura-Messenger, Righter and Sandford}]{lauretta17}
\bibinfo{author}{Lauretta, D.S.}, \bibinfo{author}{Balram-Knutson, S.S.},
  \bibinfo{author}{Beshore, E.}, \bibinfo{author}{Boynton, W.V.},
  \bibinfo{author}{Drouet~d'Aubigny, C.}, \bibinfo{author}{DellaGiustina,
  D.N.}, \bibinfo{author}{Enos, H.L.}, \bibinfo{author}{Golish, D.R.},
  \bibinfo{author}{Hergenrother, C.W.}, \bibinfo{author}{Howell, E.S.},
  \bibinfo{author}{Bennett, C.A.}, \bibinfo{author}{Morton, E.T.},
  \bibinfo{author}{Nolan, M.C.}, \bibinfo{author}{Rizk, B.},
  \bibinfo{author}{Roper, H.L.}, \bibinfo{author}{Bartels, A.E.},
  \bibinfo{author}{Bos, B.J.}, \bibinfo{author}{Dworkin, J.P.},
  \bibinfo{author}{Highsmith, D.E.}, \bibinfo{author}{Lorenz, D.A.},
  \bibinfo{author}{Lim, L.F.}, \bibinfo{author}{Mink, R.},
  \bibinfo{author}{Moreau, M.C.}, \bibinfo{author}{Nuth, J.A.},
  \bibinfo{author}{Reuter, D.C.}, \bibinfo{author}{Simon, A.A.},
  \bibinfo{author}{Bierhaus, E.B.}, \bibinfo{author}{Bryan, B.H.},
  \bibinfo{author}{Ballouz, R.}, \bibinfo{author}{Barnouin, O.S.},
  \bibinfo{author}{Binzel, R.P.}, \bibinfo{author}{Bottke, W.F.},
  \bibinfo{author}{Hamilton, V.E.}, \bibinfo{author}{Walsh, K.J.},
  \bibinfo{author}{Chesley, S.R.}, \bibinfo{author}{Christensen, P.R.},
  \bibinfo{author}{Clark, B.E.}, \bibinfo{author}{Connolly, H.C.},
  \bibinfo{author}{Crombie, M.K.}, \bibinfo{author}{Daly, M.G.},
  \bibinfo{author}{Emery, J.P.}, \bibinfo{author}{McCoy, T.J.},
  \bibinfo{author}{McMahon, J.W.}, \bibinfo{author}{Scheeres, D.J.},
  \bibinfo{author}{Messenger, S.}, \bibinfo{author}{Nakamura-Messenger, K.},
  \bibinfo{author}{Righter, K.}, \bibinfo{author}{Sandford, S.A.},
  \bibinfo{year}{2017}.
\newblock \bibinfo{title}{Osiris-rex: Sample return from asteroid (101955)
  bennu}.
\newblock \bibinfo{journal}{Space Science Reviews} \bibinfo{volume}{212},
  \bibinfo{pages}{925--984}.
\newblock \URLprefix \url{https://doi.org/10.1007/s11214-017-0405-1},
  \DOIprefix\doi{10.1007/s11214-017-0405-1}.
\bibitem[{Lee et~al.(2009)Lee, Komatitscha, Huang and Tromp}]{lee09}
\bibinfo{author}{Lee, S.J.}, \bibinfo{author}{Komatitscha, D.},
  \bibinfo{author}{Huang, B.S.}, \bibinfo{author}{Tromp, J.},
  \bibinfo{year}{2009}.
\newblock \bibinfo{title}{Effects of topography on seismic-wave propagation: An
  example from northern taiwan}.
\newblock \bibinfo{journal}{Bulletin of the Seismological Society of America}
  \bibinfo{volume}{99}, \bibinfo{pages}{314--325}.
\bibitem[{Lognonn\'e et~al.(2009)Lognonn\'e, Feuvre, Johnson and
  Weber}]{lognonne09}
\bibinfo{author}{Lognonn\'e, P.}, \bibinfo{author}{Feuvre, M.L.},
  \bibinfo{author}{Johnson, C.L.}, \bibinfo{author}{Weber, R.C.},
  \bibinfo{year}{2009}.
\newblock \bibinfo{title}{Moon meteoritic seismic hum: steady state
  prediction}.
\newblock \bibinfo{journal}{Journal of Geophysical Research (Planets)}
  \bibinfo{volume}{114}, \bibinfo{pages}{12003}.
\bibitem[{Lun et~al.(1984)Lun, Savage, Jeffrey and Chepurniy}]{lun84}
\bibinfo{author}{Lun, C.}, \bibinfo{author}{Savage, S.B.},
  \bibinfo{author}{Jeffrey, D.J.}, \bibinfo{author}{Chepurniy, N.},
  \bibinfo{year}{1984}.
\newblock \bibinfo{title}{Kinetic theories for granular flow - inelastic
  particles in couette-flow and slightly inelastic particles in a general
  flowfield}.
\newblock \bibinfo{journal}{Journal of Fluid Mechanics} \bibinfo{volume}{140},
  \bibinfo{pages}{223--256}.
\bibitem[{Matsumura et~al.(2014)Matsumura, Richardson, Michel, Schwartz and
  Ballouz}]{matsumura14}
\bibinfo{author}{Matsumura, S.}, \bibinfo{author}{Richardson, D.C.},
  \bibinfo{author}{Michel, P.}, \bibinfo{author}{Schwartz, S.R.},
  \bibinfo{author}{Ballouz, R.L.}, \bibinfo{year}{2014}.
\newblock \bibinfo{title}{The brazil nut effect and its application to
  asteroids}.
\newblock \bibinfo{journal}{Monthly Notices of the Royal Astronomical Society}
  \bibinfo{volume}{443}, \bibinfo{pages}{3368--3380}.
\bibitem[{Maurel et~al.(2017)Maurel, Ballouz, Richardson, Michel and
  Schwartz}]{maurel17}
\bibinfo{author}{Maurel, C.}, \bibinfo{author}{Ballouz, R.L.},
  \bibinfo{author}{Richardson, D.C.}, \bibinfo{author}{Michel, P.},
  \bibinfo{author}{Schwartz, S.R.}, \bibinfo{year}{2017}.
\newblock \bibinfo{title}{Numerical simulations of oscillation-driven regolith
  motion: Brazil-nut effect.}
\newblock \bibinfo{journal}{Monthly Notices of the Royal Astronomical Society}
  \bibinfo{volume}{464}, \bibinfo{pages}{2866--2881}.
\bibitem[{McGarr et~al.(1969)McGarr, Latham and Gault}]{mcgarr69}
\bibinfo{author}{McGarr, A.}, \bibinfo{author}{Latham, G.},
  \bibinfo{author}{Gault, D.}, \bibinfo{year}{1969}.
\newblock \bibinfo{title}{Meteoroid impacts as sources of seismicity on the
  moon}.
\newblock \bibinfo{journal}{Journal of Geophysical Research}
  \bibinfo{volume}{74}, \bibinfo{pages}{5981--5994}.
\bibitem[{Melosh and Ryan(1997)}]{melosh97}
\bibinfo{author}{Melosh, H.}, \bibinfo{author}{Ryan, E.}, \bibinfo{year}{1997}.
\newblock \bibinfo{title}{Note: Asteroids: Shattered but not dispersed}.
\newblock \bibinfo{journal}{Icarus} \bibinfo{volume}{129},
  \bibinfo{pages}{562--564}.
\bibitem[{Melosh(1989)}]{melosh89}
\bibinfo{author}{Melosh, H.J.}, \bibinfo{year}{1989}.
\newblock \bibinfo{title}{Impact cratering: a geologic process}.
\newblock Oxford Monographs on Geology and Geophysics,
  \bibinfo{publisher}{Oxford University Press, Oxford England}.
\bibitem[{Meschede et~al.(2011)Meschede, Myhrvold and Tromp}]{meschede11}
\bibinfo{author}{Meschede, M.A.}, \bibinfo{author}{Myhrvold, C.L.},
  \bibinfo{author}{Tromp, J.}, \bibinfo{year}{2011}.
\newblock \bibinfo{title}{Antipodal focusing of seismic waves due to large
  meteorite impacts on earth}.
\newblock \bibinfo{journal}{Geophysics Journal International}
  \bibinfo{volume}{187}, \bibinfo{pages}{529--537}.
\bibitem[{Michel et~al.(2009)Michel, O'Brien, Abe and Hirata}]{michel09}
\bibinfo{author}{Michel, P.}, \bibinfo{author}{O'Brien, D.P.},
  \bibinfo{author}{Abe, S.}, \bibinfo{author}{Hirata, N.},
  \bibinfo{year}{2009}.
\newblock \bibinfo{title}{Itokawa's cratering record as observed by hayabusa:
  Implications for its age and collisional history.}
\newblock \bibinfo{journal}{Icarus} \bibinfo{volume}{200},
  \bibinfo{pages}{503--513}.
\bibitem[{Miyamoto et~al.(2017)Miyamoto, Yano, Scheeres, Abe, Barnouin-Jha,
  Cheng, Demura, Gaskell, Hirata, Ishiguro, Michikami, Nakamura, Nakamura,
  Saito and Sasaki}]{miyamoto07}
\bibinfo{author}{Miyamoto, H.}, \bibinfo{author}{Yano, H.},
  \bibinfo{author}{Scheeres, D.J.}, \bibinfo{author}{Abe, S.},
  \bibinfo{author}{Barnouin-Jha, O.}, \bibinfo{author}{Cheng, A.F.},
  \bibinfo{author}{Demura, H.}, \bibinfo{author}{Gaskell, R.W.},
  \bibinfo{author}{Hirata, N.}, \bibinfo{author}{Ishiguro, M.},
  \bibinfo{author}{Michikami, T.}, \bibinfo{author}{Nakamura, A.M.},
  \bibinfo{author}{Nakamura, R.}, \bibinfo{author}{Saito, J.},
  \bibinfo{author}{Sasaki, S.}, \bibinfo{year}{2017}.
\newblock \bibinfo{title}{Regolith migration and sorting on asteroid itokawa}.
\newblock \bibinfo{journal}{Science} \bibinfo{volume}{316},
  \bibinfo{pages}{1011--1014}.
\bibitem[{Montagner and Roult(2008)}]{montagner08}
\bibinfo{author}{Montagner, J.}, \bibinfo{author}{Roult, G.},
  \bibinfo{year}{2008}.
\newblock \bibinfo{title}{Normal modes of the earth}.
\newblock \bibinfo{journal}{Journal of Physics: Conference Series}
  \bibinfo{volume}{118}, \bibinfo{pages}{012004}.
\bibitem[{Murdoch et~al.(2017)Murdoch, Hempel, Pou, Cadu, Garcia, Mimoun,
  Margerin and Karatekin}]{murdoch17}
\bibinfo{author}{Murdoch, N.}, \bibinfo{author}{Hempel, S.},
  \bibinfo{author}{Pou, L.}, \bibinfo{author}{Cadu, A.},
  \bibinfo{author}{Garcia, R.}, \bibinfo{author}{Mimoun, D.},
  \bibinfo{author}{Margerin, L.}, \bibinfo{author}{Karatekin, O.},
  \bibinfo{year}{2017}.
\newblock \bibinfo{title}{Probing the internal structure of the asteriod
  didymoon with a passive seismic investigation}.
\newblock \bibinfo{journal}{Planetary and Space Science} \bibinfo{volume}{144},
  \bibinfo{pages}{89--105}.
\bibitem[{Nakamura(1976)}]{nakamura76}
\bibinfo{author}{Nakamura, Y.}, \bibinfo{year}{1976}.
\newblock \bibinfo{title}{Seismic energy transmission in the lunar surface zone
  determined from signals generated by movement of lunar rovers}.
\newblock \bibinfo{journal}{Bulletin of the Seismological Society of America}
  \bibinfo{volume}{66}, \bibinfo{pages}{593--606}.
\bibitem[{{Nolan} et~al.(1992){Nolan}, {Asphaug} and {Greenberg}}]{nolan92}
\bibinfo{author}{{Nolan}, M.C.}, \bibinfo{author}{{Asphaug}, E.},
  \bibinfo{author}{{Greenberg}, R.}, \bibinfo{year}{1992}.
\newblock \bibinfo{title}{{Numerical Simulation of Impacts on Small
  Asteroids}}, in: \bibinfo{booktitle}{AAS/Division for Planetary Sciences
  Meeting Abstracts \#24}, p. \bibinfo{pages}{959}.
\bibitem[{Nolan et~al.(2013)Nolan, Magri, Howell, Benner, Giorgini,
  Hergenrother, Hudson, Lauretta, Margot, Ostro and Scheeres}]{nolan13}
\bibinfo{author}{Nolan, M.C.}, \bibinfo{author}{Magri, C.},
  \bibinfo{author}{Howell, E.S.}, \bibinfo{author}{Benner, L.A.},
  \bibinfo{author}{Giorgini, J.D.}, \bibinfo{author}{Hergenrother, C.W.},
  \bibinfo{author}{Hudson, R.S.}, \bibinfo{author}{Lauretta, D.S.},
  \bibinfo{author}{Margot, J.L.}, \bibinfo{author}{Ostro, S.J.},
  \bibinfo{author}{Scheeres, D.J.}, \bibinfo{year}{2013}.
\newblock \bibinfo{title}{Shape model and surface properties of the osiris-rex
  target asteroid (101955) bennu from radar and lightcurve observations}.
\newblock \bibinfo{journal}{Icarus} \bibinfo{volume}{226}, \bibinfo{pages}{629
  -- 640}.
\newblock \URLprefix
  \url{http://www.sciencedirect.com/science/article/pii/S0019103513002285},
  \DOIprefix\doi{https://doi.org/10.1016/j.icarus.2013.05.028}.
\bibitem[{O'Brien and Greenberg(2005)}]{obrien05}
\bibinfo{author}{O'Brien, D.P.}, \bibinfo{author}{Greenberg, R.},
  \bibinfo{year}{2005}.
\newblock \bibinfo{title}{The collisional and dynamical evolution of the main
  belt and nea size distributions}.
\newblock \bibinfo{journal}{Icarus} \bibinfo{volume}{178},
  \bibinfo{pages}{179--212}.
\bibitem[{O'Donovan et~al.(2016)O'Donovan, Ibraim, O'Sullivan, Hamlin, Wood and
  Marketos}]{odonovan16}
\bibinfo{author}{O'Donovan, J.}, \bibinfo{author}{Ibraim, E.},
  \bibinfo{author}{O'Sullivan, C.}, \bibinfo{author}{Hamlin, S.},
  \bibinfo{author}{Wood, D.M.}, \bibinfo{author}{Marketos, G.},
  \bibinfo{year}{2016}.
\newblock \bibinfo{title}{Micromechanics of seismic wave propagation in
  granular materials}.
\newblock \bibinfo{journal}{Granular Matter} \bibinfo{volume}{18},
  \bibinfo{pages}{56}.
\bibitem[{Ogawa et~al.(1980)Ogawa, Umemura and Oshima}]{ogawa80}
\bibinfo{author}{Ogawa, S.}, \bibinfo{author}{Umemura, A.},
  \bibinfo{author}{Oshima, N.}, \bibinfo{year}{1980}.
\newblock \bibinfo{title}{On the equations of fully fluidized granular
  materials}.
\newblock \bibinfo{journal}{Journal of Applied Mathematics and Physics (ZAMP)}
  \bibinfo{volume}{31}, \bibinfo{pages}{483--493}.
\bibitem[{Ouaguenouni and Roux(1997)}]{ouaguenouni97}
\bibinfo{author}{Ouaguenouni, S.}, \bibinfo{author}{Roux, J.N.},
  \bibinfo{year}{1997}.
\newblock \bibinfo{title}{Force distribution in frictionless granular packings
  at rigidity threshold}.
\newblock \bibinfo{journal}{Europhysics Letters} \bibinfo{volume}{39},
  \bibinfo{pages}{117}.
\newblock \URLprefix \url{http://stacks.iop.org/0295-5075/39/i=2/a=117}.
\bibitem[{Pereraa et~al.(2016)Pereraa, Jackson, Asphaug and Ballouz}]{perera16}
\bibinfo{author}{Pereraa, V.}, \bibinfo{author}{Jackson, A.P.},
  \bibinfo{author}{Asphaug, E.}, \bibinfo{author}{Ballouz, R.L.},
  \bibinfo{year}{2016}.
\newblock \bibinfo{title}{The spherical brazil nut effect and its significance
  to asteroids}.
\newblock \bibinfo{journal}{Icarus} \bibinfo{volume}{278},
  \bibinfo{pages}{194--203}.
\bibitem[{Pouliquen(1999)}]{pouliquen99}
\bibinfo{author}{Pouliquen, O.}, \bibinfo{year}{1999}.
\newblock \bibinfo{title}{Scaling laws in granular flows down rough inclined
  planes}.
\newblock \bibinfo{journal}{Physics of Fluids} \bibinfo{volume}{11},
  \bibinfo{pages}{542--548}.
\bibitem[{Press and Teukolsky(1977)}]{press77}
\bibinfo{author}{Press, W.}, \bibinfo{author}{Teukolsky, S.A.},
  \bibinfo{year}{1977}.
\newblock \bibinfo{title}{On formation of close binaries by two-body tidal
  capture}.
\newblock \bibinfo{journal}{Astrophysical Journal} \bibinfo{volume}{213},
  \bibinfo{pages}{183--192}.
\bibitem[{Quillen et~al.(2016a)Quillen, Giannella, Shaw and
  Ebinger}]{quillen16_crust}
\bibinfo{author}{Quillen, A.C.}, \bibinfo{author}{Giannella, D.},
  \bibinfo{author}{Shaw, J.G.}, \bibinfo{author}{Ebinger, C.},
  \bibinfo{year}{2016}a.
\newblock \bibinfo{title}{Crustal failure on icy moons from a strong tidal
  encounter}.
\newblock \bibinfo{journal}{Icarus} \bibinfo{volume}{275},
  \bibinfo{pages}{267--280}.
\bibitem[{Quillen et~al.(2016b)Quillen, Kueter-Young, Frouard and
  Ragozzine}]{quillen16_haumea}
\bibinfo{author}{Quillen, A.C.}, \bibinfo{author}{Kueter-Young, A.},
  \bibinfo{author}{Frouard, J.}, \bibinfo{author}{Ragozzine, D.},
  \bibinfo{year}{2016}b.
\newblock \bibinfo{title}{Tidal spin down rates of homogeneous triaxial
  viscoelastic bodies}.
\newblock \bibinfo{journal}{Monthly Notices of the Royal Astronomical Society}
  \bibinfo{volume}{463}, \bibinfo{pages}{1543--1553}.
\bibitem[{Quillen et~al.(2017)Quillen, Nichols-Fleming, Chen and
  Noyelles}]{quillen17_pluto}
\bibinfo{author}{Quillen, A.C.}, \bibinfo{author}{Nichols-Fleming, F.},
  \bibinfo{author}{Chen, Y.Y.}, \bibinfo{author}{Noyelles, B.},
  \bibinfo{year}{2017}.
\newblock \bibinfo{title}{Obliquity evolution of the minor satellites of pluto
  and charon}.
\newblock \bibinfo{journal}{Icarus} \bibinfo{volume}{293},
  \bibinfo{pages}{94--113}.
\bibitem[{Rein and Liu(2012)}]{rebound}
\bibinfo{author}{Rein, H.}, \bibinfo{author}{Liu, S.F.}, \bibinfo{year}{2012}.
\newblock \bibinfo{title}{Rebound: an open-source multi-purpose n-body code for
  collisional dynamics}.
\newblock \bibinfo{journal}{Astronomy and Astrophysics} \bibinfo{volume}{537},
  \bibinfo{pages}{A128}.
\bibitem[{Richardson et~al.(2009)Richardson, Michel, Walsh and
  Flynn}]{richardson09}
\bibinfo{author}{Richardson, D.}, \bibinfo{author}{Michel, P.},
  \bibinfo{author}{Walsh, K.}, \bibinfo{author}{Flynn, K.},
  \bibinfo{year}{2009}.
\newblock \bibinfo{title}{Numerical simulations of asteroids modelled as
  gravitational aggregates with cohesion}.
\newblock \bibinfo{journal}{Planetary and Space Science} \bibinfo{volume}{57},
  \bibinfo{pages}{183 -- 192}.
\newblock \URLprefix
  \url{http://www.sciencedirect.com/science/article/pii/S0032063308001037},
  \DOIprefix\doi{https://doi.org/10.1016/j.pss.2008.04.015}.
  \bibinfo{note}{catastrophic Disruption in the Solar System}.
\bibitem[{Richardson et~al.(2004)Richardson, Melosh and
  Greenberg}]{richardson04}
\bibinfo{author}{Richardson, J.E.}, \bibinfo{author}{Melosh, H.J.},
  \bibinfo{author}{Greenberg, R.}, \bibinfo{year}{2004}.
\newblock \bibinfo{title}{Impact-induced seismic activity on asteroid 433eros:
  A surface modification process}.
\newblock \bibinfo{journal}{Science} \bibinfo{volume}{306},
  \bibinfo{pages}{1526--1529}.
\bibitem[{Richardson et~al.(2005)Richardson, Melosh, Greenberg and
  O'Brien}]{richardson05}
\bibinfo{author}{Richardson, J.J.}, \bibinfo{author}{Melosh, H.J.},
  \bibinfo{author}{Greenberg, R.J.}, \bibinfo{author}{O'Brien, D.P.},
  \bibinfo{year}{2005}.
\newblock \bibinfo{title}{The global effects of impact-induced seismic activity
  on fractured asteroid surface morphology}.
\newblock \bibinfo{journal}{Icarus} \bibinfo{volume}{179},
  \bibinfo{pages}{325--349}.
\bibitem[{Robinson et~al.(2001)Robinson, Thomas, Veverka, Murchie and
  Carcich}]{robinson01}
\bibinfo{author}{Robinson, M.S.}, \bibinfo{author}{Thomas, P.C.},
  \bibinfo{author}{Veverka, J.}, \bibinfo{author}{Murchie, S.},
  \bibinfo{author}{Carcich, B.}, \bibinfo{year}{2001}.
\newblock \bibinfo{title}{The nature of ponded deposits on eros}.
\newblock \bibinfo{journal}{Nature} \bibinfo{volume}{413},
  \bibinfo{pages}{396--400}.
\newblock \URLprefix \url{http://dx.doi.org/10.1038/35096518},
  \DOIprefix\doi{10.1038/35096518}.
\bibitem[{S\'anchez and Scheeres(2011)}]{sanchez11}
\bibinfo{author}{S\'anchez, P.}, \bibinfo{author}{Scheeres, D.J.},
  \bibinfo{year}{2011}.
\newblock \bibinfo{title}{Simulating asteroid rubble piles with a
  self-gravitating soft-sphere distinct element method model}.
\newblock \bibinfo{journal}{Astrophysical Journal} \bibinfo{volume}{727},
  \bibinfo{pages}{120--134}.
\bibitem[{S\'anchez and Scheeres(2014)}]{sanchez14}
\bibinfo{author}{S\'anchez, P.}, \bibinfo{author}{Scheeres, D.J.},
  \bibinfo{year}{2014}.
\newblock \bibinfo{title}{The strength of regolith and rubble pile asteroids}.
\newblock \bibinfo{journal}{Meteoritics \& Planetary Science}
  \bibinfo{volume}{49}, \bibinfo{pages}{788--811}.
\newblock \URLprefix \url{http://dx.doi.org/10.1111/maps.12293},
  \DOIprefix\doi{10.1111/maps.12293}.
\bibitem[{Savage and Hutter(1989)}]{savage89}
\bibinfo{author}{Savage, S.}, \bibinfo{author}{Hutter, K.},
  \bibinfo{year}{1989}.
\newblock \bibinfo{title}{The motion of a finite mass of granular material down
  a rough incline}.
\newblock \bibinfo{journal}{Journal of Fluid Mechanics} \bibinfo{volume}{199},
  \bibinfo{pages}{177--215}.
\bibitem[{Savage(1988)}]{savage88}
\bibinfo{author}{Savage, S.B.}, \bibinfo{year}{1988}.
\newblock \bibinfo{title}{Streaming motions in a bed of vibrationally fluidized
  dry granular material}.
\newblock \bibinfo{journal}{Journal of Fluid Mechanics} \bibinfo{volume}{194},
  \bibinfo{pages}{457--478}.
\bibitem[{Scheeres et~al.(2006)Scheeres, Fahnestock, Ostro, Margot, Benner,
  Broschart, Bellerose, Giorgini, Nolan, Magri, Pravec, Scheirich, Rose,
  Jurgens, Jong and Suzuki}]{scheeres06}
\bibinfo{author}{Scheeres, D.}, \bibinfo{author}{Fahnestock, E.G.},
  \bibinfo{author}{Ostro, S.J.}, \bibinfo{author}{Margot, J.L.},
  \bibinfo{author}{Benner, L.A.M.}, \bibinfo{author}{Broschart, S.B.},
  \bibinfo{author}{Bellerose, J.}, \bibinfo{author}{Giorgini, J.D.},
  \bibinfo{author}{Nolan, M.C.}, \bibinfo{author}{Magri, C.},
  \bibinfo{author}{Pravec, P.}, \bibinfo{author}{Scheirich, P.},
  \bibinfo{author}{Rose, R.}, \bibinfo{author}{Jurgens, R.F.},
  \bibinfo{author}{Jong, E.M.D.}, \bibinfo{author}{Suzuki, S.},
  \bibinfo{year}{2006}.
\newblock \bibinfo{title}{Dynamical configuration of binary near-earth asteroid
  (66391) 1999 kw4}.
\newblock \bibinfo{journal}{Science} \bibinfo{volume}{314},
  \bibinfo{pages}{1280--1283}.
\bibitem[{Scheeres et~al.(2016)Scheeres, Hesar, Tardivel, Hirabayashi,
  Farnocchia, McMahon, Chesley, Barnouin, Binzel, Bottke, Daly, Emery,
  Hergenrother, Lauretta, Marshall, Michel, Nolan and Walsh}]{scheeres16}
\bibinfo{author}{Scheeres, D.}, \bibinfo{author}{Hesar, S.G.},
  \bibinfo{author}{Tardivel, S.}, \bibinfo{author}{Hirabayashi, M.},
  \bibinfo{author}{Farnocchia, D.}, \bibinfo{author}{McMahon, J.W.},
  \bibinfo{author}{Chesley, S.R.}, \bibinfo{author}{Barnouin, O.},
  \bibinfo{author}{Binzel, R.P.}, \bibinfo{author}{Bottke, W.F.},
  \bibinfo{author}{Daly, M.G.}, \bibinfo{author}{Emery, J.P.},
  \bibinfo{author}{Hergenrother, C.W.}, \bibinfo{author}{Lauretta, D.S.},
  \bibinfo{author}{Marshall, J.R.}, \bibinfo{author}{Michel, P.},
  \bibinfo{author}{Nolan, M.C.}, \bibinfo{author}{Walsh, K.J.},
  \bibinfo{year}{2016}.
\newblock \bibinfo{title}{The geophysical environment of bennu}.
\newblock \bibinfo{journal}{Icarus} \bibinfo{volume}{276},
  \bibinfo{pages}{116--140}.
\bibitem[{{Scheeres} and {S{\'a}nchez}(2018)}]{scheeres18}
\bibinfo{author}{{Scheeres}, D.J.}, \bibinfo{author}{{S{\'a}nchez}, P.},
  \bibinfo{year}{2018}.
\newblock \bibinfo{title}{{Implications of cohesive strength in asteroid
  interiors and surfaces and its measurement}}.
\newblock \bibinfo{journal}{Progress in Earth and Planetary Science}
  \bibinfo{volume}{5}, \bibinfo{pages}{25}.
\newblock \DOIprefix\doi{10.1186/s40645-018-0182-9}.
\bibitem[{Schr\"oder et~al.(2014)Schr\"oder, S, Keller, Raymond and
  Russell}]{schroder14}
\bibinfo{author}{Schr\"oder, S.E.}, \bibinfo{author}{S, M.},
  \bibinfo{author}{Keller, H.}, \bibinfo{author}{Raymond, C.},
  \bibinfo{author}{Russell, C.}, \bibinfo{year}{2014}.
\newblock \bibinfo{title}{Resolved photometry of vesta reveals physical
  properties of crater regolith}.
\newblock \bibinfo{journal}{Planetary and Space Science} \bibinfo{volume}{103},
  \bibinfo{pages}{66--81}.
\bibitem[{Schultz and Gault(1975)}]{schultz75}
\bibinfo{author}{Schultz, P.H.}, \bibinfo{author}{Gault, D.E.},
  \bibinfo{year}{1975}.
\newblock \bibinfo{title}{Seismic effects from major basin formations on the
  moon and mercury}.
\newblock \bibinfo{journal}{Moon} \bibinfo{volume}{12},
  \bibinfo{pages}{159--177}.
\bibitem[{{Schwartz} et~al.(2013){Schwartz}, {Michel} and
  {Richardson}}]{schwartz13}
\bibinfo{author}{{Schwartz}, S.R.}, \bibinfo{author}{{Michel}, P.},
  \bibinfo{author}{{Richardson}, D.C.}, \bibinfo{year}{2013}.
\newblock \bibinfo{title}{{Numerically simulating impact disruptions of
  cohesive glass bead agglomerates using the soft-sphere discrete element
  method}}.
\newblock \bibinfo{journal}{Icarus} \bibinfo{volume}{226},
  \bibinfo{pages}{67--76}.
\newblock \DOIprefix\doi{10.1016/j.icarus.2013.05.007}.
\bibitem[{Schwartz et~al.(2014)Schwartz, Michel, Richardson and
  Yano}]{schwartz14}
\bibinfo{author}{Schwartz, S.R.}, \bibinfo{author}{Michel, P.},
  \bibinfo{author}{Richardson, D.C.}, \bibinfo{author}{Yano, H.},
  \bibinfo{year}{2014}.
\newblock \bibinfo{title}{Low-speed impact simulations into regolith in support
  of asteroid sampling mechanism design i: Comparison with 1-g experiments}.
\newblock \bibinfo{journal}{Planetary and Space Science} \bibinfo{volume}{103},
  \bibinfo{pages}{174--183}.
\bibitem[{Shishkin(2007)}]{shishkin07}
\bibinfo{author}{Shishkin, N.I.}, \bibinfo{year}{2007}.
\newblock \bibinfo{title}{Seismic efficiency of a contact explosion and a high
  velocity impact}.
\newblock \bibinfo{journal}{Journal of Applied Mechanics and Technical Physics}
  \bibinfo{volume}{48}, \bibinfo{pages}{145--152}.
\bibitem[{Silbert et~al.(2003.)Silbert, Landry and Grest}]{silbert03}
\bibinfo{author}{Silbert, L.E.}, \bibinfo{author}{Landry, J.W.},
  \bibinfo{author}{Grest, G.S.}, \bibinfo{year}{2003.}
\newblock \bibinfo{title}{Granular flow down a rough inclined plane: transition
  between thin and thick piles.}
\newblock \bibinfo{journal}{Physics of Fluids} \bibinfo{volume}{15},
  \bibinfo{pages}{1--10}.
\bibitem[{Snieder and Wijk(2015)}]{snieder15}
\bibinfo{author}{Snieder, R.}, \bibinfo{author}{Wijk, K.V.},
  \bibinfo{year}{2015}.
\newblock \bibinfo{title}{A Guided Tour of Mathematical Methods for the
  Physical Sciences. 3rd ed.}
\newblock \bibinfo{publisher}{Cambridge University Press, Cambridge, UK.}
\bibitem[{Tancredi et~al.(2012)Tancredi, Maciel, Heredia, Richeri and
  Nesmachnow}]{tancredi12}
\bibinfo{author}{Tancredi, G.}, \bibinfo{author}{Maciel, A.},
  \bibinfo{author}{Heredia, L.}, \bibinfo{author}{Richeri, P.},
  \bibinfo{author}{Nesmachnow, S.}, \bibinfo{year}{2012}.
\newblock \bibinfo{title}{Granular physics in low-gravity environments using
  discrete element method}.
\newblock \bibinfo{journal}{Monthly Notices of the Royal Astronomical Society}
  \bibinfo{volume}{420}, \bibinfo{pages}{3368--3380}.
\bibitem[{Tancredi et~al.(2015)Tancredi, Roland and Bruzzone}]{tancredi15}
\bibinfo{author}{Tancredi, G.}, \bibinfo{author}{Roland, S.},
  \bibinfo{author}{Bruzzone, S.}, \bibinfo{year}{2015}.
\newblock \bibinfo{title}{Distribution of boulders and the gravity potential on
  asteroid itokawa}.
\newblock \bibinfo{journal}{Icarus} \bibinfo{volume}{247},
  \bibinfo{pages}{279--290}.
\bibitem[{Tardivel et~al.(2018)Tardivel, S\'anchez and Scheeres}]{tardivel18}
\bibinfo{author}{Tardivel, S.}, \bibinfo{author}{S\'anchez, P.},
  \bibinfo{author}{Scheeres, D.}, \bibinfo{year}{2018}.
\newblock \bibinfo{title}{Equatorial cavities on asteroids, an evidence of
  fission events}.
\newblock \bibinfo{journal}{Icarus} \bibinfo{volume}{304},
  \bibinfo{pages}{192--208}.
\bibitem[{Thomas and Robinson(2005)}]{thomas05}
\bibinfo{author}{Thomas, P.C.}, \bibinfo{author}{Robinson, M.S.},
  \bibinfo{year}{2005}.
\newblock \bibinfo{title}{Seismic resurfacing by a single impact on the
  asteroid 433 eros}.
\newblock \bibinfo{journal}{Nature} \bibinfo{volume}{436},
  \bibinfo{pages}{366--369}.
\bibitem[{Titley(1966)}]{titley66}
\bibinfo{author}{Titley, S.R.}, \bibinfo{year}{1966}.
\newblock \bibinfo{title}{Seismic energy as an agent of morphologic
  modification on the Moon.}
\newblock \bibinfo{type}{Technical Report}. United States Geological Survey,
  Flagstaff, AZ.
\bibitem[{Toks\"oz et~al.(1974)Toks\"oz, Dainty, Solomon and
  Anderson}]{toksoz74}
\bibinfo{author}{Toks\"oz, M.N.}, \bibinfo{author}{Dainty, A.M.},
  \bibinfo{author}{Solomon, S.}, \bibinfo{author}{Anderson, K.},
  \bibinfo{year}{1974}.
\newblock \bibinfo{title}{Structure of the moon}.
\newblock \bibinfo{journal}{Reviews of Geophysics} \bibinfo{volume}{12},
  \bibinfo{pages}{539--567}.
\bibitem[{Veverka et~al.(2001)Veverka, Thomas, Robinson, Murchie, Chapman,
  Bell, Harch, Merline, Bell, Bussey, Carcich, Cheng, Clark, Domingue, Dunham,
  Farquhar, Gaffey, Hawkins, Izenberg, Joseph, Kirk, Li, Lucey, Malin,
  McFadden, Miller, Owen, Peterson, Prockter, Warren, Wellnitz, Williams and
  Yeomans}]{veverka01}
\bibinfo{author}{Veverka, J.}, \bibinfo{author}{Thomas, P.C.},
  \bibinfo{author}{Robinson, M.}, \bibinfo{author}{Murchie, S.},
  \bibinfo{author}{Chapman, C.}, \bibinfo{author}{Bell, M.},
  \bibinfo{author}{Harch, A.}, \bibinfo{author}{Merline, W.J.},
  \bibinfo{author}{Bell, J.F.}, \bibinfo{author}{Bussey, B.},
  \bibinfo{author}{Carcich, B.}, \bibinfo{author}{Cheng, A.},
  \bibinfo{author}{Clark, B.}, \bibinfo{author}{Domingue, D.},
  \bibinfo{author}{Dunham, D.}, \bibinfo{author}{Farquhar, R.},
  \bibinfo{author}{Gaffey, M.J.}, \bibinfo{author}{Hawkins, E.},
  \bibinfo{author}{Izenberg, N.}, \bibinfo{author}{Joseph, J.},
  \bibinfo{author}{Kirk, R.}, \bibinfo{author}{Li, H.}, \bibinfo{author}{Lucey,
  P.}, \bibinfo{author}{Malin, M.}, \bibinfo{author}{McFadden, L.},
  \bibinfo{author}{Miller, J.K.}, \bibinfo{author}{Owen, W.M.},
  \bibinfo{author}{Peterson, C.}, \bibinfo{author}{Prockter, L.},
  \bibinfo{author}{Warren, J.}, \bibinfo{author}{Wellnitz, D.},
  \bibinfo{author}{Williams, B.G.}, \bibinfo{author}{Yeomans, D.K.},
  \bibinfo{year}{2001}.
\newblock \bibinfo{title}{Imaging of small-scale features on 433 eros from
  near: Evidence for a complex regolith}.
\newblock \bibinfo{journal}{Science} \bibinfo{volume}{292},
  \bibinfo{pages}{484--488}.
\newblock \URLprefix \url{http://science.sciencemag.org/content/292/5516/484},
  \DOIprefix\doi{10.1126/science.1058651},
  \href{http://arxiv.org/abs/http://science.sciencemag.org/content/292/5516/484.full.pdf}{{\tt
  arXiv:http://science.sciencemag.org/content/292/5516/484.full.pdf}}.
\bibitem[{Walker and Huebner(2004)}]{walker04}
\bibinfo{author}{Walker, J.}, \bibinfo{author}{Huebner, W.},
  \bibinfo{year}{2004}.
\newblock \bibinfo{title}{Seismological investigation of asteroid and comet
  interiors}, in: \bibinfo{editor}{Belton, M.J.S.} (Ed.),
  \bibinfo{booktitle}{Mitigation of Hazardous Comets and Asteroids},
  \bibinfo{publisher}{Cambridge Univ. Press, Cambridge, U. K.}. pp.
  \bibinfo{pages}{234--265}.
\bibitem[{Walsh et~al.(2008)Walsh, Richardson and Michel}]{walsh08}
\bibinfo{author}{Walsh, K.J.}, \bibinfo{author}{Richardson, D.C.},
  \bibinfo{author}{Michel, P.}, \bibinfo{year}{2008}.
\newblock \bibinfo{title}{Rotational breakup as the origin of small binary
  asteroids}.
\newblock \bibinfo{journal}{Nature} \bibinfo{volume}{454},
  \bibinfo{pages}{188--191}.
\bibitem[{Walsh et~al.(2012)Walsh, Richardson and Michel}]{walsh12}
\bibinfo{author}{Walsh, K.K.}, \bibinfo{author}{Richardson, D.C.},
  \bibinfo{author}{Michel, P.}, \bibinfo{year}{2012}.
\newblock \bibinfo{title}{Spin-up of rubble-pile asteroids: Disruption,
  satellite formation, and equilibrium shapes}.
\newblock \bibinfo{journal}{Icarus} \bibinfo{volume}{220},
  \bibinfo{pages}{514--529}.
\bibitem[{Warr et~al.(1995)Warr, Huntley and Jacques}]{warr95}
\bibinfo{author}{Warr, S.}, \bibinfo{author}{Huntley, J.M.},
  \bibinfo{author}{Jacques, T.H.}, \bibinfo{year}{1995}.
\newblock \bibinfo{title}{Fluidization of a two-dimensional granular system:
  Experimental study and scaling behavior}.
\newblock \bibinfo{journal}{Phys. Rev. E} \bibinfo{volume}{52},
  \bibinfo{pages}{5583--5595}.
\bibitem[{Wolf(1944)}]{wolf44}
\bibinfo{author}{Wolf, A.}, \bibinfo{year}{1944}.
\newblock \bibinfo{title}{The equation of motion of a geophone on the surface
  of an elastic earth}.
\newblock \bibinfo{journal}{Geophysics} \bibinfo{volume}{9},
  \bibinfo{pages}{29--34}.
\bibitem[{Yamada et~al.(2016)Yamada, Ando, Morota and Katsuragi}]{yamada16}
\bibinfo{author}{Yamada, T.M.}, \bibinfo{author}{Ando, K.},
  \bibinfo{author}{Morota, T.}, \bibinfo{author}{Katsuragi, H.},
  \bibinfo{year}{2016}.
\newblock \bibinfo{title}{Timescale of asteroid resurfacing by regolith
  convection resulting from the impact-induced global seismic shaking}.
\newblock \bibinfo{journal}{Icarus} \bibinfo{volume}{272},
  \bibinfo{pages}{165--177}.
\bibitem[{Yasui et~al.(2015)Yasui, Matsumoto and Arakawa}]{yasui15}
\bibinfo{author}{Yasui, M.}, \bibinfo{author}{Matsumoto, E.},
  \bibinfo{author}{Arakawa, M.}, \bibinfo{year}{2015}.
\newblock \bibinfo{title}{Experimental study on impact-induced seismic wave
  propagation through granular materials}.
\newblock \bibinfo{journal}{Icarus} \bibinfo{volume}{260},
  \bibinfo{pages}{320--331}.
\bibitem[{Yu et~al.(2014)Yu, Richardson, Michel, Schwartz and Ballouz}]{yu14}
\bibinfo{author}{Yu, Y.}, \bibinfo{author}{Richardson, D.C.},
  \bibinfo{author}{Michel, P.}, \bibinfo{author}{Schwartz, S.R.},
  \bibinfo{author}{Ballouz, R.L.}, \bibinfo{year}{2014}.
\newblock \bibinfo{title}{Numerical predictions of surface effects during the
  2029 close approach of asteroid 99942 apophis}.
\newblock \bibinfo{journal}{Icarus} \bibinfo{volume}{242},
  \bibinfo{pages}{82--96}.

\end{thebibliography}
\end{document}